\documentclass[letterpaper, aps, prl, twocolumn, superscriptaddress,
showpacs,nofootinbib, 10pt]{revtex4-1}

\usepackage{graphicx}%
\usepackage{dcolumn}%
\usepackage{bm}%
\usepackage[linktocpage,colorlinks,pdfusetitle,allcolors=blue]{hyperref}
\usepackage{multirow}
\usepackage{amsmath}
\usepackage{siunitx}
\usepackage{booktabs}

\usepackage{mathtools}

\DeclareSIUnit[]\admmass{\text{\ensuremath{M}}}
\DeclareSIUnit[]\sunmass{\text{\ensuremath{M_{\odot}}}}
\DeclareSIUnit[]\erg{\;\text{erg}\;}
\DeclareSIUnit[]\pc{\text{pc}}
\DeclareSIUnit[]\Mpc{\text{Mpc}}
\DeclareSIUnit[]\admsun{(\si{\admmass}\slash\SI{65}{\sunmass})}
\DeclareSIUnit[]\Gauss{G}

\DeclarePairedDelimiter\abs{\lvert}{\rvert}%

\renewcommand{\d}[1]{\ensuremath{\operatorname{d}\!{#1}}}

\begin{document}

\title{General relativistic simulations of the quasi-circular inspiral and
  merger of charged black holes: GW150914 and fundamental physics implications}

\author{Gabriele Bozzola}
\email{gabrielebozzola@arizona.edu}
\affiliation{Department of Astronomy, University of Arizona, Tucson, AZ, USA}
\author{Vasileios Paschalidis}
\email{vpaschal@arizona.edu}
\affiliation{Departments of Astronomy and Physics, University of Arizona, Tucson, AZ, USA}

\date{\today}

\begin{abstract}

  We perform general-relativistic simulations of charged black holes targeting
  GW150914. We show that the inspiral is most efficient for detecting black hole
  charge through gravitational waves and that GW150914 is compatible with having
  charge-to-mass ratio as high as \num{0.3}. Our work applies to electric and
  magnetic charge, and to theories with black holes endowed with U(1) (hidden or
  dark) charges. Using our results we place an upper bound on the deviation from
  general relativity in the dynamical, strong-field regime of the so-called
  theory of MOdified Gravity (MOG).
\end{abstract}

\maketitle

\paragraph{\textbf{Introduction}}

According to the ``no-hair''
conjecture~\cite{Israel1967,Carter1971,Robinson1974,1972CMaPh..25..152H,Hansen:1974zz,Chrusciel2012},
general relativistic black holes are described by four parameters: mass, angular
momentum, electric and magnetic charge. It is assumed, often implicitly, that
astrophysical black holes have negligible charge because of the expectation that
they would quickly discharge due to the interaction with a highly conducting
gaseous environment or by the spontaneous production of electron-positron pairs
\cite{Wald1974,Gibbons1975,Eardley1975,Hanni1982,Gong2019,Pan2019}. However,
observational data unequivocally supporting this expectation are currently
absent, and any existing constraints on black hole charge depend crucially on
the assumptions of the models employed (e.g.~\cite{Iorio2012,Zajacek2018}).
Gravitational-wave observations offer a model-independent path to constraining
the charge of astrophysical black holes. The electromagnetic fields influence
the spacetime, altering the gravitational-wave emission compared to an uncharged
binary. These deviations are accurately modeled in Einstein-Maxwell theory, and
are potentially detectable by LIGO-Virgo and future gravitational-wave
observatories. As we will discuss below, the word ``charge'' here is an umbrella
term that includes, among other things, electric or magnetic charge, dark
charge, or gravitational charge due to modifications to general relativity.

In this letter, we initiate a robust program for constraining black hole charge
by combining LIGO-Virgo observations with novel numerical relativity
simulations. Our focus here is on event GW150914~\cite{Abbott2016}.\footnote{The
  possibility that GW150914 involved charged black holes has been invoked
  \cite{Zhang2016,Liebling2016,Fraschetti2018} to explain the observation of a
  coincident electromagnetic signal by Fermi-GBM
  \cite{Connaughton2016,Connaughton2018}. This association is debated as others
  satellites did not detect the event
  \cite{Abbott2016f,Hurley2016,Evans2016,Savchenko2016}.} Using the event's sky
location and the calibrated LIGO noise, we compute the \emph{mismatch} (defined
later) between the uncharged case and various charged ones. The observed
signal-to-noise ratio sets a threshold mismatch above which two waveforms are
distinguishable
\cite{Flanagan1998b,Lindblom2008,McWilliams2010,Abbott2017h}. Hence, assuming
that the observed waveform is described by uncharged, non-spinning black holes,
we find the minimum charge that would be detectable by LIGO.

For uncharged binaries, when black hole spin is neglected and the mass-ratio is
fixed, knowing one ``mass'' parameter determines the entire gravitational
waveform. We will use here the \emph{chirp mass} $\mathcal{M}$
\cite{Maggiore2007}. In the case of inspirals of charged binaries, this
parameter can be degenerate with the charge itself
\cite{Cardoso2016b,Christiansen2020,Liu2020,Cardoso2020Erratum}. This can be
understood as follows: In Newtonian physics, gravity and electromagnetism are
both central potentials, so the electrostatic force can be accounted for by
introducing an effective Newton constant $\widetilde{G}$. Consider two bodies
with mass $m_1$, $m_2$ and charge $q_1 = \lambda_1 m_1$, $q_2 = \lambda_2 m_2$
($\lambda$ being the charge-to-mass ratio); the dynamics of the system is
indistinguishable from one with uncharged bodies with gravitational constant
$\widetilde{G} = (1 - \lambda_1 \lambda_2)G$. Since the relationship between
chirp mass and gravitational-wave frequency evolution involves Newton's
constant, introducing charges corresponds to rescaling the chirp mass while
keeping $G$ fixed. This degeneracy is broken by electromagnetic radiation
reaction and the field self-gravity.

Adopting the effective Newton constant approach, previous studies
\cite{Cardoso2016b,Christiansen2020,Liu2020,Liu2020b,Cardoso2020Erratum,Wang2020}
constructed Newtonian-based waveforms by considering the Keplerian motion of two
charged bodies and accounting for loss of energy via quadrupolar emission of
gravitational waves and dipolar emission of electromagnetic ones. The authors of
\cite{Christiansen2020} computed the bias in the binary parameters due to the
charge-chirp mass degeneracy. With similar tools, \cite{Wang2020} performed a
full Bayesian analysis with Gaussian noise to place preliminary constraints on
charge using events in the first gravitational wave transient
catalog~\cite{LIGOVirgo2018}. Alternatively, the dipole can be constrained
directly by adding a $-1$PN (Post-Newtonian) term to describe the loss of energy
due to dipole emission \cite{Wang2020}, as first done for modified theories of
gravity \cite{Barausse2016}. In \cite{Yunes2016, Barausse2016}, it was found
that the dipole can be constrained more effectively in the inspiral (also noted
in \cite{Cardoso2016b, Cardoso2020Erratum} with explicit reference to
charges). One of the main limitations of these (post)-Newtonian methods is that
they strictly apply only to the early inspiral. However, binaries like GW150914
are in the regime where numerical relativity simulations are necessary for
accurate modeling~\cite{Abbott2016}. Therefore, existing constraints on black
hole charge in events where only a few orbits to merger have been detected are
at best preliminary. Moreover, the effective Newton constant approach does not
capture the physics in cases when only one of the two components is charged, and
when the dipole moment vanishes these previous approaches do not treat
quadrupole electromagnetic emission. This is very important because as we
demonstrate here, it is binaries with near vanishing dipole moment that place
the weakest constraint on black hole charge.

A second avenue for constraining black hole charge is through the ringdown signal.
In the context of mergers of charged black holes, this was first studied in
\cite{Cardoso2016b, Cardoso2020Erratum} in the limit of small charge, using the
method of geodesic correspondence. Via a Fisher matrix analysis, it was
noticed that the ability to constrain charge depends strongly on the
signal-to-noise ratio, so GW150914 cannot be used to place strong bounds on the
charge-to-mass ratio $\lambda$ of the final black hole. However, as the authors
remarked, these results should be considered only as qualitative, since
higher-order terms in $\lambda$ were neglected.

Instead of using approximations, here we solve the full non-linear
Einstein-Maxwell equations, extracting accurate gravitational waves to overcome
the shortcomings of previous approaches. We perform numerical-relativity
simulations of black holes with (1) same charge-to-mass ratio (that we will
indicate with $\lambda^+_+$) (2) same charge-to-mass ratio but opposite sign
($\lambda^+_-$), and (3) only one charged black hole ($\lambda^+_{{\kern 0.12em}0}$).
Einstein-Maxwell theory has no intrinsic scale, so our simulations scale with
the total ADM (Arnowitt-Deser-Misner) mass of the system
\si{\admmass}~\cite{Arnowitt:1962hi}. Thus, we can explore arbitrary chirp
masses with each simulation. We compute the mismatch between gravitational
waveforms generated by charged and uncharged binaries with a range of different
masses to account for the degeneracy: black hole charge is constrained when the
mismatch is larger than a value set by the signal-to-noise ratio
\cite{Flanagan1998b,Lindblom2008,McWilliams2010,Abbott2017h} for all possible
values of the chirp mass.

An important advantage of our approach is that it furnishes a first-principles
calculation based on fundamental theories, and does not rely on particular
models. As a result, the mathematical formulation we employ has direct
fundamental physics applications. Examples are dark matter theories (e.g.,
\emph{dark electromagnetism, hidden
  sector}~\cite{Feng2009,Ackerman2009,Foot2015, Foot2015b,
  Foot2016,Agrawal:2016quu,Christiansen2020}, or \emph{mini-charged}
particles~\cite{Davidson1991,Perl1997,Davidson2000,Dubovsky2004,Dolgov2013,Vogel2014,Cardoso2016b,Gautham2019,Plestid2020}).
These theories allow black holes to be highly charged, since neutralization
arguments do not apply. Moreover, with a duality transformation
\cite{Jackson1975}, our work also constrains black hole magnetic charge (e.g.,
from primordial magnetic monopoles~\cite{Preskill1984,Stojkovic2005}). Our
simulations are also useful for the generation and calibration of
gravitational-wave template banks that target these systems.

Furthermore, our research targets theories of gravitation where gravity is also
mediated by a vector field, like the scalar-tensor-vector gravity developed in
\cite{Moffat2006} to explain ``dark matter'' phenomenology without dark matter.
This theory (also known with the acronym ``MOG''--MOdified Gravity), has been
widely studied in the past and can pass several tests, such as Solar System ones
\cite{Moffat2014} (see also~\cite{Moffat2006, Brownstein2006, Brownstein2006b,
  Armengol2016,Armengol2017,Armengol2017b, Shojai2017, Perez2017,
  Ghafourian2017, Green2017}; for a summary of the formulation, assumptions, and
successes of the theory, see \cite{Armengol2017}). MOG features a scalar field
that makes gravity stronger by increasing Newton's constant and a Proca field
that counteracts this effect in the short range. When considering systems much
smaller than the galactic scale, the vector field can be considered massless and
the scalar field becomes constant and modifies Newton's constant to
$G_{\text{eff}} = G (1 + \alpha)$. According to MOG, a body with mass $M$ has a
\emph{gravitational charge} $Q$ that is associated with the vector field and is
proportional to $M$. \emph{Moffat's prescription} sets the constant of
proportionality to $\sqrt{\alpha G_{\text{eff}} \slash (1 + \alpha)}$ so that the theory
satisfies the weak equivalence principle~\cite{Moffat2016}. In this limit, MOG
differs mathematically from Einstein-Maxwell theory only in using
$G_{\text{eff}}$ instead of $G$, and when $\alpha=0$ the theory becomes general
relativity. This rescaling gives rise to the same degeneracy in the chirp mass
and $\widetilde{G}$ that we discussed above in the case of electromagnetism: in
geometrized units, MOG solutions with mass $M_{\text{MOG}}$ and gravitational
constant $G_{\text{eff}} = 1$ are equivalent to Einstein-Maxwell solutions with
mass $M = M_{\text{MOG}} (1 + \alpha)$ and $G=1$. Hence, by scanning through all
possible values of the mass, a constraint on the charge-to-mass ratio translates
in this theory to a constraint on $Q \slash M = \sqrt{\alpha \slash (1 + \alpha)}$.

The results of this work depend on three assumptions: 1) Einstein-Maxwell
theory is the correct description of charged black holes at the energy, length,
and time scales we are investigating; 2) GW150914 is accurately modeled by
waveforms from uncharged, non-spinning binary black holes with mass ratio
$29\slash36$--the value inferred for GW150914~\cite{Abbott2016}; 3) The black
hole spin and binary mass ratio remain that of GW150914 even in the case of
non-zero charge. Spin and mass ratio may be degenerate with the charge, so the
results presented in this paper can be interpreted in two ways: if assumption 3)
holds for GW150914, then our charge limits are \emph{upper bounds} on the
charge-to-mass ratio of the binary components, otherwise, they are lower bounds
on the charge-to-mass ratio needed to leave detectable imprints in GW150914-like
events. We will explore the effects of spin and mass ratio in future works. To
further reduce the parameter space, we only consider black holes with the same
charge-to-mass ratio bracketing the possibilities. This choice also ensures the
applicability of our results to modified theories of gravity where the
charge-to-mass-ratio represents a coupling constant (as in MOG), in which case
only systems with the same charge-to-mass-ratio are relevant (in the limit we
discussed above).

\paragraph{\textbf{Methods}}
\label{sec:methods}

We employ the \texttt{Einstein Toolkit}
\cite{Loffler:2011ay,EinsteinToolkit:ascl,EinsteinToolkit:web,einsteintoolkit-zenodo}
to solve the coupled Einstein-Maxwell equations in the $3+1$ decomposition of
four-dimensional spacetime
\cite{Arnowitt:1962hi,Thorne1982,Alcubierre:2008it,Baumgarte:2010nu,Shibata2016b}.
We report the general features of our approach here and leave the details for
the Supplemental Material.

We performed simulations with charge-to-mass ratio
$\lambda \in \{ 0.01, 0.05, 0.1, 0.2, 0.3\}$ with like or opposite charge for the two
black holes (cases that we will designate as $\lambda^+_+$ and $\lambda^+_-$, where the
superscript and subscript indicate the sign of the charge of the primary and the
secondary, respectively), and only one charged black hole
($\lambda^+_{{\kern 0.12em}0}$). These cases are supplemented by an uncharged one
($\lambda^0_0$), a convergence study, and by simulations with $\lambda^+_+ = 0.4$,
$\lambda^+_{{\kern 0.12em}0} = 0.35$, and $\lambda_+^{{\kern 0.12em}0} = 0.35$.

Full non-linear evolutions of Einstein-Maxwell systems have already been
performed in the past for head-on collisions of charged black
holes~\cite{Zilhao2012,Zilhao2013}. Simulations of quasi-circular inspirals are
a non-trivial extension of that as the generation of valid initial data with the
solution of the constraint equations~\cite{Baumgarte:2010nu} is required.
In~\cite{Bozzola2019} we presented \texttt{TwoChargedPunctures}, which solves
this problem by adopting an extended Bowen-York
formalism~\cite{Bowen:1980yu,Bowen1985,Alcubierre2009} and allows the generation
of arbitrary configurations of charged black holes. We fix the initial
coordinate separation to $\SI{12.1}{\admmass}$ and we choose the black hole
initial linear momenta to yield a quasi-circular inspiral using a 2.5PN estimate
after rescaling $G$ to $\widetilde{G}$.

We evolve the spacetime and electromagnetic fields with the open-source and
well-tested \texttt{Lean} and \texttt{ProcaEvolve} codes \cite{canudacode,
  canuda, Sperhake2007,Zilhao2015}. \texttt{Lean} implements the
Baumgarte-Shapiro-Shibata-Nakamura formulation of Einstein's equation
\cite{Shibata1995,Baumgarte1998}, while \texttt{ProcaEvolve} evolves the
electromagnetic vector potential with a constraint-damping scheme for the Gauss
constraint. The evolution is on Cartesian
\texttt{Carpet}~\cite{Schnetter:2003rb} grids where the highest resolution is
approximately $\si{\admmass}\slash 65$, with $\si{\admmass}$ being the binary
ADM mass~\cite{Arnowitt:1962hi}. We extract gravitational waves based on the
Newman-Penrose formalism~\cite{Newman1962,Zilhao2015}, adopting the
fixed-frequency integration method~\cite{Reisswig2011}. We decompose the signal
into $-2$ spin weighted spherical harmonics, and focus on the dominant $l = 2$,
$m = 2$ gravitational wave mode.

Two waveforms are considered experimentally indistinguishable if their mismatch
is smaller than $1 \slash (2 \rho^2)$
\cite{Flanagan1998b,Lindblom2008,McWilliams2010,Abbott2017h}, with $\rho$ being
the signal-to-noise ratio. For GW150914, $\rho = 25.1$~\cite{Abbott2016d}, so
the threshold mismatch above which two signals are distinguishable is
approximately $\num{8e-4}$. We calculate the mismatch between strains $h_1$ and
$h_2$ as $1 - \max \mathcal{O}(h_1, h_2)$, where $\mathcal{O}(h_1, h_2)$ is the
overlap between the two signals (see Supplemental Material), and the maximum is
evaluated with respect to time-shifts, orbital-phase shifts and polarization
angles~\cite{Damour1998, Abbott2017h}. The overlap calculation is performed in
the frequency domain. We consider LIGO's noise curve at the time of GW150914
detection, and adopt the GW150914 inferred sky location. For the uncharged
signal, we set a source frame ADM mass $\si{\admmass}=\SI{65}{\sunmass}$, and a
luminosity distance of \SI{410}{\mega\pc}, corresponding to cosmological
redshift of $\approx \num{0.09}$ \cite{Planck2018}. In the Supplemental Material
we discuss how different choices for these parameters affect the results. To
account for the charge-chirp mass degeneracy, we compute the mismatch between
gravitational waves from uncharged black holes and the ones from charged systems
with different chirp masses $\mathcal{M}$. To vary the chirp mass, we rescale
\si{\admmass} by a factor that we indicate with $\mathcal{M}\slash
\mathcal{M}_{00}$, where $\mathcal{M}_{00}$ is the chirp mass of the uncharged
simulation. We estimate the error on the mismatch by comparing simulations at
different resolutions.

\paragraph{\textbf{Results and Discussion}}
\label{sec:results}

The mismatch between a charged and the uncharged binary grows with the
charge-to-mass ratio $\lambda$. So, we may place an upper bound on the charge by
finding the value of $\lambda$ at which the minimum mismatch (as we vary the chirp
mass) is larger than \num{8e-4}. We find that, assuming negligible spin and mass
ratio of $29\slash 36$, GW150914 constrains $\lambda$ to be smaller than
\begin{equation}
  \label{eq:constraints-results}
  \lambda^+_+ = 0.4\,, \quad \lambda^+_- = 0.2\,, \quad \text{and} \quad  \lambda^+_{{\kern 0.12em}0} = 0.35\,.
\end{equation}
Regardless of the value of the spin and the mass ratio,
Equation~\eqref{eq:constraints-results} provide lower bounds on $\lambda$ needed to
have detectable effects in GW150914-like events.

In our simulations we always endow the more massive black hole with positive
charge. Since the mass asymmetry of the system is small, we expect our
conclusions to remain the same in the opposite case. The simulation with
$\lambda_+^{{\kern 0.10em}0} = 0.35$ confirms this expectation: the computed minimum
mismatch differs by $\SI{10}{\percent}$ from the $\lambda^+_{{\kern 0.12em}0 } = 0.35$
case. Thus, the effect of the mass asymmetry is small.

\begin{figure}[htbp]
  \hspace{-0.3cm}
  \includegraphics{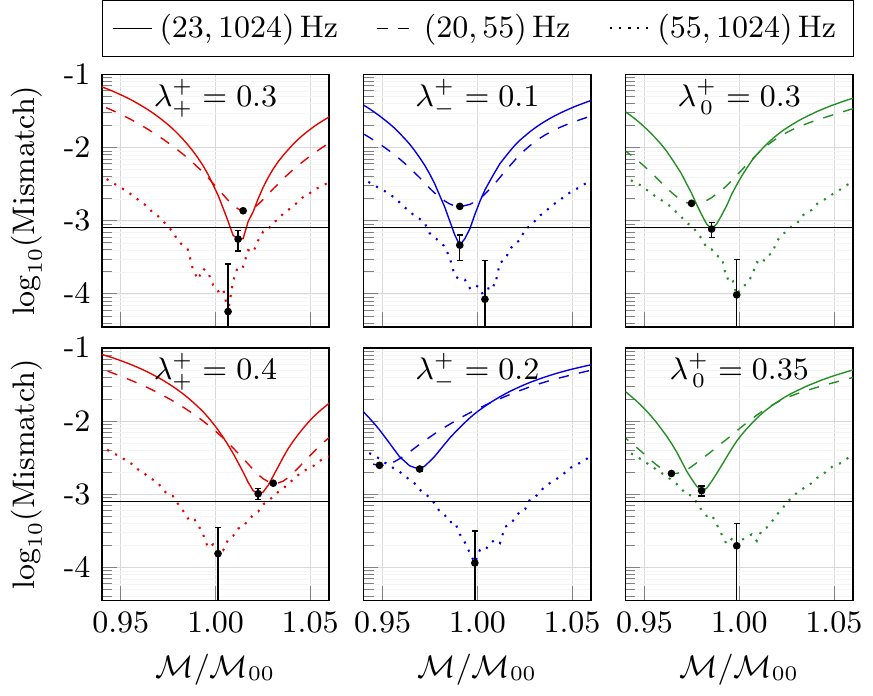}
  \caption{Mismatch between the strains from uncharged black holes and from
    charged ones with chirp mass rescaled by
    $\mathcal{M}\slash \mathcal{M}_{00}$. Solid curves are the mismatch
    including all available frequencies (the entire signal), dashed ones are
    only restricting to the frequency range $ (23,55)\,\si{\Hz}$ (inspiral) and
    dotted ones have frequencies restricted to $ (55,1024)\,\si{\Hz}$ (merger
    and ringdown). The solid horizontal line is the detection threshold for
    GW150914 (\num{8e-4}). The top (bottom) row is the largest (smallest) value
    of $\lambda$ (in our survey) compatible (incompatible) with GW150914. Red curves
    (left panels) are for the simulation with $\lambda^+_+ = 0.3$ (top) and
    $\lambda^+_+ = 0.4$ (bottom), blue (central panels) for $\lambda^+_- = 0.1$ and
    $\lambda^+_- = 0.2$, and green (middle panels) for $\lambda^+_{{\kern 0.12em}0} = 0.3$
    and $\lambda^+_{{\kern 0.12em}0} = 0.35$. The error bars shown are estimated
    comparing the standard resolution simulation with the one at higher
    resolution. We report the error bar only at minimum mismatch, but each point
    along the curve has the same level of error.}
  \label{fig:mismatch_chirp_mass_all}
\end{figure}

In Figure~\ref{fig:mismatch_chirp_mass_all}, we show the mismatch between the
uncharged simulation and charged ones as a function of the rescaling factor
$\mathcal{M}\slash \mathcal{M}_{00}$ for the chirp mass. The figure has three
sets of curves. Solid curves represent the mismatch computed on the entirety of
the signal (i.e., all frequencies are included). In the top panels, these curves
have minima below the threshold mismatch (horizontal solid line) for some value
of $\mathcal{M}\slash \mathcal{M}_{00}\vert_{\text{min}}$. Thus, gravitational
waves from these charged configurations are indistinguishable from the signal
that we adopt as true for GW150914. The opposite holds in the bottom panels.
Therefore, under the assumptions of our study, GW150914 is compatible with
involving charged black holes with $Q \slash M$ up to about \num{0.3}. The noise
curve adopted plays an important role: if instead of the realistic one, we
consider the Zero-Detuned-High-Power noise curve~\cite{ligozdhp}, the mismatch
increases by a factor of about 3, making the top panels in
Figure~\ref{fig:mismatch_chirp_mass_all} incompatible with the observation, and
hence distinguishable. Thus, it is important to use the realistic noise in these
calculations.

Figure~\ref{fig:mismatch_chirp_mass_all} reports two additional sets of curves:
dashed lines, representing the mismatch computed including frequencies below
\SI{55}{\Hz}, and dotted ones for frequencies above \SI{55}{\Hz}. In other
words, the dashed and dotted curves are the mismatch that would be computed if
we had detected \emph{only} the inspiral or \emph{only} the plunge and merger
phases. The frequency of \SI{55}{\Hz} marks conventionally the end of the
\emph{inspiral} phase~\cite{Abbott2016o}.  Including a larger range of
frequencies, decreases the minimum mismatch (from dashed lines to solid). Hence,
previous studies focusing only on the inspiral overestimate the mismatch and the
bias in the extracted chirp mass.

\begin{figure}[htbp]
  \hspace*{-0.3cm}
  \includegraphics{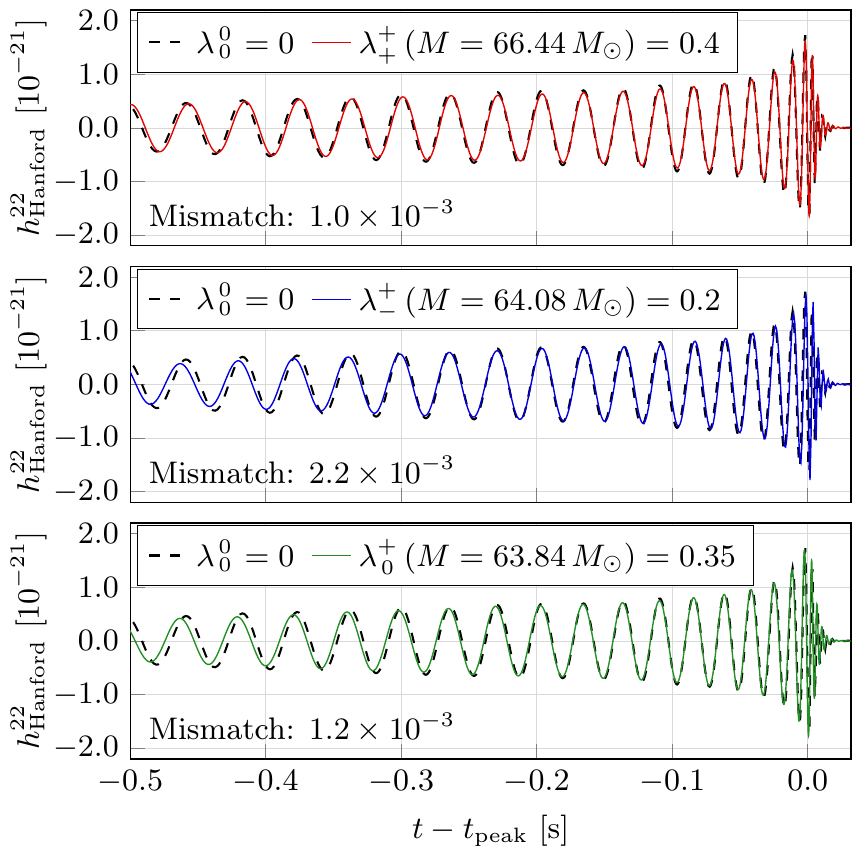}
  \caption{Comparison between the $(2,2)$ mode of the detector-response strain
    for Hanford for the simulations with no charges (dashed curves) and the ones
    with it, but with chirp mass $\mathcal{M}$ rescaled with respect to
    $\mathcal{M}_{00} = \SI{28.095}{\sunmass}$. Time and phase shifts are
    applied to minimize the mismatch between the two signals. All of these
    waveforms have two-detectors mismatch larger than the detection threshold
    for GW150914 of \num{8e-4}, mostly coming from the inspiral phase.}
  \label{fig:gwsignal_comparison}
\end{figure}

Figure~\ref{fig:mismatch_chirp_mass_all} shows that the mismatch is
significantly higher in the inspiral, suggesting that it is the dominant
contribution in the overall mismatch. Figure~\ref{fig:gwsignal_comparison}
further emphasizes this conclusion: we plot the strain the Hanford detector
would observe, if there was no noise, i.e.,
$h^{22}_{\text{Hanford}} = F_{\times} h_{\times}^{22} + F_{+} h_{+}^{22}$, where $F$ is
the detector antenna pattern~\cite{Abbott2016d}. The dashed curves represent
GW150914 and the solid ones are the strains from the charged simulations,
rescaled and shifted to maximize the overlap. The plot shows that the greatest
difference between charged and uncharged black holes arises in the earlier
inspiral. Thus, signals that stay for a longer duration in LIGO-Virgo bands
allow for stronger constraints on the charge. All waveforms in
Figure~\ref{fig:gwsignal_comparison} have mismatch with GW150914 larger than
\num{8e-4}, hence the corresponding charge configurations are incompatible with
GW150914.

One of the reasons why the merger+ringdown phase of the signal is not as
informative as the inspiral is that the properties of the final black holes do
not depend strongly on the initial charge configuration. In all our simulations,
the mass of the final black hole is the same to within \SI{1}{\percent}
($M_{\text{final}} \approx \SI{0.96}{\admmass}$), and the dimensionless spin
differs by at most \SI{6}{\percent} ($a_{\text{final}}\slash M_{\text{final}}
\approx \num{0.66}$). In particular, in our opposite charge cases, the final
mass and spin have sub-percent differences with respect to the uncharged case,
and, as expected from relativistic estimates, the case with same charge has a
lower spin~\cite{Jaiakson2017}. This result agrees with \cite{Cardoso2016b,
  Cardoso2020Erratum}: a large charge or a large signal-to-noise ratio is
required to extract the charge information from the ringdown.

Our full non-linear study supports previous results that were obtained with
parametrized methods. Constraints on the dipolar gravitational-wave emission
were placed in \cite{Yunes2016, Barausse2016} using Fisher matrix analysis based
on phenomenological waveform models. Translated into an upper bound on the
normalized electric dipole, the constraint becomes
$\zeta = \abs{\lambda_1 - \lambda_2}\slash \sqrt{1 - \lambda_1 \lambda_2} \lesssim 0.31 $ \cite{Cardoso2016b, Wang2020}.
Our work shows that $\zeta < \num{0.3}$ (from the case with
$\lambda^+_{{\kern 0.12em}0} = 0.3$). However, our work goes further by placing a
constraint on the individual black hole charge.

Our results can also be applied to the so-called theory of MOdified Gravity
(MOG) \cite{Moffat2015}. At scales relevant for compact binary mergers, this
theory replaces Newton's constant $G \to G_{\text{eff}}$, and postulates the
existence of a \emph{gravitational charge} $Q = \sqrt{\alpha
  G_{\text{eff}}\slash(1 + \alpha)} M$. The difference in Newton's constant is
degenerate with a change in chirp mass, which we thoroughly
explored. Figure~\ref{fig:mismatch_chirp_mass_all} shows that when $\lambda^+_+
= 0.4$ no matter how the chirp mass is changed, it is not possible to reconcile
GW150914 with the merger of charged black holes with $\lambda^+_+ = 0.4$.
Hence, our study directly constrains $\alpha \lesssim 0.19$. This implies that
the theory cannot deviate much from general relativity in the strong field,
under the assumptions made in this work.

\paragraph{\textbf{Conclusions}}
\label{sec:conclusions}

In this letter, we presented fully self-consistent general relativistic
simulations of the inspiral and merger of charged non-spinning black holes with
mass ratio $29\slash36$. We considered cases where both black holes are charged with
the same charge-to-mass ratio ($\lambda^+_+$), opposite charge-to-mass ratio
($\lambda^+_-$), and only one black hole charged ($\lambda^+_{{\kern 0.12em}0}$). By
comparing waveforms from uncharged systems to those from charged ones, we
addressed the charge-chirp mass degeneracy and found that, assuming non-spinning
black holes with mass ratio of $29\slash36$ for GW150914, $\lambda$ has to be smaller than:
\begin{equation}
  \label{eq:constraints-conclustion}
  \lambda^+_+ = 0.4\,, \quad \lambda^+_- = 0.2\,, \quad \text{and} \quad  \lambda^+_{{\kern 0.12em}0} = 0.35\,.
\end{equation}
These results hold under the assumption that spin and mass-ratio play a
secondary role. Independently of that,
Equation~\eqref{eq:constraints-conclustion} provides a lower bound on the
charge-to-mass ratio needed to leave measurable effects on the gravitational
waves from GW150914-like events.

We found that the inspiral is the most constraining part of the signal for
charge (Figures~\ref{fig:mismatch_chirp_mass_all},
\ref{fig:gwsignal_comparison}). So, low-mass binaries, having more orbits in
LIGO-Virgo bands, will likely yield tighter bounds on black hole charge. Our
full non-linear analysis confirms that it is challenging to constrain charge
from the ringdown phase of merging charged black holes~\cite{Cardoso2016b,
  Cardoso2020Erratum}.

The bounds found in this study do not apply only to electric charge, but they
can be directly translated to constraints on modified theories of gravity and
exotic astrophysical scenarios, e.g., dark matter models~\cite{Cardoso2016b}, or
primordial magnetic monopoles~\cite{Preskill1984}. In this work, we applied our
findings to Moffat's scalar-vector-tensor gravity (SVTG or
MOG)~\cite{Moffat2006} and constrained its $\alpha$ parameter to $\alpha \lesssim 0.19$ (note
that $\alpha = 0$ is general relativity). Here, we did not consider the effects of
black hole spin and the binary mass ratio and including these parameters can
introduce degeneracies that make the constraint less stringent. Applications to
lower-mass black hole binary detections may be able to constrain this theory
significantly in the strong field, dynamical regime.

In the future, we will consider systems with spinning black holes, different
mass-ratios, and asymmetric charge-to-mass ratio. With a large enough bank of
simulations, we will produce surrogate models (e.g.~\cite{Varma:2018mmi}) to perform
full parameter estimation of GW150914 and other LIGO-Virgo events.

\paragraph{Acknowledgments} { \small We thank M.\ Zilh{\~a}o for help on
  \texttt{ProcaEvolve}, J.\ R.\ Westernacher-Schneider for useful discussions,
  and D.\ Brown, V.\ Cardoso, J.\ Moffat and U.\ Sperhake for comments on the
  manuscript. We also wish to thank D.\ Brown for discussions on
  gravitational-wave data analysis. We are grateful to the developers and
  maintainers of the open-source codes that we used: their work was essential to
  the research presented here. This work was in part supported by NSF Grant
  PHY-1912619 to the University of Arizona. We acknowledge the hospitality of
  the Kavli Institute for Theoretical Physics (KITP), where part of the work was
  conducted. KITP is partially supported by the NSF grant No.\ PHY-1748958.
  Computational resources were provided by the Extreme Science and Engineering
  Discovery Environment (XSEDE) under grant number TG-PHY190020. XSEDE is
  supported by the NSF grant No.\ ACI-1548562. Simulations were performed on
  \texttt{Comet}, and \texttt{Stampede2}, which is funded by the NSF through
  award ACI-1540931. }

\subsubsection*{\textbf{Supplemental material}}

\paragraph{\textbf{Details of the numerical methods}}

We generate constraint-satisfying initial data with
\texttt{TwoChargedPunctures}, which can build arbitrary configurations of
charged binary black holes. The values of the initial black hole linear momenta
are chosen to yield a quasi-circular inspiral. To do so, we first use a 2.5
post-Newtonian expression to determine the values required to generate a
quasi-circular inspiral in the uncharged case. Next, for given charge-to-mass
ratios $\lambda_1$ and $\lambda_2$, we rescale $G$ to $\tilde{G}$, by multiplying the linear
momenta with $\sqrt{1 - \lambda_1 \lambda_2}$ (since they are proportional to $\sqrt{G}$).
For the initial orbital separation chosen, and the charge-to-mass ratios
explored, this choice yields near quasi-circular inspirals: estimating the
eccentricity with the method described in \cite{Pfeiffer2007} or in the Appendix
of \cite{Tsokaros2019}, the maximum eccentricity after the first orbit is
\num{0.005}, except in the $\lambda^+_- = 0.3$ case, where it is \num{0.014}. Our
experiments show that our method for setting the initial black hole linear
momenta must be modified to achieve very low eccentricity in simulations with
black holes that have close to extremal and opposite charges (i.e.,~large
$\lambda^+_-$), in which case eccentricity-reduction methods or more sophisticated
post-Newtonian expansions that include the electromagnetic fields may be
required.

For the time integration of the Einstein-Maxwell equations we use the method of
lines with a fourth-order Runge-Kutta scheme. The spacetime evolution is
performed adopting sixth-order finite-differences with the \texttt{Lean} code
\cite{Sperhake2007}, which is based on the Baumgarte-Shapiro-Shibata-Nakamura
(BSSN) formulation \cite{Shibata1995,Baumgarte1998} of the Einstein equations,
and exploits the puncture approach for the black-hole singularities. Apparent
horizons are located with \texttt{AHFinderDirect}
\cite{Thornburg:1995cp,Thornburg:2003sf}, and their physical properties
\cite{Ashtekar2000,Ashtekar2004} are computed with
\texttt{QuasiLocalMeasuresEM}--a version of \texttt{QuasiLocalMeasures}
\cite{Dreyer:2002mx} we extended to implement the formalism of quasi-isolated
horizons in full Einstein-Maxwell theory \cite{Bozzola2019}. Maxwell's equations
are evolved also using sixth-order finite differences with the
\texttt{ProcaEvolve} code \cite{Zilhao2015}, which is designed to keep the
magnetic and electric fields divergenceless. We adopt the Lorenz gauge for the
electromagnetic sector, and the $1+\log$ and $\Gamma$-freezing gauge conditions
for the lapse function and shift
vector~\cite{Alcubierre:2002kk,van-Meter2006,Hinder:2013oqa}. To improve the
stability of the simulation, we add seventh-order Kreiss-Oliger dissipation
\cite{Kreiss:1973aa} to all evolved variables and we introduce an extra
parabolic term to the equations for the evolution of the electric field to
further dissipate violations of Gauss's constraint. The choice of the
dissipation parameters is critical to ensure long-term evolutions, in
particular, high dissipation is needed near the black holes, and but stability
near the outer boundary requires lower Kreiss-Oliger dissipation. We will
present the details in an upcoming paper that will provide an in-depth
discussion on the formalism.  Both \texttt{Lean} and \texttt{ProcaEvolve} are
part of the \texttt{Canuda} open-source suite \cite{canudacode} and have already
been tested and used in previous studies (e.g.~\cite{Sperhake2007,Zilhao2015}).

We employ \texttt{Carpet}~\cite{Schnetter:2003rb} for moving-box adaptive
mesh-refinement, and use nine refinement levels. The outer boundary is placed at
$\SI{1033}{\admmass}$ where we impose outgoing-wave boundary conditions. We
performed selected simulations with outer boundary twice as far, and found that
all reported quantities are invariant with the outer boundary location to within
one part in \num{e8}. Sixth-order accurate evolutions with such grid
configurations are computationally expensive. Most of our simulations were
performed on ten nodes on the \texttt{Stampede2} (480 Skylake Intel CPUs, 1.92
TB of volatile memory and comparable storage) and took up to several weeks of
wall time to be completed. Evolutions for different $\lambda$ were run
concurrently and, as expected, the cases with black holes charged with the same
sign required significantly more time.

The extraction of gravitational and electromagnetic waves is performed at ten
different spatial radii in the range $(\SI{45.19}{\admmass},
\SI{192.74}{\admmass})$. In this work we report quantities at the extraction
radius $\SI{111.69}{\admmass}$. We checked that our results do not depend on the
extraction radius, and small differences from different radii are taken into
account in our error budget. We remove the first period from the extracted
signals as it contains \emph{junk} radiation from the initial data
\cite{Brandt1995, Gleiser1998}.

We calculate the mismatch between strains $h_1$ and $h_2$ as~\cite{Damour1998,
  Abbott2017h}
\begin{equation}
  \label{eq:mismatch}
  \text{mismatch}(h_1, h_2) = 1 - \max \mathcal{O}(h_1, h_2)\,,
\end{equation}
with the maximum evaluated with respect to time-shifts, orbital-phase shifts and
polarization angles. Here, $\mathcal{O}(h_1, h_2)$ is the overlap between $h_1$
and $h_2$
\begin{equation}
  \label{eq:overlap}
  \mathcal{O}(h_1, h_2) = \frac{(h_1, h_2)}{\sqrt{(h_1, h_1) (h_2, h_2)}}
\end{equation}
with $(h_1, h_2)$ being the two-detector noise-weighted inner product between
the two signals in the frequency domain $\tilde{h}_1(f)$ and $\tilde{h}_2(f)$,
\cite{Harry2011}
\begin{equation}
  \label{eq:inner-produced}
  (h_1, h_2) = \sum_{\substack{\text{Hanford} \\ \text{Livingston}}} \left[4 \mathrm{Re } \int_{f_{\text{min}}}^{f_{\text{max}}}
    \frac{\tilde{h}_1(f) \tilde{h}_2^{\star}(f)}{S_n(f)} \d f \right]\,,
\end{equation}
where $S_n(f)$ is the power spectral noise, and an asterisk denotes complex
conjugation. We crop the waveforms to ensure that they all end at the same time
after merger. Then we apply a Tukey window to the time series with parameter 0.1
so that the signal goes smoothly to zero. We pad numerical waveforms with zeros
so that all frequency series have the same length, and we take discrete Fourier
transforms. We consider three choices for the combination $(f_{\text{min}},
f_{\text{max}})$: $(23, 1024)~\si{\Hz}$ to include the entire signal; $(23,
55)~\si{\Hz}$ to take into account only the ``inspiral'' (at least six orbits)
and $(55, 1024)~\si{\Hz}$ for the plunge and post-merger phases (corresponding
to approximately the last two cycles). We choose these frequencies following the
LIGO-Virgo collaboration in identifying the first part as \emph{inspiral}, and
the second is what LIGO-Virgo further splits in \emph{intermediate} +
\emph{merger and ringdown} \cite{Abbott2016o}. This second group of frequencies
is in the most sensitive range for LIGO. The lowest frequency in our simulations
is approximately $\SI{23}{\Hz}$. For $S_n(f)$, we employ the calibrated noise
registered in coincidence with GW150914 (downloaded from the Gravitational Waves
Open Science Center~\cite{GW150914}). We use the inferred sky location of the
source (right ascension: $8\,\mathrm{h}$, declination: \SI{-70}{\degree}, UTC
time: 09:50:45.39 September 14 2015) and the corresponding gravitational-wave
antenna pattern of the two detectors~\cite{Abbott2016d}.

\paragraph{\textbf{Error budget and convergence}}

Our simulations exhibit excellent conservation of total energy, total angular
momentum, and total charge. Summing up the mass of the final black hole, and the
energies carried away by gravitational and electromagnetic waves, we find the
initial ADM energy to within 1 part in \num{2e4}. Similarly, angular momentum is
conserved to within 1 part in \num{7e3}. In these calculations we also
extrapolate waves to spatial infinity following \cite{Hinder:2013oqa} and
include all harmonic modes up to $l = 8$. Results are nearly invariant if a
finite extraction radius is considered instead. For energy and angular momentum
radiated we use the Newman-Penrose scalars \cite{Ruiz2008,Ashtekar2017}. Charge
is conserved to a high degree of accuracy: if $Q_1, Q_2$ are the initial horizon
charges and $Q_{\text{final}}$ the final black hole charge computed by
\texttt{QuasiLocalMeasuresEM}~\cite{Bozzola2019}, we find that $
\abs{Q_{\text{final}} - (Q_1+ Q_2)}\slash(\abs{Q_1} + \abs{Q_2}) \le
\num{2e-5}$.

For the case $\lambda^+_- = 0.3$ we performed a convergence study by considering
resolutions \SI{25}{\percent} higher ($\si{\admmass}\slash81$) and lower
($\si{\admmass}\slash52$) compared to the canonical one. Among our cases,
$\lambda^+_- = 0.3$ exhibits the highest velocities, and strongest emission of
energy and angular momentum in gravitational and electromagnetic waves. The
high-resolution simulation is also used to provide an estimate for the error of
the standard resolution simulations. The conserved quantities reported in the
previous paragraph improve by a factor of $\approx 2$ for the simulation at
higher resolution. %

\begin{figure}[htbp]
  \hspace{-0.3cm}
  \includegraphics{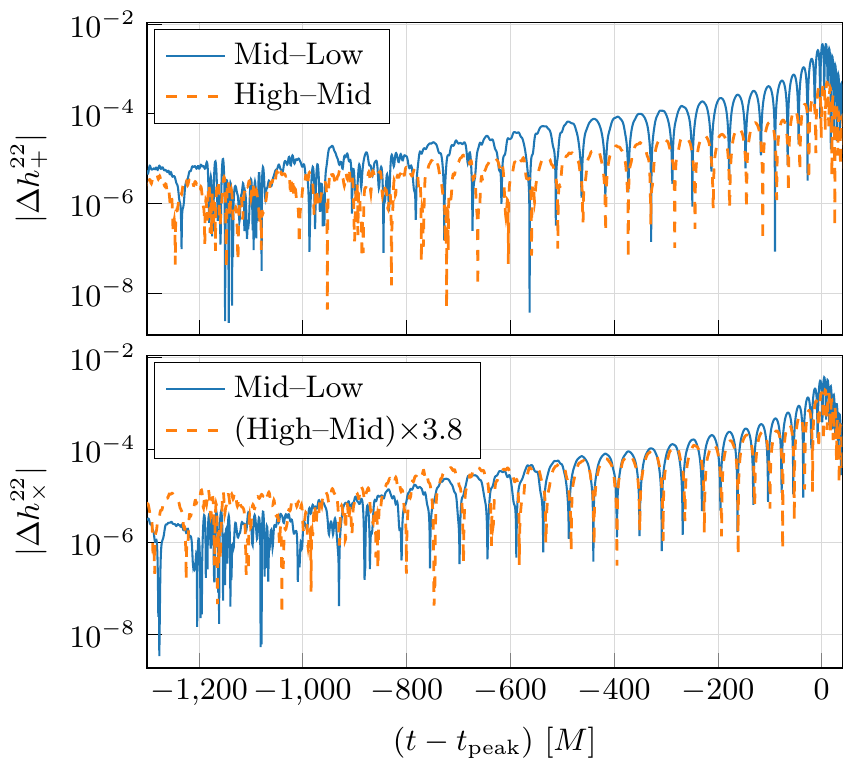}
  \caption{Self-convergence of the plus and cross polarization of the strain for
    simulations with $\lambda^+_- = 0.3$. The blue solid (orange dashed) lines are the
    absolute value of the difference between the strain at medium and low (high
    and medium) resolution. In the bottom panel we rescale the difference
    between the high and medium resolutions assuming sixth order convergence,
    i.e., by a factor of $(1.25)^6 \approx 3.8$, where $1.25$ is the ratio between the
    resolutions.}
  \label{fig:convergence}
\end{figure}

We show convergence more formally in Figure~\ref{fig:convergence}, where we
report the absolute value of the difference of $h^{22}_+$ between different
resolutions (and similarly for $h^{22}_\times$). Early on we observe the
well-known resolution-dependent high-frequency
noise~\cite{Zlochower2012,Etienne:2014tia} due to reflection/diffraction
phenomena across refinement-level boundaries.  After an initial noise-dominated
phase, the difference between the two higher resolution simulations (orange
dashed curve) becomes smaller than the one between the two lower-resolution runs
(blue solid line), demonstrating self-convergence.

We estimate the error of the mismatch by finding the maximum mismatch between
the simulation with standard resolution and the one with higher resolution with
respect to changing the extraction radius, the cutoff frequency for the
fixed-frequency integration, and the amount of signal cropped at the beginning
of the simulation to remove ``junk'' radiation. We find an error of \num{1.5e-4}
for the total signal, \num{3e-5} for frequencies up to \SI{55}{\Hz}, and
\num{2e-4} for frequencies above \SI{55}{\Hz}. These numbers are well below the
LIGO GW150914 threshold mismatch of \num{8e-4} for distinguishing two different
waveforms. The minimum of the mismatch when only considering high frequencies
alone is of the same order as our error, which explains why the dotted curves in
Figure~\ref{fig:mismatch_chirp_mass_all} are noisier compared to the other
ones. This systematic error in our simulation prevents us from estimating what
signal-to-noise ratio would be needed to extract charge information from the
ringdown phase (see dotted lines in Figure~\ref{fig:mismatch_chirp_mass_all}).

To confirm that the mismatch we compute is due to the presence of charge and not
the residual initial eccentricity, we use \texttt{EccentricFD}
\cite{Huerta2014}--a non-spinning frequency-domain, inspiral-only template
available in \texttt{PyCBC} \citep{PyCBC,Biwer2019}. Focusing on the inspiral
(up to \SI{55}{\Hz}), we find that the values of eccentricity we measure in our
simulations ($\approx \num{0.005}$) produce mismatches that are at least one
order of magnitude smaller than the ones we reported in the main text. Even for
the largest eccentricity we measure ($\num{0.014}$), the computed mismatch
remains subdominant ($\approx \num{2e-4}$). Therefore, this assures us that for
the large values of $\lambda$ in our survey, the mismatch is due to black hole
charge and not to the initial eccentricity.

\paragraph{\textbf{Confidence levels}}

In the main text, we represent GW150914 as an uncharged binary black hole with
total mass \SI{65}{\sunmass}, at a luminosity distance of \SI{410}{\Mpc}, and
use for the signal-to-noise ratio $\rho$ the value \SI{25.1}. These are the most
probable parameters for the event according to the Bayesian analysis performed
by the LIGO-Virgo collaboration~\cite{Abbott2016d}. Moreover, we adopted as
threshold mismatch for distinguishing two signals the standard choice of
$1\slash (2\rho^{2})$. Using Equation~(18) in~\cite{Baird2013} (with one degree of
freedom $k = 1$ since we compare charged configurations with an uncharged one,
and maximize the overlap over all other parameters), a mismatch of
$1\slash (2\rho^{2})$ corresponds to a \SI{68}{\percent} confidence level. The most
probable parameters represent neither the worst nor the best case scenario for
distinguishing black hole charge. The worst-case-scenario is when both the
number of gravitational wave cycles in LIGO's sensitivity band and the
signal-to-noise ratio are minimized: for GW150914 this happens when the distance
is \SI{570}{\Mpc} and the total mass is \SI{69.5}{\sunmass}, which are the
largest values in LIGO's \SI{90}{\percent} confidence levels. The minimum
signal-to-noise ratio reported by LIGO is $\rho = 23.4$. Even under these
conditions, our constraint on $\lambda^+_-$ remains unchanged, but for the
$\lambda^{+}_{+} = 0.4$ case, the threshold mismatch for distinguishing charge must
reduce to \SI{6.8e-4}, making the confidence level of our constraint (based on
Equation~(18) in \cite{Baird2013}) \SI{61}{\percent}.

\bibliography{bibliography,einsteintoolkit}

\def\prd{Phys. Rev. D}\def\prl{Phys. Rev. Lett.}\def\apjl{Astrophys. J.
  Lett.}\def\apjs{Astrophys. J. Suppl.}\def\apj{Astrophys. J.}\def\aj{Astron.
  J.}\def\aap{Astron. Astrophys.}\def\aaps{Astron. Astrophys.
  Suppl.}\def\araa{Ann. Rev. Astron. Astrophys.}\def\adp{Ann.
  Phy.}\def\cqg{Classical Quant. Grav.}\def\mnras{Mon. Not. R. Astron.
  Soc.}\def\physrep{Phys. Rep.}\def\nat{Nat.}\def\pasj{Pub. Astron. Soc. of
  Jap.}\def\jcap{J. Cosm. Astropart. Phys.}\def\apss{Astrophys. Spac. Sci.}
\begin{thebibliography}{122}%
\makeatletter
\providecommand \@ifxundefined [1]{%
 \@ifx{#1\undefined}
}%
\providecommand \@ifnum [1]{%
 \ifnum #1\expandafter \@firstoftwo
 \else \expandafter \@secondoftwo
 \fi
}%
\providecommand \@ifx [1]{%
 \ifx #1\expandafter \@firstoftwo
 \else \expandafter \@secondoftwo
 \fi
}%
\providecommand \natexlab [1]{#1}%
\providecommand \enquote  [1]{``#1''}%
\providecommand \bibnamefont  [1]{#1}%
\providecommand \bibfnamefont [1]{#1}%
\providecommand \citenamefont [1]{#1}%
\providecommand \href@noop [0]{\@secondoftwo}%
\providecommand \href [0]{\begingroup \@sanitize@url \@href}%
\providecommand \@href[1]{\@@startlink{#1}\@@href}%
\providecommand \@@href[1]{\endgroup#1\@@endlink}%
\providecommand \@sanitize@url [0]{\catcode `\\12\catcode `\$12\catcode
  `\&12\catcode `\#12\catcode `\^12\catcode `\_12\catcode `\%12\relax}%
\providecommand \@@startlink[1]{}%
\providecommand \@@endlink[0]{}%
\providecommand \url  [0]{\begingroup\@sanitize@url \@url }%
\providecommand \@url [1]{\endgroup\@href {#1}{\urlprefix }}%
\providecommand \urlprefix  [0]{URL }%
\providecommand \Eprint [0]{\href }%
\providecommand \doibase [0]{https://doi.org/}%
\providecommand \selectlanguage [0]{\@gobble}%
\providecommand \bibinfo  [0]{\@secondoftwo}%
\providecommand \bibfield  [0]{\@secondoftwo}%
\providecommand \translation [1]{[#1]}%
\providecommand \BibitemOpen [0]{}%
\providecommand \bibitemStop [0]{}%
\providecommand \bibitemNoStop [0]{.\EOS\space}%
\providecommand \EOS [0]{\spacefactor3000\relax}%
\providecommand \BibitemShut  [1]{\csname bibitem#1\endcsname}%
\let\auto@bib@innerbib\@empty
%</preamble>
\bibitem [{\citenamefont {{Israel}}(1967)}]{Israel1967}%
  \BibitemOpen
  \bibfield  {author} {\bibinfo {author} {\bibfnamefont {W.}~\bibnamefont
  {{Israel}}},\ }\bibfield  {title} {\bibinfo {title} {{Event Horizons in
  Static Vacuum Space-Times}},\ }\href
  {https://doi.org/10.1103/PhysRev.164.1776} {\bibfield  {journal} {\bibinfo
  {journal} {Physical Review}\ }\textbf {\bibinfo {volume} {164}},\ \bibinfo
  {pages} {1776} (\bibinfo {year} {1967})}\BibitemShut {NoStop}%
\bibitem [{\citenamefont {{Carter}}(1971)}]{Carter1971}%
  \BibitemOpen
  \bibfield  {author} {\bibinfo {author} {\bibfnamefont {B.}~\bibnamefont
  {{Carter}}},\ }\bibfield  {title} {\bibinfo {title} {{Axisymmetric Black Hole
  Has Only Two Degrees of Freedom}},\ }\href
  {https://doi.org/10.1103/PhysRevLett.26.331} {\bibfield  {journal} {\bibinfo
  {journal} {Physical Review Letters}\ }\textbf {\bibinfo {volume} {26}},\
  \bibinfo {pages} {331} (\bibinfo {year} {1971})}\BibitemShut {NoStop}%
\bibitem [{\citenamefont {{Robinson}}(1974)}]{Robinson1974}%
  \BibitemOpen
  \bibfield  {author} {\bibinfo {author} {\bibfnamefont {D.~C.}\ \bibnamefont
  {{Robinson}}},\ }\bibfield  {title} {\bibinfo {title} {{Classification of
  black holes with electromagnetic fields}},\ }\href
  {https://doi.org/10.1103/PhysRevD.10.458} {\bibfield  {journal} {\bibinfo
  {journal} {\prd}\ }\textbf {\bibinfo {volume} {10}},\ \bibinfo {pages} {458}
  (\bibinfo {year} {1974})}\BibitemShut {NoStop}%
\bibitem [{\citenamefont {{Hawking}}(1972)}]{1972CMaPh..25..152H}%
  \BibitemOpen
  \bibfield  {author} {\bibinfo {author} {\bibfnamefont {S.~W.}\ \bibnamefont
  {{Hawking}}},\ }\bibfield  {title} {\bibinfo {title} {{Black holes in general
  relativity}},\ }\href {https://doi.org/10.1007/BF01877517} {\bibfield
  {journal} {\bibinfo  {journal} {Communications in Mathematical Physics}\
  }\textbf {\bibinfo {volume} {25}},\ \bibinfo {pages} {152} (\bibinfo {year}
  {1972})}\BibitemShut {NoStop}%
\bibitem [{\citenamefont {Hansen}(1974)}]{Hansen:1974zz}%
  \BibitemOpen
  \bibfield  {author} {\bibinfo {author} {\bibfnamefont {R.}~\bibnamefont
  {Hansen}},\ }\bibfield  {title} {\bibinfo {title} {{Multipole moments of
  stationary space-times}},\ }\href {https://doi.org/10.1063/1.1666501}
  {\bibfield  {journal} {\bibinfo  {journal} {J. Math. Phys.}\ }\textbf
  {\bibinfo {volume} {15}},\ \bibinfo {pages} {46} (\bibinfo {year}
  {1974})}\BibitemShut {NoStop}%
\bibitem [{\citenamefont {{Chru{\'s}ciel}}\ \emph {et~al.}(2012)\citenamefont
  {{Chru{\'s}ciel}}, \citenamefont {{Costa}},\ and\ \citenamefont
  {{Heusler}}}]{Chrusciel2012}%
  \BibitemOpen
  \bibfield  {author} {\bibinfo {author} {\bibfnamefont {P.~T.}\ \bibnamefont
  {{Chru{\'s}ciel}}}, \bibinfo {author} {\bibfnamefont {J.~L.}\ \bibnamefont
  {{Costa}}},\ and\ \bibinfo {author} {\bibfnamefont {M.}~\bibnamefont
  {{Heusler}}},\ }\bibfield  {title} {\bibinfo {title} {{Stationary Black
  Holes: Uniqueness and Beyond}},\ }\href {https://doi.org/10.12942/lrr-2012-7}
  {\bibfield  {journal} {\bibinfo  {journal} {Living Reviews in Relativity}\
  }\textbf {\bibinfo {volume} {15}},\ \bibinfo {eid} {7} (\bibinfo {year}
  {2012})},\ \Eprint {https://arxiv.org/abs/1205.6112} {arXiv:1205.6112
  [gr-qc]} \BibitemShut {NoStop}%
\bibitem [{\citenamefont {{Wald}}(1974)}]{Wald1974}%
  \BibitemOpen
  \bibfield  {author} {\bibinfo {author} {\bibfnamefont {R.~M.}\ \bibnamefont
  {{Wald}}},\ }\bibfield  {title} {\bibinfo {title} {{Black hole in a uniform
  magnetic field}},\ }\href {https://doi.org/10.1103/PhysRevD.10.1680}
  {\bibfield  {journal} {\bibinfo  {journal} {\prd}\ }\textbf {\bibinfo
  {volume} {10}},\ \bibinfo {pages} {1680} (\bibinfo {year}
  {1974})}\BibitemShut {NoStop}%
\bibitem [{\citenamefont {{Gibbons}}(1975)}]{Gibbons1975}%
  \BibitemOpen
  \bibfield  {author} {\bibinfo {author} {\bibfnamefont {G.~W.}\ \bibnamefont
  {{Gibbons}}},\ }\bibfield  {title} {\bibinfo {title} {{Vacuum polarization
  and the spontaneous loss of charge by black holes}},\ }\href
  {https://doi.org/10.1007/BF01609829} {\bibfield  {journal} {\bibinfo
  {journal} {Communications in Mathematical Physics}\ }\textbf {\bibinfo
  {volume} {44}},\ \bibinfo {pages} {245} (\bibinfo {year} {1975})}\BibitemShut
  {NoStop}%
\bibitem [{\citenamefont {{Eardley}}\ and\ \citenamefont
  {{Press}}(1975)}]{Eardley1975}%
  \BibitemOpen
  \bibfield  {author} {\bibinfo {author} {\bibfnamefont {D.~M.}\ \bibnamefont
  {{Eardley}}}\ and\ \bibinfo {author} {\bibfnamefont {W.~H.}\ \bibnamefont
  {{Press}}},\ }\bibfield  {title} {\bibinfo {title} {{Astrophysical processes
  near black holes}},\ }\href
  {https://doi.org/10.1146/annurev.aa.13.090175.002121} {\bibfield  {journal}
  {\bibinfo  {journal} {\araa}\ }\textbf {\bibinfo {volume} {13}},\ \bibinfo
  {pages} {381} (\bibinfo {year} {1975})}\BibitemShut {NoStop}%
\bibitem [{\citenamefont {{Hanni}}(1982)}]{Hanni1982}%
  \BibitemOpen
  \bibfield  {author} {\bibinfo {author} {\bibfnamefont {R.~S.}\ \bibnamefont
  {{Hanni}}},\ }\bibfield  {title} {\bibinfo {title} {{Limits on the charge of
  a collapsed object}},\ }\href {https://doi.org/10.1103/PhysRevD.25.2509}
  {\bibfield  {journal} {\bibinfo  {journal} {\prd}\ }\textbf {\bibinfo
  {volume} {25}},\ \bibinfo {pages} {2509} (\bibinfo {year}
  {1982})}\BibitemShut {NoStop}%
\bibitem [{\citenamefont {{Gong}}\ \emph {et~al.}(2019)\citenamefont {{Gong}},
  \citenamefont {{Cao}}, \citenamefont {{Gao}},\ and\ \citenamefont
  {{Zhang}}}]{Gong2019}%
  \BibitemOpen
  \bibfield  {author} {\bibinfo {author} {\bibfnamefont {Y.}~\bibnamefont
  {{Gong}}}, \bibinfo {author} {\bibfnamefont {Z.}~\bibnamefont {{Cao}}},
  \bibinfo {author} {\bibfnamefont {H.}~\bibnamefont {{Gao}}},\ and\ \bibinfo
  {author} {\bibfnamefont {B.}~\bibnamefont {{Zhang}}},\ }\bibfield  {title}
  {\bibinfo {title} {{On neutralization of charged black holes}},\ }\href
  {https://doi.org/10.1093/mnras/stz1904} {\bibfield  {journal} {\bibinfo
  {journal} {\mnras}\ }\textbf {\bibinfo {volume} {488}},\ \bibinfo {pages}
  {2722} (\bibinfo {year} {2019})},\ \Eprint {https://arxiv.org/abs/1907.05239}
  {arXiv:1907.05239 [gr-qc]} \BibitemShut {NoStop}%
\bibitem [{\citenamefont {{Pan}}\ and\ \citenamefont {{Yang}}(2019)}]{Pan2019}%
  \BibitemOpen
  \bibfield  {author} {\bibinfo {author} {\bibfnamefont {Z.}~\bibnamefont
  {{Pan}}}\ and\ \bibinfo {author} {\bibfnamefont {H.}~\bibnamefont {{Yang}}},\
  }\bibfield  {title} {\bibinfo {title} {{Black hole discharge:
  Very-high-energy gamma rays from black hole-neutron star mergers}},\ }\href
  {https://doi.org/10.1103/PhysRevD.100.043025} {\bibfield  {journal} {\bibinfo
   {journal} {\prd}\ }\textbf {\bibinfo {volume} {100}},\ \bibinfo {eid}
  {043025} (\bibinfo {year} {2019})},\ \Eprint
  {https://arxiv.org/abs/1905.04775} {arXiv:1905.04775 [astro-ph.HE]}
  \BibitemShut {NoStop}%
\bibitem [{\citenamefont {{Iorio}}(2012)}]{Iorio2012}%
  \BibitemOpen
  \bibfield  {author} {\bibinfo {author} {\bibfnamefont {L.}~\bibnamefont
  {{Iorio}}},\ }\bibfield  {title} {\bibinfo {title} {{Constraining the
  electric charges of some astronomical bodies in Reissner-Nordstr{\"o}m
  spacetimes and generic r $^{-2}$-type power-law potentials from orbital
  motions}},\ }\href {https://doi.org/10.1007/s10714-012-1365-0} {\bibfield
  {journal} {\bibinfo  {journal} {General Relativity and Gravitation}\ }\textbf
  {\bibinfo {volume} {44}},\ \bibinfo {pages} {1753} (\bibinfo {year}
  {2012})},\ \Eprint {https://arxiv.org/abs/1112.3520} {arXiv:1112.3520
  [gr-qc]} \BibitemShut {NoStop}%
\bibitem [{\citenamefont {{Zaja{\v c}ek}}\ \emph {et~al.}(2018)\citenamefont
  {{Zaja{\v c}ek}}, \citenamefont {{Tursunov}}, \citenamefont {{Eckart}},\ and\
  \citenamefont {{Britzen}}}]{Zajacek2018}%
  \BibitemOpen
  \bibfield  {author} {\bibinfo {author} {\bibfnamefont {M.}~\bibnamefont
  {{Zaja{\v c}ek}}}, \bibinfo {author} {\bibfnamefont {A.}~\bibnamefont
  {{Tursunov}}}, \bibinfo {author} {\bibfnamefont {A.}~\bibnamefont
  {{Eckart}}},\ and\ \bibinfo {author} {\bibfnamefont {S.}~\bibnamefont
  {{Britzen}}},\ }\bibfield  {title} {\bibinfo {title} {{On the charge of the
  Galactic centre black hole}},\ }\href {https://doi.org/10.1093/mnras/sty2182}
  {\bibfield  {journal} {\bibinfo  {journal} {\mnras}\ }\textbf {\bibinfo
  {volume} {480}},\ \bibinfo {pages} {4408} (\bibinfo {year} {2018})},\ \Eprint
  {https://arxiv.org/abs/1808.07327} {arXiv:1808.07327} \BibitemShut {NoStop}%
\bibitem [{\citenamefont {Abbott}\ \emph {et~al.}(2016)\citenamefont {Abbott}
  \emph {et~al.}}]{Abbott2016}%
  \BibitemOpen
  \bibfield  {author} {\bibinfo {author} {\bibfnamefont {B.~P.}\ \bibnamefont
  {Abbott}} \emph {et~al.} (\bibinfo {collaboration} {LIGO Scientific
  Collaboration and Virgo Collaboration}),\ }\bibfield  {title} {\bibinfo
  {title} {Observation of gravitational waves from a binary black hole
  merger},\ }\href {https://doi.org/10.1103/PhysRevLett.116.061102} {\bibfield
  {journal} {\bibinfo  {journal} {\prl}\ }\textbf {\bibinfo {volume} {116}},\
  \bibinfo {pages} {061102} (\bibinfo {year} {2016})},\ \Eprint
  {https://arxiv.org/abs/1602.03837} {1602.03837} \BibitemShut {NoStop}%
\bibitem [{\citenamefont {{Zhang}}(2016)}]{Zhang2016}%
  \BibitemOpen
  \bibfield  {author} {\bibinfo {author} {\bibfnamefont {B.}~\bibnamefont
  {{Zhang}}},\ }\bibfield  {title} {\bibinfo {title} {{Mergers of Charged Black
  Holes: Gravitational-wave Events, Short Gamma-Ray Bursts, and Fast Radio
  Bursts}},\ }\href {https://doi.org/10.3847/2041-8205/827/2/L31} {\bibfield
  {journal} {\bibinfo  {journal} {\apjl}\ }\textbf {\bibinfo {volume} {827}},\
  \bibinfo {eid} {L31} (\bibinfo {year} {2016})},\ \Eprint
  {https://arxiv.org/abs/1602.04542} {arXiv:1602.04542 [astro-ph.HE]}
  \BibitemShut {NoStop}%
\bibitem [{\citenamefont {{Liebling}}\ and\ \citenamefont
  {{Palenzuela}}(2016)}]{Liebling2016}%
  \BibitemOpen
  \bibfield  {author} {\bibinfo {author} {\bibfnamefont {S.~L.}\ \bibnamefont
  {{Liebling}}}\ and\ \bibinfo {author} {\bibfnamefont {C.}~\bibnamefont
  {{Palenzuela}}},\ }\bibfield  {title} {\bibinfo {title} {{Electromagnetic
  luminosity of the coalescence of charged black hole binaries}},\ }\href
  {https://doi.org/10.1103/PhysRevD.94.064046} {\bibfield  {journal} {\bibinfo
  {journal} {\prd}\ }\textbf {\bibinfo {volume} {94}},\ \bibinfo {eid} {064046}
  (\bibinfo {year} {2016})},\ \Eprint {https://arxiv.org/abs/1607.02140}
  {arXiv:1607.02140 [gr-qc]} \BibitemShut {NoStop}%
\bibitem [{\citenamefont {{Fraschetti}}(2018)}]{Fraschetti2018}%
  \BibitemOpen
  \bibfield  {author} {\bibinfo {author} {\bibfnamefont {F.}~\bibnamefont
  {{Fraschetti}}},\ }\bibfield  {title} {\bibinfo {title} {{Possible role of
  magnetic reconnection in the electromagnetic counterpart of binary black hole
  merger}},\ }\href {https://doi.org/10.1088/1475-7516/2018/04/054} {\bibfield
  {journal} {\bibinfo  {journal} {Journ. Cosm. Astrop. Phys.}\ }\textbf
  {\bibinfo {volume} {4}},\ \bibinfo {eid} {054} (\bibinfo {year} {2018})},\
  \Eprint {https://arxiv.org/abs/1603.01950} {arXiv:1603.01950 [astro-ph.HE]}
  \BibitemShut {NoStop}%
\bibitem [{\citenamefont {{Connaughton}}\ \emph {et~al.}(2016)\citenamefont
  {{Connaughton}}, \citenamefont {{Burns}}, \citenamefont {{Goldstein}},
  \citenamefont {{Blackburn}}, \citenamefont {{Briggs}}, \citenamefont
  {{Zhang}}, \citenamefont {{Camp}}, \citenamefont {{Christensen}},
  \citenamefont {{Hui}}, \citenamefont {{Jenke}}, \citenamefont {{Littenberg}},
  \citenamefont {{McEnery}}, \citenamefont {{Racusin}}, \citenamefont
  {{Shawhan}}, \citenamefont {{Singer}}, \citenamefont {{Veitch}},
  \citenamefont {{Wilson-Hodge}}, \citenamefont {{Bhat}}, \citenamefont
  {{Bissaldi}}, \citenamefont {{Cleveland}}, \citenamefont {{Fitzpatrick}},
  \citenamefont {{Giles}}, \citenamefont {{Gibby}}, \citenamefont {{von
  Kienlin}}, \citenamefont {{Kippen}}, \citenamefont {{McBreen}}, \citenamefont
  {{Mailyan}}, \citenamefont {{Meegan}}, \citenamefont {{Paciesas}},
  \citenamefont {{Preece}}, \citenamefont {{Roberts}}, \citenamefont
  {{Sparke}}, \citenamefont {{Stanbro}}, \citenamefont {{Toelge}},\ and\
  \citenamefont {{Veres}}}]{Connaughton2016}%
  \BibitemOpen
  \bibfield  {author} {\bibinfo {author} {\bibfnamefont {V.}~\bibnamefont
  {{Connaughton}}}, \bibinfo {author} {\bibfnamefont {E.}~\bibnamefont
  {{Burns}}}, \bibinfo {author} {\bibfnamefont {A.}~\bibnamefont
  {{Goldstein}}}, \bibinfo {author} {\bibfnamefont {L.}~\bibnamefont
  {{Blackburn}}}, \bibinfo {author} {\bibfnamefont {M.~S.}\ \bibnamefont
  {{Briggs}}}, \bibinfo {author} {\bibfnamefont {B.-B.}\ \bibnamefont
  {{Zhang}}}, \bibinfo {author} {\bibfnamefont {J.}~\bibnamefont {{Camp}}},
  \bibinfo {author} {\bibfnamefont {N.}~\bibnamefont {{Christensen}}}, \bibinfo
  {author} {\bibfnamefont {C.~M.}\ \bibnamefont {{Hui}}}, \bibinfo {author}
  {\bibfnamefont {P.}~\bibnamefont {{Jenke}}}, \bibinfo {author} {\bibfnamefont
  {T.}~\bibnamefont {{Littenberg}}}, \bibinfo {author} {\bibfnamefont {J.~E.}\
  \bibnamefont {{McEnery}}}, \bibinfo {author} {\bibfnamefont {J.}~\bibnamefont
  {{Racusin}}}, \bibinfo {author} {\bibfnamefont {P.}~\bibnamefont
  {{Shawhan}}}, \bibinfo {author} {\bibfnamefont {L.}~\bibnamefont {{Singer}}},
  \bibinfo {author} {\bibfnamefont {J.}~\bibnamefont {{Veitch}}}, \bibinfo
  {author} {\bibfnamefont {C.~A.}\ \bibnamefont {{Wilson-Hodge}}}, \bibinfo
  {author} {\bibfnamefont {P.~N.}\ \bibnamefont {{Bhat}}}, \bibinfo {author}
  {\bibfnamefont {E.}~\bibnamefont {{Bissaldi}}}, \bibinfo {author}
  {\bibfnamefont {W.}~\bibnamefont {{Cleveland}}}, \bibinfo {author}
  {\bibfnamefont {G.}~\bibnamefont {{Fitzpatrick}}}, \bibinfo {author}
  {\bibfnamefont {M.~M.}\ \bibnamefont {{Giles}}}, \bibinfo {author}
  {\bibfnamefont {M.~H.}\ \bibnamefont {{Gibby}}}, \bibinfo {author}
  {\bibfnamefont {A.}~\bibnamefont {{von Kienlin}}}, \bibinfo {author}
  {\bibfnamefont {R.~M.}\ \bibnamefont {{Kippen}}}, \bibinfo {author}
  {\bibfnamefont {S.}~\bibnamefont {{McBreen}}}, \bibinfo {author}
  {\bibfnamefont {B.}~\bibnamefont {{Mailyan}}}, \bibinfo {author}
  {\bibfnamefont {C.~A.}\ \bibnamefont {{Meegan}}}, \bibinfo {author}
  {\bibfnamefont {W.~S.}\ \bibnamefont {{Paciesas}}}, \bibinfo {author}
  {\bibfnamefont {R.~D.}\ \bibnamefont {{Preece}}}, \bibinfo {author}
  {\bibfnamefont {O.~J.}\ \bibnamefont {{Roberts}}}, \bibinfo {author}
  {\bibfnamefont {L.}~\bibnamefont {{Sparke}}}, \bibinfo {author}
  {\bibfnamefont {M.}~\bibnamefont {{Stanbro}}}, \bibinfo {author}
  {\bibfnamefont {K.}~\bibnamefont {{Toelge}}},\ and\ \bibinfo {author}
  {\bibfnamefont {P.}~\bibnamefont {{Veres}}},\ }\bibfield  {title} {\bibinfo
  {title} {{Fermi GBM Observations of LIGO Gravitational-wave Event
  GW150914}},\ }\href {https://doi.org/10.3847/2041-8205/826/1/L6} {\bibfield
  {journal} {\bibinfo  {journal} {\apjl}\ }\textbf {\bibinfo {volume} {826}},\
  \bibinfo {eid} {L6} (\bibinfo {year} {2016})},\ \Eprint
  {https://arxiv.org/abs/1602.03920} {arXiv:1602.03920 [astro-ph.HE]}
  \BibitemShut {NoStop}%
\bibitem [{\citenamefont {{Connaughton}}\ \emph {et~al.}(2018)\citenamefont
  {{Connaughton}}, \citenamefont {{Burns}}, \citenamefont {{Goldstein}},
  \citenamefont {{Blackburn}}, \citenamefont {{Briggs}}, \citenamefont
  {{Christensen}}, \citenamefont {{Hui}}, \citenamefont {{Kocevski}},
  \citenamefont {{Littenberg}}, \citenamefont {{McEnery}}, \citenamefont
  {{Racusin}}, \citenamefont {{Shawhan}}, \citenamefont {{Veitch}},
  \citenamefont {{Wilson-Hodge}}, \citenamefont {{Bhat}}, \citenamefont
  {{Bissaldi}}, \citenamefont {{Cleveland}}, \citenamefont {{Giles}},
  \citenamefont {{Gibby}}, \citenamefont {{von Kienlin}}, \citenamefont
  {{Kippen}}, \citenamefont {{McBreen}}, \citenamefont {{Meegan}},
  \citenamefont {{Paciesas}}, \citenamefont {{Preece}}, \citenamefont
  {{Roberts}}, \citenamefont {{Stanbro}},\ and\ \citenamefont
  {{Veres}}}]{Connaughton2018}%
  \BibitemOpen
  \bibfield  {author} {\bibinfo {author} {\bibfnamefont {V.}~\bibnamefont
  {{Connaughton}}}, \bibinfo {author} {\bibfnamefont {E.}~\bibnamefont
  {{Burns}}}, \bibinfo {author} {\bibfnamefont {A.}~\bibnamefont
  {{Goldstein}}}, \bibinfo {author} {\bibfnamefont {L.}~\bibnamefont
  {{Blackburn}}}, \bibinfo {author} {\bibfnamefont {M.~S.}\ \bibnamefont
  {{Briggs}}}, \bibinfo {author} {\bibfnamefont {N.}~\bibnamefont
  {{Christensen}}}, \bibinfo {author} {\bibfnamefont {C.~M.}\ \bibnamefont
  {{Hui}}}, \bibinfo {author} {\bibfnamefont {D.}~\bibnamefont {{Kocevski}}},
  \bibinfo {author} {\bibfnamefont {T.}~\bibnamefont {{Littenberg}}}, \bibinfo
  {author} {\bibfnamefont {J.~E.}\ \bibnamefont {{McEnery}}}, \bibinfo {author}
  {\bibfnamefont {J.}~\bibnamefont {{Racusin}}}, \bibinfo {author}
  {\bibfnamefont {P.}~\bibnamefont {{Shawhan}}}, \bibinfo {author}
  {\bibfnamefont {J.}~\bibnamefont {{Veitch}}}, \bibinfo {author}
  {\bibfnamefont {C.~A.}\ \bibnamefont {{Wilson-Hodge}}}, \bibinfo {author}
  {\bibfnamefont {P.~N.}\ \bibnamefont {{Bhat}}}, \bibinfo {author}
  {\bibfnamefont {E.}~\bibnamefont {{Bissaldi}}}, \bibinfo {author}
  {\bibfnamefont {W.}~\bibnamefont {{Cleveland}}}, \bibinfo {author}
  {\bibfnamefont {M.~M.}\ \bibnamefont {{Giles}}}, \bibinfo {author}
  {\bibfnamefont {M.~H.}\ \bibnamefont {{Gibby}}}, \bibinfo {author}
  {\bibfnamefont {A.}~\bibnamefont {{von Kienlin}}}, \bibinfo {author}
  {\bibfnamefont {R.~M.}\ \bibnamefont {{Kippen}}}, \bibinfo {author}
  {\bibfnamefont {S.}~\bibnamefont {{McBreen}}}, \bibinfo {author}
  {\bibfnamefont {C.~A.}\ \bibnamefont {{Meegan}}}, \bibinfo {author}
  {\bibfnamefont {W.~S.}\ \bibnamefont {{Paciesas}}}, \bibinfo {author}
  {\bibfnamefont {R.~D.}\ \bibnamefont {{Preece}}}, \bibinfo {author}
  {\bibfnamefont {O.~J.}\ \bibnamefont {{Roberts}}}, \bibinfo {author}
  {\bibfnamefont {M.}~\bibnamefont {{Stanbro}}},\ and\ \bibinfo {author}
  {\bibfnamefont {P.}~\bibnamefont {{Veres}}},\ }\bibfield  {title} {\bibinfo
  {title} {{On the Interpretation of the Fermi-GBM Transient Observed in
  Coincidence with LIGO Gravitational-wave Event GW150914}},\ }\href
  {https://doi.org/10.3847/2041-8213/aaa4f2} {\bibfield  {journal} {\bibinfo
  {journal} {\apjl}\ }\textbf {\bibinfo {volume} {853}},\ \bibinfo {eid} {L9}
  (\bibinfo {year} {2018})},\ \Eprint {https://arxiv.org/abs/1801.02305}
  {arXiv:1801.02305 [astro-ph.HE]} \BibitemShut {NoStop}%
\bibitem [{\citenamefont {{Abbott}}\ \emph
  {et~al.}(2016{\natexlab{a}})\citenamefont {{Abbott}}, \citenamefont
  {{Abbott}}, \citenamefont {{Abbott}}, \citenamefont {{Abernathy}},
  \citenamefont {{Acernese}}, \citenamefont {{Ackley}}, \citenamefont
  {{Adams}}, \citenamefont {{Adams}}, \citenamefont {{Addesso}}, \citenamefont
  {{Adhikari}},\ and\ \citenamefont {et~al.}}]{Abbott2016f}%
  \BibitemOpen
  \bibfield  {author} {\bibinfo {author} {\bibfnamefont {B.~P.}\ \bibnamefont
  {{Abbott}}}, \bibinfo {author} {\bibfnamefont {R.}~\bibnamefont {{Abbott}}},
  \bibinfo {author} {\bibfnamefont {T.~D.}\ \bibnamefont {{Abbott}}}, \bibinfo
  {author} {\bibfnamefont {M.~R.}\ \bibnamefont {{Abernathy}}}, \bibinfo
  {author} {\bibfnamefont {F.}~\bibnamefont {{Acernese}}}, \bibinfo {author}
  {\bibfnamefont {K.}~\bibnamefont {{Ackley}}}, \bibinfo {author}
  {\bibfnamefont {C.}~\bibnamefont {{Adams}}}, \bibinfo {author} {\bibfnamefont
  {T.}~\bibnamefont {{Adams}}}, \bibinfo {author} {\bibfnamefont
  {P.}~\bibnamefont {{Addesso}}}, \bibinfo {author} {\bibfnamefont {R.~X.}\
  \bibnamefont {{Adhikari}}},\ and\ \bibinfo {author} {\bibnamefont {et~al.}},\
  }\bibfield  {title} {\bibinfo {title} {{Localization and Broadband Follow-up
  of the Gravitational-wave Transient GW150914}},\ }\href
  {https://doi.org/10.3847/2041-8205/826/1/L13} {\bibfield  {journal} {\bibinfo
   {journal} {\apjl}\ }\textbf {\bibinfo {volume} {826}},\ \bibinfo {eid} {L13}
  (\bibinfo {year} {2016}{\natexlab{a}})},\ \Eprint
  {https://arxiv.org/abs/1602.08492} {arXiv:1602.08492 [astro-ph.HE]}
  \BibitemShut {NoStop}%
\bibitem [{\citenamefont {{Hurley}}\ \emph {et~al.}(2016)\citenamefont
  {{Hurley}}, \citenamefont {{Svinkin}}, \citenamefont {{Aptekar}},
  \citenamefont {{Golenetskii}}, \citenamefont {{Frederiks}}, \citenamefont
  {{Boynton}}, \citenamefont {{Mitrofanov}}, \citenamefont {{Golovin}},
  \citenamefont {{Kozyrev}}, \citenamefont {{Litvak}}, \citenamefont {{Sanin}},
  \citenamefont {{Rau}}, \citenamefont {{von Kienlin}}, \citenamefont
  {{Zhang}}, \citenamefont {{Connaughton}}, \citenamefont {{Meegan}},
  \citenamefont {{Cline}},\ and\ \citenamefont {{Gehrels}}}]{Hurley2016}%
  \BibitemOpen
  \bibfield  {author} {\bibinfo {author} {\bibfnamefont {K.}~\bibnamefont
  {{Hurley}}}, \bibinfo {author} {\bibfnamefont {D.~S.}\ \bibnamefont
  {{Svinkin}}}, \bibinfo {author} {\bibfnamefont {R.~L.}\ \bibnamefont
  {{Aptekar}}}, \bibinfo {author} {\bibfnamefont {S.~V.}\ \bibnamefont
  {{Golenetskii}}}, \bibinfo {author} {\bibfnamefont {D.~D.}\ \bibnamefont
  {{Frederiks}}}, \bibinfo {author} {\bibfnamefont {W.}~\bibnamefont
  {{Boynton}}}, \bibinfo {author} {\bibfnamefont {I.~G.}\ \bibnamefont
  {{Mitrofanov}}}, \bibinfo {author} {\bibfnamefont {D.~V.}\ \bibnamefont
  {{Golovin}}}, \bibinfo {author} {\bibfnamefont {A.~S.}\ \bibnamefont
  {{Kozyrev}}}, \bibinfo {author} {\bibfnamefont {M.~L.}\ \bibnamefont
  {{Litvak}}}, \bibinfo {author} {\bibfnamefont {A.~B.}\ \bibnamefont
  {{Sanin}}}, \bibinfo {author} {\bibfnamefont {A.}~\bibnamefont {{Rau}}},
  \bibinfo {author} {\bibfnamefont {A.}~\bibnamefont {{von Kienlin}}}, \bibinfo
  {author} {\bibfnamefont {X.}~\bibnamefont {{Zhang}}}, \bibinfo {author}
  {\bibfnamefont {V.}~\bibnamefont {{Connaughton}}}, \bibinfo {author}
  {\bibfnamefont {C.}~\bibnamefont {{Meegan}}}, \bibinfo {author}
  {\bibfnamefont {T.}~\bibnamefont {{Cline}}},\ and\ \bibinfo {author}
  {\bibfnamefont {N.}~\bibnamefont {{Gehrels}}},\ }\bibfield  {title} {\bibinfo
  {title} {{The Interplanetary Network Response to LIGO GW150914}},\ }\href
  {https://doi.org/10.3847/2041-8205/829/1/L12} {\bibfield  {journal} {\bibinfo
   {journal} {\apjl}\ }\textbf {\bibinfo {volume} {829}},\ \bibinfo {eid} {L12}
  (\bibinfo {year} {2016})}\BibitemShut {NoStop}%
\bibitem [{\citenamefont {{Evans}}\ \emph {et~al.}(2016)\citenamefont
  {{Evans}}, \citenamefont {{Kennea}}, \citenamefont {{Barthelmy}},
  \citenamefont {{Beardmore}}, \citenamefont {{Burrows}}, \citenamefont
  {{Campana}}, \citenamefont {{Cenko}}, \citenamefont {{Gehrels}},
  \citenamefont {{Giommi}}, \citenamefont {{Gronwall}}, \citenamefont
  {{Marshall}}, \citenamefont {{Malesani}}, \citenamefont {{Markwardt}},
  \citenamefont {{Mingo}}, \citenamefont {{Nousek}}, \citenamefont {{O'Brien}},
  \citenamefont {{Osborne}}, \citenamefont {{Pagani}}, \citenamefont {{Page}},
  \citenamefont {{Palmer}}, \citenamefont {{Perri}}, \citenamefont {{Racusin}},
  \citenamefont {{Siegel}}, \citenamefont {{Sbarufatti}},\ and\ \citenamefont
  {{Tagliaferri}}}]{Evans2016}%
  \BibitemOpen
  \bibfield  {author} {\bibinfo {author} {\bibfnamefont {P.~A.}\ \bibnamefont
  {{Evans}}}, \bibinfo {author} {\bibfnamefont {J.~A.}\ \bibnamefont
  {{Kennea}}}, \bibinfo {author} {\bibfnamefont {S.~D.}\ \bibnamefont
  {{Barthelmy}}}, \bibinfo {author} {\bibfnamefont {A.~P.}\ \bibnamefont
  {{Beardmore}}}, \bibinfo {author} {\bibfnamefont {D.~N.}\ \bibnamefont
  {{Burrows}}}, \bibinfo {author} {\bibfnamefont {S.}~\bibnamefont
  {{Campana}}}, \bibinfo {author} {\bibfnamefont {S.~B.}\ \bibnamefont
  {{Cenko}}}, \bibinfo {author} {\bibfnamefont {N.}~\bibnamefont {{Gehrels}}},
  \bibinfo {author} {\bibfnamefont {P.}~\bibnamefont {{Giommi}}}, \bibinfo
  {author} {\bibfnamefont {C.}~\bibnamefont {{Gronwall}}}, \bibinfo {author}
  {\bibfnamefont {F.~E.}\ \bibnamefont {{Marshall}}}, \bibinfo {author}
  {\bibfnamefont {D.}~\bibnamefont {{Malesani}}}, \bibinfo {author}
  {\bibfnamefont {C.~B.}\ \bibnamefont {{Markwardt}}}, \bibinfo {author}
  {\bibfnamefont {B.}~\bibnamefont {{Mingo}}}, \bibinfo {author} {\bibfnamefont
  {J.~A.}\ \bibnamefont {{Nousek}}}, \bibinfo {author} {\bibfnamefont {P.~T.}\
  \bibnamefont {{O'Brien}}}, \bibinfo {author} {\bibfnamefont {J.~P.}\
  \bibnamefont {{Osborne}}}, \bibinfo {author} {\bibfnamefont {C.}~\bibnamefont
  {{Pagani}}}, \bibinfo {author} {\bibfnamefont {K.~L.}\ \bibnamefont
  {{Page}}}, \bibinfo {author} {\bibfnamefont {D.~M.}\ \bibnamefont
  {{Palmer}}}, \bibinfo {author} {\bibfnamefont {M.}~\bibnamefont {{Perri}}},
  \bibinfo {author} {\bibfnamefont {J.~L.}\ \bibnamefont {{Racusin}}}, \bibinfo
  {author} {\bibfnamefont {M.~H.}\ \bibnamefont {{Siegel}}}, \bibinfo {author}
  {\bibfnamefont {B.}~\bibnamefont {{Sbarufatti}}},\ and\ \bibinfo {author}
  {\bibfnamefont {G.}~\bibnamefont {{Tagliaferri}}},\ }\bibfield  {title}
  {\bibinfo {title} {{Swift follow-up of the gravitational wave source
  GW150914}},\ }\href {https://doi.org/10.1093/mnrasl/slw065} {\bibfield
  {journal} {\bibinfo  {journal} {\mnras}\ }\textbf {\bibinfo {volume} {460}},\
  \bibinfo {pages} {L40} (\bibinfo {year} {2016})},\ \Eprint
  {https://arxiv.org/abs/1602.03868} {arXiv:1602.03868 [astro-ph.HE]}
  \BibitemShut {NoStop}%
\bibitem [{\citenamefont {{Savchenko}}\ \emph {et~al.}(2016)\citenamefont
  {{Savchenko}}, \citenamefont {{Ferrigno}}, \citenamefont {{Mereghetti}},
  \citenamefont {{Natalucci}}, \citenamefont {{Bazzano}}, \citenamefont
  {{Bozzo}}, \citenamefont {{Brandt}}, \citenamefont {{Courvoisier}},
  \citenamefont {{Diehl}}, \citenamefont {{Hanlon}}, \citenamefont {{von
  Kienlin}}, \citenamefont {{Kuulkers}}, \citenamefont {{Laurent}},
  \citenamefont {{Lebrun}}, \citenamefont {{Roques}}, \citenamefont
  {{Ubertini}},\ and\ \citenamefont {{Weidenspointner}}}]{Savchenko2016}%
  \BibitemOpen
  \bibfield  {author} {\bibinfo {author} {\bibfnamefont {V.}~\bibnamefont
  {{Savchenko}}}, \bibinfo {author} {\bibfnamefont {C.}~\bibnamefont
  {{Ferrigno}}}, \bibinfo {author} {\bibfnamefont {S.}~\bibnamefont
  {{Mereghetti}}}, \bibinfo {author} {\bibfnamefont {L.}~\bibnamefont
  {{Natalucci}}}, \bibinfo {author} {\bibfnamefont {A.}~\bibnamefont
  {{Bazzano}}}, \bibinfo {author} {\bibfnamefont {E.}~\bibnamefont {{Bozzo}}},
  \bibinfo {author} {\bibfnamefont {S.}~\bibnamefont {{Brandt}}}, \bibinfo
  {author} {\bibfnamefont {T.~J.-L.}\ \bibnamefont {{Courvoisier}}}, \bibinfo
  {author} {\bibfnamefont {R.}~\bibnamefont {{Diehl}}}, \bibinfo {author}
  {\bibfnamefont {L.}~\bibnamefont {{Hanlon}}}, \bibinfo {author}
  {\bibfnamefont {A.}~\bibnamefont {{von Kienlin}}}, \bibinfo {author}
  {\bibfnamefont {E.}~\bibnamefont {{Kuulkers}}}, \bibinfo {author}
  {\bibfnamefont {P.}~\bibnamefont {{Laurent}}}, \bibinfo {author}
  {\bibfnamefont {F.}~\bibnamefont {{Lebrun}}}, \bibinfo {author}
  {\bibfnamefont {J.~P.}\ \bibnamefont {{Roques}}}, \bibinfo {author}
  {\bibfnamefont {P.}~\bibnamefont {{Ubertini}}},\ and\ \bibinfo {author}
  {\bibfnamefont {G.}~\bibnamefont {{Weidenspointner}}},\ }\bibfield  {title}
  {\bibinfo {title} {{INTEGRAL Upper Limits on Gamma-Ray Emission Associated
  with the Gravitational Wave Event GW150914}},\ }\href
  {https://doi.org/10.3847/2041-8205/820/2/L36} {\bibfield  {journal} {\bibinfo
   {journal} {\apjl}\ }\textbf {\bibinfo {volume} {820}},\ \bibinfo {eid} {L36}
  (\bibinfo {year} {2016})},\ \Eprint {https://arxiv.org/abs/1602.04180}
  {arXiv:1602.04180 [astro-ph.HE]} \BibitemShut {NoStop}%
\bibitem [{\citenamefont {{Flanagan}}\ and\ \citenamefont
  {{Hughes}}(1998)}]{Flanagan1998b}%
  \BibitemOpen
  \bibfield  {author} {\bibinfo {author} {\bibfnamefont {{\'E}.~{\'E}.}\
  \bibnamefont {{Flanagan}}}\ and\ \bibinfo {author} {\bibfnamefont {S.~A.}\
  \bibnamefont {{Hughes}}},\ }\bibfield  {title} {\bibinfo {title} {{Measuring
  gravitational waves from binary black hole coalescences. II. The waves'
  information and its extraction, with and without templates}},\ }\href
  {https://doi.org/10.1103/PhysRevD.57.4566} {\bibfield  {journal} {\bibinfo
  {journal} {\prd}\ }\textbf {\bibinfo {volume} {57}},\ \bibinfo {pages} {4566}
  (\bibinfo {year} {1998})},\ \Eprint {https://arxiv.org/abs/gr-qc/9710129}
  {gr-qc/9710129} \BibitemShut {NoStop}%
\bibitem [{\citenamefont {{Lindblom}}\ \emph {et~al.}(2008)\citenamefont
  {{Lindblom}}, \citenamefont {{Owen}},\ and\ \citenamefont
  {{Brown}}}]{Lindblom2008}%
  \BibitemOpen
  \bibfield  {author} {\bibinfo {author} {\bibfnamefont {L.}~\bibnamefont
  {{Lindblom}}}, \bibinfo {author} {\bibfnamefont {B.~J.}\ \bibnamefont
  {{Owen}}},\ and\ \bibinfo {author} {\bibfnamefont {D.~A.}\ \bibnamefont
  {{Brown}}},\ }\bibfield  {title} {\bibinfo {title} {{Model waveform accuracy
  standards for gravitational wave data analysis}},\ }\href
  {https://doi.org/10.1103/PhysRevD.78.124020} {\bibfield  {journal} {\bibinfo
  {journal} {\prd}\ }\textbf {\bibinfo {volume} {78}},\ \bibinfo {eid} {124020}
  (\bibinfo {year} {2008})},\ \Eprint {https://arxiv.org/abs/0809.3844}
  {arXiv:0809.3844 [gr-qc]} \BibitemShut {NoStop}%
\bibitem [{\citenamefont {{McWilliams}}\ \emph {et~al.}(2010)\citenamefont
  {{McWilliams}}, \citenamefont {{Kelly}},\ and\ \citenamefont
  {{Baker}}}]{McWilliams2010}%
  \BibitemOpen
  \bibfield  {author} {\bibinfo {author} {\bibfnamefont {S.~T.}\ \bibnamefont
  {{McWilliams}}}, \bibinfo {author} {\bibfnamefont {B.~J.}\ \bibnamefont
  {{Kelly}}},\ and\ \bibinfo {author} {\bibfnamefont {J.~G.}\ \bibnamefont
  {{Baker}}},\ }\bibfield  {title} {\bibinfo {title} {{Observing mergers of
  nonspinning black-hole binaries}},\ }\href
  {https://doi.org/10.1103/PhysRevD.82.024014} {\bibfield  {journal} {\bibinfo
  {journal} {\prd}\ }\textbf {\bibinfo {volume} {82}},\ \bibinfo {eid} {024014}
  (\bibinfo {year} {2010})},\ \Eprint {https://arxiv.org/abs/1004.0961}
  {arXiv:1004.0961 [gr-qc]} \BibitemShut {NoStop}%
\bibitem [{\citenamefont {{Abbott}}\ \emph {et~al.}(2017)\citenamefont
  {{Abbott}}, \citenamefont {{Abbott}}, \citenamefont {{Abbott}}, \citenamefont
  {{Abernathy}}, \citenamefont {{Acernese}}, \citenamefont {{Ackley}},
  \citenamefont {{Adams}}, \citenamefont {{Adams}}, \citenamefont {{Addesso}},
  \citenamefont {{Adhikari}},\ and\ \citenamefont {et~al.}}]{Abbott2017h}%
  \BibitemOpen
  \bibfield  {author} {\bibinfo {author} {\bibfnamefont {B.~P.}\ \bibnamefont
  {{Abbott}}}, \bibinfo {author} {\bibfnamefont {R.}~\bibnamefont {{Abbott}}},
  \bibinfo {author} {\bibfnamefont {T.~D.}\ \bibnamefont {{Abbott}}}, \bibinfo
  {author} {\bibfnamefont {M.~R.}\ \bibnamefont {{Abernathy}}}, \bibinfo
  {author} {\bibfnamefont {F.}~\bibnamefont {{Acernese}}}, \bibinfo {author}
  {\bibfnamefont {K.}~\bibnamefont {{Ackley}}}, \bibinfo {author}
  {\bibfnamefont {C.}~\bibnamefont {{Adams}}}, \bibinfo {author} {\bibfnamefont
  {T.}~\bibnamefont {{Adams}}}, \bibinfo {author} {\bibfnamefont
  {P.}~\bibnamefont {{Addesso}}}, \bibinfo {author} {\bibfnamefont {R.~X.}\
  \bibnamefont {{Adhikari}}},\ and\ \bibinfo {author} {\bibnamefont {et~al.}},\
  }\bibfield  {title} {\bibinfo {title} {{Effects of waveform model systematics
  on the interpretation of GW150914}},\ }\href
  {https://doi.org/10.1088/1361-6382/aa6854} {\bibfield  {journal} {\bibinfo
  {journal} {Classical and Quantum Gravity}\ }\textbf {\bibinfo {volume}
  {34}},\ \bibinfo {eid} {104002} (\bibinfo {year} {2017})},\ \Eprint
  {https://arxiv.org/abs/1611.07531} {arXiv:1611.07531 [gr-qc]} \BibitemShut
  {NoStop}%
\bibitem [{\citenamefont {Maggiore}(2007)}]{Maggiore2007}%
  \BibitemOpen
  \bibfield  {author} {\bibinfo {author} {\bibfnamefont {M.}~\bibnamefont
  {Maggiore}},\ }\href@noop {} {\emph {\bibinfo {title} {{Gravitational Waves.
  Vol. 1: Theory and Experiments}}}},\ Oxford Master Series in Physics\
  (\bibinfo  {publisher} {Oxford University Press},\ \bibinfo {year}
  {2007})\BibitemShut {NoStop}%
\bibitem [{\citenamefont {{Cardoso}}\ \emph {et~al.}(2016)\citenamefont
  {{Cardoso}}, \citenamefont {{Macedo}}, \citenamefont {{Pani}},\ and\
  \citenamefont {{Ferrari}}}]{Cardoso2016b}%
  \BibitemOpen
  \bibfield  {author} {\bibinfo {author} {\bibfnamefont {V.}~\bibnamefont
  {{Cardoso}}}, \bibinfo {author} {\bibfnamefont {C.~F.~B.}\ \bibnamefont
  {{Macedo}}}, \bibinfo {author} {\bibfnamefont {P.}~\bibnamefont {{Pani}}},\
  and\ \bibinfo {author} {\bibfnamefont {V.}~\bibnamefont {{Ferrari}}},\
  }\bibfield  {title} {\bibinfo {title} {{Black holes and gravitational waves
  in models of minicharged dark matter}},\ }\href
  {https://doi.org/10.1088/1475-7516/2016/05/054} {\bibfield  {journal}
  {\bibinfo  {journal} {\jcap}\ }\textbf {\bibinfo {volume} {5}},\ \bibinfo
  {eid} {054} (\bibinfo {year} {2016})},\ \Eprint
  {https://arxiv.org/abs/1604.07845} {arXiv:1604.07845 [hep-ph]} \BibitemShut
  {NoStop}%
\bibitem [{\citenamefont {{Christiansen}}\ \emph {et~al.}(2020)\citenamefont
  {{Christiansen}}, \citenamefont {{Beltr{\'a}n Jim{\'e}nez}},\ and\
  \citenamefont {{Mota}}}]{Christiansen2020}%
  \BibitemOpen
  \bibfield  {author} {\bibinfo {author} {\bibfnamefont {{\O}.}~\bibnamefont
  {{Christiansen}}}, \bibinfo {author} {\bibfnamefont {J.}~\bibnamefont
  {{Beltr{\'a}n Jim{\'e}nez}}},\ and\ \bibinfo {author} {\bibfnamefont {D.~F.}\
  \bibnamefont {{Mota}}},\ }\bibfield  {title} {\bibinfo {title} {{Charged
  Black Hole Mergers: Orbit Circularisation and Chirp Mass Bias}},\ }\href@noop
  {} {\bibfield  {journal} {\bibinfo  {journal} {arXiv e-prints}\ ,\ \bibinfo
  {eid} {arXiv:2003.11452}} (\bibinfo {year} {2020})},\ \Eprint
  {https://arxiv.org/abs/2003.11452} {arXiv:2003.11452 [gr-qc]} \BibitemShut
  {NoStop}%
\bibitem [{\citenamefont {{Liu}}\ \emph
  {et~al.}(2020{\natexlab{a}})\citenamefont {{Liu}}, \citenamefont {{Guo}},
  \citenamefont {{Cai}},\ and\ \citenamefont {{Kim}}}]{Liu2020}%
  \BibitemOpen
  \bibfield  {author} {\bibinfo {author} {\bibfnamefont {L.}~\bibnamefont
  {{Liu}}}, \bibinfo {author} {\bibfnamefont {Z.-K.}\ \bibnamefont {{Guo}}},
  \bibinfo {author} {\bibfnamefont {R.-G.}\ \bibnamefont {{Cai}}},\ and\
  \bibinfo {author} {\bibfnamefont {S.~P.}\ \bibnamefont {{Kim}}},\ }\bibfield
  {title} {\bibinfo {title} {{Merger rate distribution of primordial black hole
  binaries with electric charges}},\ }\href@noop {} {\bibfield  {journal}
  {\bibinfo  {journal} {arXiv e-prints}\ ,\ \bibinfo {eid} {arXiv:2001.02984}}
  (\bibinfo {year} {2020}{\natexlab{a}})},\ \Eprint
  {https://arxiv.org/abs/2001.02984} {arXiv:2001.02984 [astro-ph.CO]}
  \BibitemShut {NoStop}%
\bibitem [{\citenamefont {{Cardoso}}\ \emph {et~al.}(2020)\citenamefont
  {{Cardoso}}, \citenamefont {{Macedo}}, \citenamefont {{Pani}},\ and\
  \citenamefont {{Ferrari}}}]{Cardoso2020Erratum}%
  \BibitemOpen
  \bibfield  {author} {\bibinfo {author} {\bibfnamefont {V.}~\bibnamefont
  {{Cardoso}}}, \bibinfo {author} {\bibfnamefont {C.~F.~B.}\ \bibnamefont
  {{Macedo}}}, \bibinfo {author} {\bibfnamefont {P.}~\bibnamefont {{Pani}}},\
  and\ \bibinfo {author} {\bibfnamefont {V.}~\bibnamefont {{Ferrari}}},\
  }\bibfield  {title} {\bibinfo {title} {{Erratum: Black holes and
  gravitational waves in models of minicharged dark matter Erratum: Black holes
  and gravitational waves in models of minicharged dark matter}},\ }\href
  {https://doi.org/10.1088/1475-7516/2020/04/E01} {\bibfield  {journal}
  {\bibinfo  {journal} {\jcap}\ }\textbf {\bibinfo {volume} {2020}},\ \bibinfo
  {eid} {E01} (\bibinfo {year} {2020})}\BibitemShut {NoStop}%
\bibitem [{\citenamefont {{Liu}}\ \emph
  {et~al.}(2020{\natexlab{b}})\citenamefont {{Liu}}, \citenamefont
  {{Christiansen}}, \citenamefont {{Guo}}, \citenamefont {{Cai}},\ and\
  \citenamefont {{Kim}}}]{Liu2020b}%
  \BibitemOpen
  \bibfield  {author} {\bibinfo {author} {\bibfnamefont {L.}~\bibnamefont
  {{Liu}}}, \bibinfo {author} {\bibfnamefont {{\O}.}~\bibnamefont
  {{Christiansen}}}, \bibinfo {author} {\bibfnamefont {Z.-K.}\ \bibnamefont
  {{Guo}}}, \bibinfo {author} {\bibfnamefont {R.-G.}\ \bibnamefont {{Cai}}},\
  and\ \bibinfo {author} {\bibfnamefont {S.~P.}\ \bibnamefont {{Kim}}},\
  }\bibfield  {title} {\bibinfo {title} {{Gravitational and electromagnetic
  radiation from binary black holes with electric and magnetic charges:
  Circular orbits on a cone}},\ }\href
  {https://doi.org/10.1103/PhysRevD.102.103520} {\bibfield  {journal} {\bibinfo
   {journal} {\prd}\ }\textbf {\bibinfo {volume} {102}},\ \bibinfo {eid}
  {103520} (\bibinfo {year} {2020}{\natexlab{b}})},\ \Eprint
  {https://arxiv.org/abs/2008.02326} {arXiv:2008.02326 [gr-qc]} \BibitemShut
  {NoStop}%
\bibitem [{\citenamefont {{Wang}}\ \emph {et~al.}(2020)\citenamefont {{Wang}},
  \citenamefont {{Li}}, \citenamefont {{Jiang}}, \citenamefont {{Hu}},\ and\
  \citenamefont {{Fan}}}]{Wang2020}%
  \BibitemOpen
  \bibfield  {author} {\bibinfo {author} {\bibfnamefont {H.-T.}\ \bibnamefont
  {{Wang}}}, \bibinfo {author} {\bibfnamefont {P.-C.}\ \bibnamefont {{Li}}},
  \bibinfo {author} {\bibfnamefont {J.-L.}\ \bibnamefont {{Jiang}}}, \bibinfo
  {author} {\bibfnamefont {Y.-M.}\ \bibnamefont {{Hu}}},\ and\ \bibinfo
  {author} {\bibfnamefont {Y.-Z.}\ \bibnamefont {{Fan}}},\ }\bibfield  {title}
  {\bibinfo {title} {{Post-Newtonian waveform for charged binary black hole
  inspirals and analysis with GWTC-1 events}},\ }\href@noop {} {\bibfield
  {journal} {\bibinfo  {journal} {arXiv e-prints}\ ,\ \bibinfo {eid}
  {arXiv:2004.12421}} (\bibinfo {year} {2020})},\ \Eprint
  {https://arxiv.org/abs/2004.12421} {arXiv:2004.12421 [gr-qc]} \BibitemShut
  {NoStop}%
\bibitem [{\citenamefont {{Abbott}}\ \emph {et~al.}(2019)\citenamefont
  {{Abbott}}, \citenamefont {{Abbott}}, \citenamefont {{Abbott}}, \citenamefont
  {{Abraham}}, \citenamefont {{Acernese}}, \citenamefont {{Ackley}},
  \citenamefont {{Adams}}, \citenamefont {{Adhikari}}, \citenamefont {{Adya}},
  \citenamefont {{Affeldt}},\ and\ \citenamefont {et~al.}}]{LIGOVirgo2018}%
  \BibitemOpen
  \bibfield  {author} {\bibinfo {author} {\bibfnamefont {B.~P.}\ \bibnamefont
  {{Abbott}}}, \bibinfo {author} {\bibfnamefont {R.}~\bibnamefont {{Abbott}}},
  \bibinfo {author} {\bibfnamefont {T.~D.}\ \bibnamefont {{Abbott}}}, \bibinfo
  {author} {\bibfnamefont {S.}~\bibnamefont {{Abraham}}}, \bibinfo {author}
  {\bibfnamefont {F.}~\bibnamefont {{Acernese}}}, \bibinfo {author}
  {\bibfnamefont {K.}~\bibnamefont {{Ackley}}}, \bibinfo {author}
  {\bibfnamefont {C.}~\bibnamefont {{Adams}}}, \bibinfo {author} {\bibfnamefont
  {R.~X.}\ \bibnamefont {{Adhikari}}}, \bibinfo {author} {\bibfnamefont
  {V.~B.}\ \bibnamefont {{Adya}}}, \bibinfo {author} {\bibfnamefont
  {C.}~\bibnamefont {{Affeldt}}},\ and\ \bibinfo {author} {\bibnamefont
  {et~al.}},\ }\bibfield  {title} {\bibinfo {title} {{GWTC-1: A
  Gravitational-Wave Transient Catalog of Compact Binary Mergers Observed by
  LIGO and Virgo during the First and Second Observing Runs}},\ }\href
  {https://doi.org/10.1103/PhysRevX.9.031040} {\bibfield  {journal} {\bibinfo
  {journal} {Physical Review X}\ }\textbf {\bibinfo {volume} {9}},\ \bibinfo
  {eid} {031040} (\bibinfo {year} {2019})},\ \Eprint
  {https://arxiv.org/abs/1811.12907} {arXiv:1811.12907 [astro-ph.HE]}
  \BibitemShut {NoStop}%
\bibitem [{\citenamefont {{Barausse}}\ \emph {et~al.}(2016)\citenamefont
  {{Barausse}}, \citenamefont {{Yunes}},\ and\ \citenamefont
  {{Chamberlain}}}]{Barausse2016}%
  \BibitemOpen
  \bibfield  {author} {\bibinfo {author} {\bibfnamefont {E.}~\bibnamefont
  {{Barausse}}}, \bibinfo {author} {\bibfnamefont {N.}~\bibnamefont
  {{Yunes}}},\ and\ \bibinfo {author} {\bibfnamefont {K.}~\bibnamefont
  {{Chamberlain}}},\ }\bibfield  {title} {\bibinfo {title} {{Theory-Agnostic
  Constraints on Black-Hole Dipole Radiation with Multiband Gravitational-Wave
  Astrophysics}},\ }\href {https://doi.org/10.1103/PhysRevLett.116.241104}
  {\bibfield  {journal} {\bibinfo  {journal} {\prl}\ }\textbf {\bibinfo
  {volume} {116}},\ \bibinfo {eid} {241104} (\bibinfo {year} {2016})},\ \Eprint
  {https://arxiv.org/abs/1603.04075} {arXiv:1603.04075 [gr-qc]} \BibitemShut
  {NoStop}%
\bibitem [{\citenamefont {{Yunes}}\ \emph {et~al.}(2016)\citenamefont
  {{Yunes}}, \citenamefont {{Yagi}},\ and\ \citenamefont
  {{Pretorius}}}]{Yunes2016}%
  \BibitemOpen
  \bibfield  {author} {\bibinfo {author} {\bibfnamefont {N.}~\bibnamefont
  {{Yunes}}}, \bibinfo {author} {\bibfnamefont {K.}~\bibnamefont {{Yagi}}},\
  and\ \bibinfo {author} {\bibfnamefont {F.}~\bibnamefont {{Pretorius}}},\
  }\bibfield  {title} {\bibinfo {title} {{Theoretical physics implications of
  the binary black-hole mergers GW150914 and GW151226}},\ }\href
  {https://doi.org/10.1103/PhysRevD.94.084002} {\bibfield  {journal} {\bibinfo
  {journal} {\prd}\ }\textbf {\bibinfo {volume} {94}},\ \bibinfo {eid} {084002}
  (\bibinfo {year} {2016})},\ \Eprint {https://arxiv.org/abs/1603.08955}
  {arXiv:1603.08955 [gr-qc]} \BibitemShut {NoStop}%
\bibitem [{\citenamefont {Arnowitt}\ \emph {et~al.}(2008)\citenamefont
  {Arnowitt}, \citenamefont {Deser},\ and\ \citenamefont
  {Misner}}]{Arnowitt:1962hi}%
  \BibitemOpen
  \bibfield  {author} {\bibinfo {author} {\bibfnamefont {R.~L.}\ \bibnamefont
  {Arnowitt}}, \bibinfo {author} {\bibfnamefont {S.}~\bibnamefont {Deser}},\
  and\ \bibinfo {author} {\bibfnamefont {C.~W.}\ \bibnamefont {Misner}},\
  }\bibfield  {title} {\bibinfo {title} {The dynamics of general relativity},\
  }\href {https://doi.org/10.1007/s10714-008-0661-1} {\bibfield  {journal}
  {\bibinfo  {journal} {General Relativity and Gravitation}\ }\textbf {\bibinfo
  {volume} {40}},\ \bibinfo {pages} {1997} (\bibinfo {year} {2008})},\ \Eprint
  {https://arxiv.org/abs/arXiv:gr-qc/0405109} {arXiv:gr-qc/0405109}
  \BibitemShut {NoStop}%
%%CITATION = GR-QC/0405109;%%
\bibitem [{\citenamefont {{Feng}}\ \emph {et~al.}(2009)\citenamefont {{Feng}},
  \citenamefont {{Kaplinghat}}, \citenamefont {{Tu}},\ and\ \citenamefont
  {{Yu}}}]{Feng2009}%
  \BibitemOpen
  \bibfield  {author} {\bibinfo {author} {\bibfnamefont {J.~L.}\ \bibnamefont
  {{Feng}}}, \bibinfo {author} {\bibfnamefont {M.}~\bibnamefont
  {{Kaplinghat}}}, \bibinfo {author} {\bibfnamefont {H.}~\bibnamefont {{Tu}}},\
  and\ \bibinfo {author} {\bibfnamefont {H.-B.}\ \bibnamefont {{Yu}}},\
  }\bibfield  {title} {\bibinfo {title} {{Hidden charged dark matter}},\ }\href
  {https://doi.org/10.1088/1475-7516/2009/07/004} {\bibfield  {journal}
  {\bibinfo  {journal} {\jcap}\ }\textbf {\bibinfo {volume} {2009}},\ \bibinfo
  {eid} {004} (\bibinfo {year} {2009})},\ \Eprint
  {https://arxiv.org/abs/0905.3039} {arXiv:0905.3039 [hep-ph]} \BibitemShut
  {NoStop}%
\bibitem [{\citenamefont {{Ackerman}}\ \emph {et~al.}(2009)\citenamefont
  {{Ackerman}}, \citenamefont {{Buckley}}, \citenamefont {{Carroll}},\ and\
  \citenamefont {{Kamionkowski}}}]{Ackerman2009}%
  \BibitemOpen
  \bibfield  {author} {\bibinfo {author} {\bibfnamefont {L.}~\bibnamefont
  {{Ackerman}}}, \bibinfo {author} {\bibfnamefont {M.~R.}\ \bibnamefont
  {{Buckley}}}, \bibinfo {author} {\bibfnamefont {S.~M.}\ \bibnamefont
  {{Carroll}}},\ and\ \bibinfo {author} {\bibfnamefont {M.}~\bibnamefont
  {{Kamionkowski}}},\ }\bibfield  {title} {\bibinfo {title} {{Dark matter and
  dark radiation}},\ }\href {https://doi.org/10.1103/PhysRevD.79.023519}
  {\bibfield  {journal} {\bibinfo  {journal} {\prd}\ }\textbf {\bibinfo
  {volume} {79}},\ \bibinfo {eid} {023519} (\bibinfo {year} {2009})},\ \Eprint
  {https://arxiv.org/abs/0810.5126} {arXiv:0810.5126 [hep-ph]} \BibitemShut
  {NoStop}%
\bibitem [{\citenamefont {{Foot}}\ and\ \citenamefont
  {{Vagnozzi}}(2015{\natexlab{a}})}]{Foot2015}%
  \BibitemOpen
  \bibfield  {author} {\bibinfo {author} {\bibfnamefont {R.}~\bibnamefont
  {{Foot}}}\ and\ \bibinfo {author} {\bibfnamefont {S.}~\bibnamefont
  {{Vagnozzi}}},\ }\bibfield  {title} {\bibinfo {title} {{Dissipative hidden
  sector dark matter}},\ }\href {https://doi.org/10.1103/PhysRevD.91.023512}
  {\bibfield  {journal} {\bibinfo  {journal} {\prd}\ }\textbf {\bibinfo
  {volume} {91}},\ \bibinfo {eid} {023512} (\bibinfo {year}
  {2015}{\natexlab{a}})},\ \Eprint {https://arxiv.org/abs/1409.7174}
  {arXiv:1409.7174 [hep-ph]} \BibitemShut {NoStop}%
\bibitem [{\citenamefont {{Foot}}\ and\ \citenamefont
  {{Vagnozzi}}(2015{\natexlab{b}})}]{Foot2015b}%
  \BibitemOpen
  \bibfield  {author} {\bibinfo {author} {\bibfnamefont {R.}~\bibnamefont
  {{Foot}}}\ and\ \bibinfo {author} {\bibfnamefont {S.}~\bibnamefont
  {{Vagnozzi}}},\ }\bibfield  {title} {\bibinfo {title} {{Diurnal modulation
  signal from dissipative hidden sector dark matter}},\ }\href
  {https://doi.org/10.1016/j.physletb.2015.06.063} {\bibfield  {journal}
  {\bibinfo  {journal} {Physics Letters B}\ }\textbf {\bibinfo {volume}
  {748}},\ \bibinfo {pages} {61} (\bibinfo {year} {2015}{\natexlab{b}})},\
  \Eprint {https://arxiv.org/abs/1412.0762} {arXiv:1412.0762 [hep-ph]}
  \BibitemShut {NoStop}%
\bibitem [{\citenamefont {{Foot}}\ and\ \citenamefont
  {{Vagnozzi}}(2016)}]{Foot2016}%
  \BibitemOpen
  \bibfield  {author} {\bibinfo {author} {\bibfnamefont {R.}~\bibnamefont
  {{Foot}}}\ and\ \bibinfo {author} {\bibfnamefont {S.}~\bibnamefont
  {{Vagnozzi}}},\ }\bibfield  {title} {\bibinfo {title} {{Solving the
  small-scale structure puzzles with dissipative dark matter}},\ }\href
  {https://doi.org/10.1088/1475-7516/2016/07/013} {\bibfield  {journal}
  {\bibinfo  {journal} {\jcap}\ }\textbf {\bibinfo {volume} {2016}},\ \bibinfo
  {eid} {013} (\bibinfo {year} {2016})},\ \Eprint
  {https://arxiv.org/abs/1602.02467} {arXiv:1602.02467 [astro-ph.CO]}
  \BibitemShut {NoStop}%
\bibitem [{\citenamefont {Agrawal}\ \emph {et~al.}(2017)\citenamefont
  {Agrawal}, \citenamefont {Cyr-Racine}, \citenamefont {Randall},\ and\
  \citenamefont {Scholtz}}]{Agrawal:2016quu}%
  \BibitemOpen
  \bibfield  {author} {\bibinfo {author} {\bibfnamefont {P.}~\bibnamefont
  {Agrawal}}, \bibinfo {author} {\bibfnamefont {F.-Y.}\ \bibnamefont
  {Cyr-Racine}}, \bibinfo {author} {\bibfnamefont {L.}~\bibnamefont
  {Randall}},\ and\ \bibinfo {author} {\bibfnamefont {J.}~\bibnamefont
  {Scholtz}},\ }\bibfield  {title} {\bibinfo {title} {{Make Dark Matter Charged
  Again}},\ }\href {https://doi.org/10.1088/1475-7516/2017/05/022} {\bibfield
  {journal} {\bibinfo  {journal} {JCAP}\ }\textbf {\bibinfo {volume} {05}},\
  \bibinfo {pages} {022}},\ \Eprint {https://arxiv.org/abs/1610.04611}
  {arXiv:1610.04611 [hep-ph]} \BibitemShut {NoStop}%
\bibitem [{\citenamefont {{Davidson}}\ \emph {et~al.}(1991)\citenamefont
  {{Davidson}}, \citenamefont {{Campbell}},\ and\ \citenamefont
  {{Bailey}}}]{Davidson1991}%
  \BibitemOpen
  \bibfield  {author} {\bibinfo {author} {\bibfnamefont {S.}~\bibnamefont
  {{Davidson}}}, \bibinfo {author} {\bibfnamefont {B.}~\bibnamefont
  {{Campbell}}},\ and\ \bibinfo {author} {\bibfnamefont {D.}~\bibnamefont
  {{Bailey}}},\ }\bibfield  {title} {\bibinfo {title} {{Limits on particles of
  small electric charge}},\ }\href {https://doi.org/10.1103/PhysRevD.43.2314}
  {\bibfield  {journal} {\bibinfo  {journal} {\prd}\ }\textbf {\bibinfo
  {volume} {43}},\ \bibinfo {pages} {2314} (\bibinfo {year}
  {1991})}\BibitemShut {NoStop}%
\bibitem [{\citenamefont {{Perl}}\ and\ \citenamefont
  {{Lee}}(1997)}]{Perl1997}%
  \BibitemOpen
  \bibfield  {author} {\bibinfo {author} {\bibfnamefont {M.~L.}\ \bibnamefont
  {{Perl}}}\ and\ \bibinfo {author} {\bibfnamefont {E.~R.}\ \bibnamefont
  {{Lee}}},\ }\bibfield  {title} {\bibinfo {title} {{The search for elementary
  particles with fractional electric charge and the philosophy of speculative
  experiments}},\ }\href {https://doi.org/10.1119/1.18641} {\bibfield
  {journal} {\bibinfo  {journal} {American Journal of Physics}\ }\textbf
  {\bibinfo {volume} {65}},\ \bibinfo {pages} {698} (\bibinfo {year}
  {1997})}\BibitemShut {NoStop}%
\bibitem [{\citenamefont {{Davidson}}\ \emph {et~al.}(2000)\citenamefont
  {{Davidson}}, \citenamefont {{Hannestad}},\ and\ \citenamefont
  {{Raffelt}}}]{Davidson2000}%
  \BibitemOpen
  \bibfield  {author} {\bibinfo {author} {\bibfnamefont {S.}~\bibnamefont
  {{Davidson}}}, \bibinfo {author} {\bibfnamefont {S.}~\bibnamefont
  {{Hannestad}}},\ and\ \bibinfo {author} {\bibfnamefont {G.}~\bibnamefont
  {{Raffelt}}},\ }\bibfield  {title} {\bibinfo {title} {{Updated bounds on
  milli-charged particles}},\ }\href
  {https://doi.org/10.1088/1126-6708/2000/05/003} {\bibfield  {journal}
  {\bibinfo  {journal} {Journal of High Energy Physics}\ }\textbf {\bibinfo
  {volume} {5}},\ \bibinfo {eid} {003} (\bibinfo {year} {2000})},\ \Eprint
  {https://arxiv.org/abs/hep-ph/0001179} {hep-ph/0001179} \BibitemShut
  {NoStop}%
\bibitem [{\citenamefont {{Dubovsky}}\ \emph {et~al.}(2004)\citenamefont
  {{Dubovsky}}, \citenamefont {{Gorbunov}},\ and\ \citenamefont
  {{Rubtsov}}}]{Dubovsky2004}%
  \BibitemOpen
  \bibfield  {author} {\bibinfo {author} {\bibfnamefont {S.~L.}\ \bibnamefont
  {{Dubovsky}}}, \bibinfo {author} {\bibfnamefont {D.~S.}\ \bibnamefont
  {{Gorbunov}}},\ and\ \bibinfo {author} {\bibfnamefont {G.~I.}\ \bibnamefont
  {{Rubtsov}}},\ }\bibfield  {title} {\bibinfo {title} {{Narrowing the window
  for millicharged particles by CMB anisotropy}},\ }\href
  {https://doi.org/10.1134/1.1675909} {\bibfield  {journal} {\bibinfo
  {journal} {Soviet Journal of Experimental and Theoretical Physics Letters}\
  }\textbf {\bibinfo {volume} {79}},\ \bibinfo {pages} {1} (\bibinfo {year}
  {2004})},\ \Eprint {https://arxiv.org/abs/hep-ph/0311189}
  {arXiv:hep-ph/0311189 [astro-ph]} \BibitemShut {NoStop}%
\bibitem [{\citenamefont {{Dolgov}}\ \emph {et~al.}(2013)\citenamefont
  {{Dolgov}}, \citenamefont {{Dubovsky}}, \citenamefont {{Rubtsov}},\ and\
  \citenamefont {{Tkachev}}}]{Dolgov2013}%
  \BibitemOpen
  \bibfield  {author} {\bibinfo {author} {\bibfnamefont {A.~D.}\ \bibnamefont
  {{Dolgov}}}, \bibinfo {author} {\bibfnamefont {S.~L.}\ \bibnamefont
  {{Dubovsky}}}, \bibinfo {author} {\bibfnamefont {G.~I.}\ \bibnamefont
  {{Rubtsov}}},\ and\ \bibinfo {author} {\bibfnamefont {I.~I.}\ \bibnamefont
  {{Tkachev}}},\ }\bibfield  {title} {\bibinfo {title} {{Constraints on
  millicharged particles from Planck data}},\ }\href
  {https://doi.org/10.1103/PhysRevD.88.117701} {\bibfield  {journal} {\bibinfo
  {journal} {\prd}\ }\textbf {\bibinfo {volume} {88}},\ \bibinfo {eid} {117701}
  (\bibinfo {year} {2013})},\ \Eprint {https://arxiv.org/abs/1310.2376}
  {arXiv:1310.2376 [hep-ph]} \BibitemShut {NoStop}%
\bibitem [{\citenamefont {{Vogel}}\ and\ \citenamefont
  {{Redondo}}(2014)}]{Vogel2014}%
  \BibitemOpen
  \bibfield  {author} {\bibinfo {author} {\bibfnamefont {H.}~\bibnamefont
  {{Vogel}}}\ and\ \bibinfo {author} {\bibfnamefont {J.}~\bibnamefont
  {{Redondo}}},\ }\bibfield  {title} {\bibinfo {title} {{Dark radiation
  constraints on minicharged particles in models with a hidden photon}},\
  }\href {https://doi.org/10.1088/1475-7516/2014/02/029} {\bibfield  {journal}
  {\bibinfo  {journal} {\jcap}\ }\textbf {\bibinfo {volume} {2014}},\ \bibinfo
  {eid} {029} (\bibinfo {year} {2014})},\ \Eprint
  {https://arxiv.org/abs/1311.2600} {arXiv:1311.2600 [hep-ph]} \BibitemShut
  {NoStop}%
\bibitem [{\citenamefont {{Gautham A.}}\ and\ \citenamefont
  {{Sethi}}(2020)}]{Gautham2019}%
  \BibitemOpen
  \bibfield  {author} {\bibinfo {author} {\bibfnamefont {P.}~\bibnamefont
  {{Gautham A.}}}\ and\ \bibinfo {author} {\bibfnamefont {S.}~\bibnamefont
  {{Sethi}}},\ }\bibfield  {title} {\bibinfo {title} {{Cosmological
  implications of electromagnetically interacting dark matter: milli-charged
  particles and atoms with singly and doubly charged dark matter}},\ }\href
  {https://doi.org/10.1088/1475-7516/2020/03/039} {\bibfield  {journal}
  {\bibinfo  {journal} {\jcap}\ }\textbf {\bibinfo {volume} {2020}},\ \bibinfo
  {eid} {039} (\bibinfo {year} {2020})},\ \Eprint
  {https://arxiv.org/abs/1910.04779} {arXiv:1910.04779 [astro-ph.CO]}
  \BibitemShut {NoStop}%
\bibitem [{\citenamefont {{Plestid}}\ \emph {et~al.}(2020)\citenamefont
  {{Plestid}}, \citenamefont {{Takhistov}}, \citenamefont {{Tsai}},
  \citenamefont {{Bringmann}}, \citenamefont {{Kusenko}},\ and\ \citenamefont
  {{Pospelov}}}]{Plestid2020}%
  \BibitemOpen
  \bibfield  {author} {\bibinfo {author} {\bibfnamefont {R.}~\bibnamefont
  {{Plestid}}}, \bibinfo {author} {\bibfnamefont {V.}~\bibnamefont
  {{Takhistov}}}, \bibinfo {author} {\bibfnamefont {Y.-D.}\ \bibnamefont
  {{Tsai}}}, \bibinfo {author} {\bibfnamefont {T.}~\bibnamefont {{Bringmann}}},
  \bibinfo {author} {\bibfnamefont {A.}~\bibnamefont {{Kusenko}}},\ and\
  \bibinfo {author} {\bibfnamefont {M.}~\bibnamefont {{Pospelov}}},\ }\bibfield
   {title} {\bibinfo {title} {{New Constraints on Millicharged Particles from
  Cosmic-ray Production}},\ }\href@noop {} {\bibfield  {journal} {\bibinfo
  {journal} {arXiv e-prints}\ ,\ \bibinfo {eid} {arXiv:2002.11732}} (\bibinfo
  {year} {2020})},\ \Eprint {https://arxiv.org/abs/2002.11732}
  {arXiv:2002.11732 [hep-ph]} \BibitemShut {NoStop}%
\bibitem [{\citenamefont {{Jackson}}(1975)}]{Jackson1975}%
  \BibitemOpen
  \bibfield  {author} {\bibinfo {author} {\bibfnamefont {J.~D.}\ \bibnamefont
  {{Jackson}}},\ }\href@noop {} {\emph {\bibinfo {title} {{Classical
  electrodynamics}}}}\ (\bibinfo  {publisher} {Wiley},\ \bibinfo {year}
  {1975})\BibitemShut {NoStop}%
\bibitem [{\citenamefont {{Preskill}}(1984)}]{Preskill1984}%
  \BibitemOpen
  \bibfield  {author} {\bibinfo {author} {\bibfnamefont {J.}~\bibnamefont
  {{Preskill}}},\ }\bibfield  {title} {\bibinfo {title} {{Magnetic
  Monopoles}},\ }\href {https://doi.org/10.1146/annurev.ns.34.120184.002333}
  {\bibfield  {journal} {\bibinfo  {journal} {Annual Review of Nuclear and
  Particle Science}\ }\textbf {\bibinfo {volume} {34}},\ \bibinfo {pages} {461}
  (\bibinfo {year} {1984})}\BibitemShut {NoStop}%
\bibitem [{\citenamefont {{Stojkovic}}\ and\ \citenamefont
  {{Freese}}(2005)}]{Stojkovic2005}%
  \BibitemOpen
  \bibfield  {author} {\bibinfo {author} {\bibfnamefont {D.}~\bibnamefont
  {{Stojkovic}}}\ and\ \bibinfo {author} {\bibfnamefont {K.}~\bibnamefont
  {{Freese}}},\ }\bibfield  {title} {\bibinfo {title} {{A black hole solution
  to the cosmological monopole problem}},\ }\href
  {https://doi.org/10.1016/j.physletb.2004.12.019} {\bibfield  {journal}
  {\bibinfo  {journal} {Physics Letters B}\ }\textbf {\bibinfo {volume}
  {606}},\ \bibinfo {pages} {251} (\bibinfo {year} {2005})},\ \Eprint
  {https://arxiv.org/abs/hep-ph/0403248} {hep-ph/0403248} \BibitemShut
  {NoStop}%
\bibitem [{\citenamefont {Moffat}(2006)}]{Moffat2006}%
  \BibitemOpen
  \bibfield  {author} {\bibinfo {author} {\bibfnamefont {J.~W.}\ \bibnamefont
  {Moffat}},\ }\bibfield  {title} {\bibinfo {title}
  {Scalar{\textendash}tensor{\textendash}vector gravity theory},\ }\href
  {https://doi.org/10.1088/1475-7516/2006/03/004} {\bibfield  {journal}
  {\bibinfo  {journal} {Journal of Cosmology and Astroparticle Physics.}\
  }\textbf {\bibinfo {volume} {2006}},\ \bibinfo {pages} {004} (\bibinfo {year}
  {2006})}\BibitemShut {NoStop}%
\bibitem [{\citenamefont {{Moffat}}(2014)}]{Moffat2014}%
  \BibitemOpen
  \bibfield  {author} {\bibinfo {author} {\bibfnamefont {J.~W.}\ \bibnamefont
  {{Moffat}}},\ }\bibfield  {title} {\bibinfo {title} {{Scalar and Vector Field
  Constraints, Deflection of Light and Lensing in Modified Gravity (MOG)}},\
  }\href@noop {} {\bibfield  {journal} {\bibinfo  {journal} {arXiv e-prints}\
  ,\ \bibinfo {eid} {arXiv:1410.2464}} (\bibinfo {year} {2014})},\ \Eprint
  {https://arxiv.org/abs/1410.2464} {arXiv:1410.2464 [gr-qc]} \BibitemShut
  {NoStop}%
\bibitem [{\citenamefont {{Brownstein}}\ and\ \citenamefont
  {{Moffat}}(2006{\natexlab{a}})}]{Brownstein2006}%
  \BibitemOpen
  \bibfield  {author} {\bibinfo {author} {\bibfnamefont {J.~R.}\ \bibnamefont
  {{Brownstein}}}\ and\ \bibinfo {author} {\bibfnamefont {J.~W.}\ \bibnamefont
  {{Moffat}}},\ }\bibfield  {title} {\bibinfo {title} {{Galaxy Rotation Curves
  without Nonbaryonic Dark Matter}},\ }\href {https://doi.org/10.1086/498208}
  {\bibfield  {journal} {\bibinfo  {journal} {\apj}\ }\textbf {\bibinfo
  {volume} {636}},\ \bibinfo {pages} {721} (\bibinfo {year}
  {2006}{\natexlab{a}})},\ \Eprint {https://arxiv.org/abs/astro-ph/0506370}
  {arXiv:astro-ph/0506370 [astro-ph]} \BibitemShut {NoStop}%
\bibitem [{\citenamefont {{Brownstein}}\ and\ \citenamefont
  {{Moffat}}(2006{\natexlab{b}})}]{Brownstein2006b}%
  \BibitemOpen
  \bibfield  {author} {\bibinfo {author} {\bibfnamefont {J.~R.}\ \bibnamefont
  {{Brownstein}}}\ and\ \bibinfo {author} {\bibfnamefont {J.~W.}\ \bibnamefont
  {{Moffat}}},\ }\bibfield  {title} {\bibinfo {title} {{Galaxy cluster masses
  without non-baryonic dark matter}},\ }\href
  {https://doi.org/10.1111/j.1365-2966.2006.09996.x} {\bibfield  {journal}
  {\bibinfo  {journal} {\mnras}\ }\textbf {\bibinfo {volume} {367}},\ \bibinfo
  {pages} {527} (\bibinfo {year} {2006}{\natexlab{b}})},\ \Eprint
  {https://arxiv.org/abs/astro-ph/0507222} {arXiv:astro-ph/0507222 [astro-ph]}
  \BibitemShut {NoStop}%
\bibitem [{\citenamefont {{Lopez Armengol}}\ and\ \citenamefont
  {{Romero}}(2016)}]{Armengol2016}%
  \BibitemOpen
  \bibfield  {author} {\bibinfo {author} {\bibfnamefont {F.~G.}\ \bibnamefont
  {{Lopez Armengol}}}\ and\ \bibinfo {author} {\bibfnamefont {G.~E.}\
  \bibnamefont {{Romero}}},\ }\bibfield  {title} {\bibinfo {title}
  {{Scalar-Tensor-Vector Gravity: solutions with matter content}},\ }\href@noop
  {} {\bibfield  {journal} {\bibinfo  {journal} {Boletin de la Asociacion
  Argentina de Astronomia La Plata Argentina}\ }\textbf {\bibinfo {volume}
  {58}},\ \bibinfo {pages} {231} (\bibinfo {year} {2016})}\BibitemShut
  {NoStop}%
\bibitem [{\citenamefont {{Lopez Armengol}}\ and\ \citenamefont
  {{Romero}}(2017{\natexlab{a}})}]{Armengol2017}%
  \BibitemOpen
  \bibfield  {author} {\bibinfo {author} {\bibfnamefont {F.~G.}\ \bibnamefont
  {{Lopez Armengol}}}\ and\ \bibinfo {author} {\bibfnamefont {G.~E.}\
  \bibnamefont {{Romero}}},\ }\bibfield  {title} {\bibinfo {title} {{Neutron
  stars in Scalar-Tensor-Vector Gravity}},\ }\href
  {https://doi.org/10.1007/s10714-017-2184-0} {\bibfield  {journal} {\bibinfo
  {journal} {General Relativity and Gravitation}\ }\textbf {\bibinfo {volume}
  {49}},\ \bibinfo {eid} {27} (\bibinfo {year} {2017}{\natexlab{a}})},\ \Eprint
  {https://arxiv.org/abs/1611.05721} {arXiv:1611.05721 [gr-qc]} \BibitemShut
  {NoStop}%
\bibitem [{\citenamefont {{Lopez Armengol}}\ and\ \citenamefont
  {{Romero}}(2017{\natexlab{b}})}]{Armengol2017b}%
  \BibitemOpen
  \bibfield  {author} {\bibinfo {author} {\bibfnamefont {F.~G.}\ \bibnamefont
  {{Lopez Armengol}}}\ and\ \bibinfo {author} {\bibfnamefont {G.~E.}\
  \bibnamefont {{Romero}}},\ }\bibfield  {title} {\bibinfo {title} {{Effects of
  Scalar-Tensor-Vector Gravity on relativistic jets}},\ }\href
  {https://doi.org/10.1007/s10509-017-3197-6} {\bibfield  {journal} {\bibinfo
  {journal} {\apss}\ }\textbf {\bibinfo {volume} {362}},\ \bibinfo {eid} {214}
  (\bibinfo {year} {2017}{\natexlab{b}})},\ \Eprint
  {https://arxiv.org/abs/1611.09918} {arXiv:1611.09918 [astro-ph.HE]}
  \BibitemShut {NoStop}%
\bibitem [{\citenamefont {{Shojai}}\ \emph {et~al.}(2017)\citenamefont
  {{Shojai}}, \citenamefont {{Cheraghchi}},\ and\ \citenamefont {{Bouzari
  Nezhad}}}]{Shojai2017}%
  \BibitemOpen
  \bibfield  {author} {\bibinfo {author} {\bibfnamefont {F.}~\bibnamefont
  {{Shojai}}}, \bibinfo {author} {\bibfnamefont {S.}~\bibnamefont
  {{Cheraghchi}}},\ and\ \bibinfo {author} {\bibfnamefont {H.}~\bibnamefont
  {{Bouzari Nezhad}}},\ }\bibfield  {title} {\bibinfo {title} {{On the
  gravitational instability in the Newtonian limit of MOG}},\ }\href
  {https://doi.org/10.1016/j.physletb.2017.04.029} {\bibfield  {journal}
  {\bibinfo  {journal} {Physics Letters B}\ }\textbf {\bibinfo {volume}
  {770}},\ \bibinfo {pages} {43} (\bibinfo {year} {2017})},\ \Eprint
  {https://arxiv.org/abs/1704.04161} {arXiv:1704.04161 [gr-qc]} \BibitemShut
  {NoStop}%
\bibitem [{\citenamefont {{P{\'e}rez}}\ \emph {et~al.}(2017)\citenamefont
  {{P{\'e}rez}}, \citenamefont {{Armengol}},\ and\ \citenamefont
  {{Romero}}}]{Perez2017}%
  \BibitemOpen
  \bibfield  {author} {\bibinfo {author} {\bibfnamefont {D.}~\bibnamefont
  {{P{\'e}rez}}}, \bibinfo {author} {\bibfnamefont {F.~G.~L.}\ \bibnamefont
  {{Armengol}}},\ and\ \bibinfo {author} {\bibfnamefont {G.~E.}\ \bibnamefont
  {{Romero}}},\ }\bibfield  {title} {\bibinfo {title} {{Accretion disks around
  black holes in scalar-tensor-vector gravity}},\ }\href
  {https://doi.org/10.1103/PhysRevD.95.104047} {\bibfield  {journal} {\bibinfo
  {journal} {\prd}\ }\textbf {\bibinfo {volume} {95}},\ \bibinfo {eid} {104047}
  (\bibinfo {year} {2017})},\ \Eprint {https://arxiv.org/abs/1705.02713}
  {arXiv:1705.02713 [astro-ph.HE]} \BibitemShut {NoStop}%
\bibitem [{\citenamefont {{Ghafourian}}\ and\ \citenamefont
  {{Roshan}}(2017)}]{Ghafourian2017}%
  \BibitemOpen
  \bibfield  {author} {\bibinfo {author} {\bibfnamefont {N.}~\bibnamefont
  {{Ghafourian}}}\ and\ \bibinfo {author} {\bibfnamefont {M.}~\bibnamefont
  {{Roshan}}},\ }\bibfield  {title} {\bibinfo {title} {{Global stability of
  self-gravitating discs in modified gravity}},\ }\href
  {https://doi.org/10.1093/mnras/stx661} {\bibfield  {journal} {\bibinfo
  {journal} {\mnras}\ }\textbf {\bibinfo {volume} {468}},\ \bibinfo {pages}
  {4450} (\bibinfo {year} {2017})}\BibitemShut {NoStop}%
\bibitem [{\citenamefont {{Green}}\ \emph {et~al.}(2018)\citenamefont
  {{Green}}, \citenamefont {{Moffat}},\ and\ \citenamefont
  {{Toth}}}]{Green2017}%
  \BibitemOpen
  \bibfield  {author} {\bibinfo {author} {\bibfnamefont {M.~A.}\ \bibnamefont
  {{Green}}}, \bibinfo {author} {\bibfnamefont {J.~W.}\ \bibnamefont
  {{Moffat}}},\ and\ \bibinfo {author} {\bibfnamefont {V.~T.}\ \bibnamefont
  {{Toth}}},\ }\bibfield  {title} {\bibinfo {title} {{Modified gravity (MOG),
  the speed of gravitational radiation and the event GW170817/GRB170817A}},\
  }\href {https://doi.org/10.1016/j.physletb.2018.03.015} {\bibfield  {journal}
  {\bibinfo  {journal} {Physics Letters B}\ }\textbf {\bibinfo {volume}
  {780}},\ \bibinfo {pages} {300} (\bibinfo {year} {2018})},\ \Eprint
  {https://arxiv.org/abs/1710.11177} {arXiv:1710.11177 [gr-qc]} \BibitemShut
  {NoStop}%
\bibitem [{\citenamefont {{Moffat}}(2016)}]{Moffat2016}%
  \BibitemOpen
  \bibfield  {author} {\bibinfo {author} {\bibfnamefont {J.~W.}\ \bibnamefont
  {{Moffat}}},\ }\bibfield  {title} {\bibinfo {title} {{LIGO GW150914 and
  GW151226 gravitational wave detection and generalized gravitation theory
  (MOG)}},\ }\href {https://doi.org/10.1016/j.physletb.2016.10.082} {\bibfield
  {journal} {\bibinfo  {journal} {Physics Letters B}\ }\textbf {\bibinfo
  {volume} {763}},\ \bibinfo {pages} {427} (\bibinfo {year} {2016})},\ \Eprint
  {https://arxiv.org/abs/1603.05225} {arXiv:1603.05225 [gr-qc]} \BibitemShut
  {NoStop}%
\bibitem [{\citenamefont {L{\"{o}}ffler}\ \emph {et~al.}(2012)\citenamefont
  {L{\"{o}}ffler}, \citenamefont {Faber}, \citenamefont {Bentivegna},
  \citenamefont {Bode}, \citenamefont {Diener}, \citenamefont {Haas},
  \citenamefont {Hinder}, \citenamefont {Mundim}, \citenamefont {Ott},
  \citenamefont {Schnetter}, \citenamefont {Allen}, \citenamefont
  {Campanelli},\ and\ \citenamefont {Laguna}}]{Loffler:2011ay}%
  \BibitemOpen
  \bibfield  {author} {\bibinfo {author} {\bibfnamefont {F.}~\bibnamefont
  {L{\"{o}}ffler}}, \bibinfo {author} {\bibfnamefont {J.}~\bibnamefont
  {Faber}}, \bibinfo {author} {\bibfnamefont {E.}~\bibnamefont {Bentivegna}},
  \bibinfo {author} {\bibfnamefont {T.}~\bibnamefont {Bode}}, \bibinfo {author}
  {\bibfnamefont {P.}~\bibnamefont {Diener}}, \bibinfo {author} {\bibfnamefont
  {R.}~\bibnamefont {Haas}}, \bibinfo {author} {\bibfnamefont {I.}~\bibnamefont
  {Hinder}}, \bibinfo {author} {\bibfnamefont {B.~C.}\ \bibnamefont {Mundim}},
  \bibinfo {author} {\bibfnamefont {C.~D.}\ \bibnamefont {Ott}}, \bibinfo
  {author} {\bibfnamefont {E.}~\bibnamefont {Schnetter}}, \bibinfo {author}
  {\bibfnamefont {G.}~\bibnamefont {Allen}}, \bibinfo {author} {\bibfnamefont
  {M.}~\bibnamefont {Campanelli}},\ and\ \bibinfo {author} {\bibfnamefont
  {P.}~\bibnamefont {Laguna}},\ }\bibfield  {title} {\bibinfo {title} {{{T}he
  {E}instein {T}oolkit: {A} {C}ommunity {C}omputational {I}nfrastructure for
  {R}elativistic {A}strophysics}},\ }\href
  {https://doi.org/doi:10.1088/0264-9381/29/11/115001} {\bibfield  {journal}
  {\bibinfo  {journal} {Class. Quantum Grav.}\ }\textbf {\bibinfo {volume}
  {29}},\ \bibinfo {pages} {115001} (\bibinfo {year} {2012})},\ \Eprint
  {https://arxiv.org/abs/arXiv:1111.3344 [gr-qc]} {arXiv:1111.3344 [gr-qc]}
  \BibitemShut {NoStop}%
\bibitem [{\citenamefont {{Collaborative
  Effort}}(2011)}]{EinsteinToolkit:ascl}%
  \BibitemOpen
  \bibfield  {author} {\bibinfo {author} {\bibnamefont {{Collaborative
  Effort}}},\ }\href@noop {} {\bibinfo {title} {{Einstein Toolkit for
  Relativistic Astrophysics}}},\ \bibinfo {howpublished} {Astrophysics Source
  Code Library} (\bibinfo {year} {2011}),\ \Eprint
  {https://arxiv.org/abs/ascl:1102.014} {ascl:1102.014} \BibitemShut {NoStop}%
\bibitem [{EinsteinToolkit()}]{EinsteinToolkit:web}%
  \BibitemOpen
  EinsteinToolkit,\ \href {http://einsteintoolkit.org/} {\bibinfo {title}
  {{Einstein Toolkit}: Open software for relativistic astrophysics}} (\bibinfo
  {year} {2019})\BibitemShut {NoStop}%
\bibitem [{\citenamefont {Babiuc-Hamilton}\ \emph {et~al.}(2019)\citenamefont
  {Babiuc-Hamilton}, \citenamefont {Brandt}, \citenamefont {Diener},
  \citenamefont {Elley}, \citenamefont {Etienne}, \citenamefont {Ficarra},
  \citenamefont {Haas}, \citenamefont {Witek}, \citenamefont {Alcubierre},
  \citenamefont {Alic}, \citenamefont {Allen}, \citenamefont {Ansorg},
  \citenamefont {Baiotti}, \citenamefont {Benger}, \citenamefont {Bentivegna},
  \citenamefont {Bernuzzi}, \citenamefont {Bode}, \citenamefont {Bruegmann},
  \citenamefont {Corvino}, \citenamefont {Pietri}, \citenamefont {Dimmelmeier},
  \citenamefont {Dooley}, \citenamefont {Dorband}, \citenamefont {Khamra},
  \citenamefont {Faber}, \citenamefont {Font}, \citenamefont {Frieben},
  \citenamefont {Giacomazzo}, \citenamefont {Goodale}, \citenamefont
  {Gundlach}, \citenamefont {Hawke}, \citenamefont {Hawley}, \citenamefont
  {Hinder}, \citenamefont {Husa}, \citenamefont {Iyer}, \citenamefont
  {Kellermann}, \citenamefont {Knapp}, \citenamefont {Koppitz}, \citenamefont
  {Lanferman}, \citenamefont {Löffler}, \citenamefont {Masso}, \citenamefont
  {Menger}, \citenamefont {Merzky}, \citenamefont {Miller}, \citenamefont
  {Moesta}, \citenamefont {Montero}, \citenamefont {Mundim}, \citenamefont
  {Nerozzi}, \citenamefont {Ott}, \citenamefont {Paruchuri}, \citenamefont
  {Pollney}, \citenamefont {Radice}, \citenamefont {Radke}, \citenamefont
  {Reisswig}, \citenamefont {Rezzolla}, \citenamefont {Rideout}, \citenamefont
  {Ripeanu}, \citenamefont {Schnetter}, \citenamefont {Schutz}, \citenamefont
  {Seidel}, \citenamefont {Seidel}, \citenamefont {Shalf}, \citenamefont
  {Sperhake}, \citenamefont {Stergioulas}, \citenamefont {Suen}, \citenamefont
  {Szilagyi}, \citenamefont {Takahashi}, \citenamefont {Thomas}, \citenamefont
  {Thornburg}, \citenamefont {Tobias}, \citenamefont {Tonita}, \citenamefont
  {Walker}, \citenamefont {Wan}, \citenamefont {Wardell}, \citenamefont
  {Zilhão}, \citenamefont {Zink},\ and\ \citenamefont
  {Zlochower}}]{einsteintoolkit-zenodo}%
  \BibitemOpen
  \bibfield  {author} {\bibinfo {author} {\bibfnamefont {M.}~\bibnamefont
  {Babiuc-Hamilton}}, \bibinfo {author} {\bibfnamefont {S.~R.}\ \bibnamefont
  {Brandt}}, \bibinfo {author} {\bibfnamefont {P.}~\bibnamefont {Diener}},
  \bibinfo {author} {\bibfnamefont {M.}~\bibnamefont {Elley}}, \bibinfo
  {author} {\bibfnamefont {Z.}~\bibnamefont {Etienne}}, \bibinfo {author}
  {\bibfnamefont {G.}~\bibnamefont {Ficarra}}, \bibinfo {author} {\bibfnamefont
  {R.}~\bibnamefont {Haas}}, \bibinfo {author} {\bibfnamefont {H.}~\bibnamefont
  {Witek}}, \bibinfo {author} {\bibfnamefont {M.}~\bibnamefont {Alcubierre}},
  \bibinfo {author} {\bibfnamefont {D.}~\bibnamefont {Alic}}, \bibinfo {author}
  {\bibfnamefont {G.}~\bibnamefont {Allen}}, \bibinfo {author} {\bibfnamefont
  {M.}~\bibnamefont {Ansorg}}, \bibinfo {author} {\bibfnamefont
  {L.}~\bibnamefont {Baiotti}}, \bibinfo {author} {\bibfnamefont
  {W.}~\bibnamefont {Benger}}, \bibinfo {author} {\bibfnamefont
  {E.}~\bibnamefont {Bentivegna}}, \bibinfo {author} {\bibfnamefont
  {S.}~\bibnamefont {Bernuzzi}}, \bibinfo {author} {\bibfnamefont
  {T.}~\bibnamefont {Bode}}, \bibinfo {author} {\bibfnamefont {B.}~\bibnamefont
  {Bruegmann}}, \bibinfo {author} {\bibfnamefont {G.}~\bibnamefont {Corvino}},
  \bibinfo {author} {\bibfnamefont {R.~D.}\ \bibnamefont {Pietri}}, \bibinfo
  {author} {\bibfnamefont {H.}~\bibnamefont {Dimmelmeier}}, \bibinfo {author}
  {\bibfnamefont {R.}~\bibnamefont {Dooley}}, \bibinfo {author} {\bibfnamefont
  {N.}~\bibnamefont {Dorband}}, \bibinfo {author} {\bibfnamefont {Y.~E.}\
  \bibnamefont {Khamra}}, \bibinfo {author} {\bibfnamefont {J.}~\bibnamefont
  {Faber}}, \bibinfo {author} {\bibfnamefont {T.}~\bibnamefont {Font}},
  \bibinfo {author} {\bibfnamefont {J.}~\bibnamefont {Frieben}}, \bibinfo
  {author} {\bibfnamefont {B.}~\bibnamefont {Giacomazzo}}, \bibinfo {author}
  {\bibfnamefont {T.}~\bibnamefont {Goodale}}, \bibinfo {author} {\bibfnamefont
  {C.}~\bibnamefont {Gundlach}}, \bibinfo {author} {\bibfnamefont
  {I.}~\bibnamefont {Hawke}}, \bibinfo {author} {\bibfnamefont
  {S.}~\bibnamefont {Hawley}}, \bibinfo {author} {\bibfnamefont
  {I.}~\bibnamefont {Hinder}}, \bibinfo {author} {\bibfnamefont
  {S.}~\bibnamefont {Husa}}, \bibinfo {author} {\bibfnamefont {S.}~\bibnamefont
  {Iyer}}, \bibinfo {author} {\bibfnamefont {T.}~\bibnamefont {Kellermann}},
  \bibinfo {author} {\bibfnamefont {A.}~\bibnamefont {Knapp}}, \bibinfo
  {author} {\bibfnamefont {M.}~\bibnamefont {Koppitz}}, \bibinfo {author}
  {\bibfnamefont {G.}~\bibnamefont {Lanferman}}, \bibinfo {author}
  {\bibfnamefont {F.}~\bibnamefont {Löffler}}, \bibinfo {author}
  {\bibfnamefont {J.}~\bibnamefont {Masso}}, \bibinfo {author} {\bibfnamefont
  {L.}~\bibnamefont {Menger}}, \bibinfo {author} {\bibfnamefont
  {A.}~\bibnamefont {Merzky}}, \bibinfo {author} {\bibfnamefont
  {M.}~\bibnamefont {Miller}}, \bibinfo {author} {\bibfnamefont
  {P.}~\bibnamefont {Moesta}}, \bibinfo {author} {\bibfnamefont
  {P.}~\bibnamefont {Montero}}, \bibinfo {author} {\bibfnamefont
  {B.}~\bibnamefont {Mundim}}, \bibinfo {author} {\bibfnamefont
  {A.}~\bibnamefont {Nerozzi}}, \bibinfo {author} {\bibfnamefont
  {C.}~\bibnamefont {Ott}}, \bibinfo {author} {\bibfnamefont {R.}~\bibnamefont
  {Paruchuri}}, \bibinfo {author} {\bibfnamefont {D.}~\bibnamefont {Pollney}},
  \bibinfo {author} {\bibfnamefont {D.}~\bibnamefont {Radice}}, \bibinfo
  {author} {\bibfnamefont {T.}~\bibnamefont {Radke}}, \bibinfo {author}
  {\bibfnamefont {C.}~\bibnamefont {Reisswig}}, \bibinfo {author}
  {\bibfnamefont {L.}~\bibnamefont {Rezzolla}}, \bibinfo {author}
  {\bibfnamefont {D.}~\bibnamefont {Rideout}}, \bibinfo {author} {\bibfnamefont
  {M.}~\bibnamefont {Ripeanu}}, \bibinfo {author} {\bibfnamefont
  {E.}~\bibnamefont {Schnetter}}, \bibinfo {author} {\bibfnamefont
  {B.}~\bibnamefont {Schutz}}, \bibinfo {author} {\bibfnamefont
  {E.}~\bibnamefont {Seidel}}, \bibinfo {author} {\bibfnamefont
  {E.}~\bibnamefont {Seidel}}, \bibinfo {author} {\bibfnamefont
  {J.}~\bibnamefont {Shalf}}, \bibinfo {author} {\bibfnamefont
  {U.}~\bibnamefont {Sperhake}}, \bibinfo {author} {\bibfnamefont
  {N.}~\bibnamefont {Stergioulas}}, \bibinfo {author} {\bibfnamefont {W.-M.}\
  \bibnamefont {Suen}}, \bibinfo {author} {\bibfnamefont {B.}~\bibnamefont
  {Szilagyi}}, \bibinfo {author} {\bibfnamefont {R.}~\bibnamefont {Takahashi}},
  \bibinfo {author} {\bibfnamefont {M.}~\bibnamefont {Thomas}}, \bibinfo
  {author} {\bibfnamefont {J.}~\bibnamefont {Thornburg}}, \bibinfo {author}
  {\bibfnamefont {M.}~\bibnamefont {Tobias}}, \bibinfo {author} {\bibfnamefont
  {A.}~\bibnamefont {Tonita}}, \bibinfo {author} {\bibfnamefont
  {P.}~\bibnamefont {Walker}}, \bibinfo {author} {\bibfnamefont {M.-B.}\
  \bibnamefont {Wan}}, \bibinfo {author} {\bibfnamefont {B.}~\bibnamefont
  {Wardell}}, \bibinfo {author} {\bibfnamefont {M.}~\bibnamefont {Zilhão}},
  \bibinfo {author} {\bibfnamefont {B.}~\bibnamefont {Zink}},\ and\ \bibinfo
  {author} {\bibfnamefont {Y.}~\bibnamefont {Zlochower}},\ }\href
  {https://doi.org/10.5281/zenodo.3522086} {\bibinfo {title} {The einstein
  toolkit}} (\bibinfo {year} {2019}),\ \bibinfo {note} {to find out more, visit
  http://einsteintoolkit.org}\BibitemShut {NoStop}%
\bibitem [{\citenamefont {{Thorne}}\ and\ \citenamefont
  {{MacDonald}}(1982)}]{Thorne1982}%
  \BibitemOpen
  \bibfield  {author} {\bibinfo {author} {\bibfnamefont {K.~S.}\ \bibnamefont
  {{Thorne}}}\ and\ \bibinfo {author} {\bibfnamefont {D.}~\bibnamefont
  {{MacDonald}}},\ }\bibfield  {title} {\bibinfo {title} {{Electrodynamics in
  Curved Spacetime - 3+1 Formulation}},\ }\href
  {https://doi.org/10.1093/mnras/198.2.339} {\bibfield  {journal} {\bibinfo
  {journal} {\mnras}\ }\textbf {\bibinfo {volume} {198}},\ \bibinfo {pages}
  {339} (\bibinfo {year} {1982})}\BibitemShut {NoStop}%
\bibitem [{\citenamefont {{Alcubierre}}(2008)}]{Alcubierre:2008it}%
  \BibitemOpen
  \bibfield  {author} {\bibinfo {author} {\bibfnamefont {M.}~\bibnamefont
  {{Alcubierre}}},\ }\href@noop {} {\emph {\bibinfo {title} {{Introduction to
  3+1 Numerical Relativity}}}}\ (\bibinfo  {publisher} {Oxford University
  Press, UK},\ \bibinfo {year} {2008})\BibitemShut {NoStop}%
\bibitem [{\citenamefont {{Baumgarte}}\ and\ \citenamefont
  {{Shapiro}}(2010)}]{Baumgarte:2010nu}%
  \BibitemOpen
  \bibfield  {author} {\bibinfo {author} {\bibfnamefont {T.~W.}\ \bibnamefont
  {{Baumgarte}}}\ and\ \bibinfo {author} {\bibfnamefont {S.~L.}\ \bibnamefont
  {{Shapiro}}},\ }\href@noop {} {\emph {\bibinfo {title} {{Numerical
  Relativity: Solving Einstein's Equations on the Computer}}}}\ (\bibinfo
  {publisher} {Cambridge University Press},\ \bibinfo {year}
  {2010})\BibitemShut {NoStop}%
\bibitem [{\citenamefont {{Shibata}}(2016)}]{Shibata2016b}%
  \BibitemOpen
  \bibfield  {author} {\bibinfo {author} {\bibfnamefont {M.}~\bibnamefont
  {{Shibata}}},\ }\href {https://doi.org/10.1142/9692} {\emph {\bibinfo {title}
  {Numerical Relativity}}}\ (\bibinfo  {publisher} {World Scientific Publishing
  Co},\ \bibinfo {year} {2016})\BibitemShut {NoStop}%
\bibitem [{\citenamefont {{Zilh{\~a}o}}\ \emph {et~al.}(2012)\citenamefont
  {{Zilh{\~a}o}}, \citenamefont {{Cardoso}}, \citenamefont {{Herdeiro}},
  \citenamefont {{Lehner}},\ and\ \citenamefont {{Sperhake}}}]{Zilhao2012}%
  \BibitemOpen
  \bibfield  {author} {\bibinfo {author} {\bibfnamefont {M.}~\bibnamefont
  {{Zilh{\~a}o}}}, \bibinfo {author} {\bibfnamefont {V.}~\bibnamefont
  {{Cardoso}}}, \bibinfo {author} {\bibfnamefont {C.}~\bibnamefont
  {{Herdeiro}}}, \bibinfo {author} {\bibfnamefont {L.}~\bibnamefont
  {{Lehner}}},\ and\ \bibinfo {author} {\bibfnamefont {U.}~\bibnamefont
  {{Sperhake}}},\ }\bibfield  {title} {\bibinfo {title} {{Collisions of charged
  black holes}},\ }\href {https://doi.org/10.1103/PhysRevD.85.124062}
  {\bibfield  {journal} {\bibinfo  {journal} {\prd}\ }\textbf {\bibinfo
  {volume} {85}},\ \bibinfo {eid} {124062} (\bibinfo {year} {2012})},\ \Eprint
  {https://arxiv.org/abs/1205.1063} {arXiv:1205.1063 [gr-qc]} \BibitemShut
  {NoStop}%
\bibitem [{\citenamefont {{Zilh{\~a}o}}\ \emph {et~al.}(2014)\citenamefont
  {{Zilh{\~a}o}}, \citenamefont {{Cardoso}}, \citenamefont {{Herdeiro}},
  \citenamefont {{Lehner}},\ and\ \citenamefont {{Sperhake}}}]{Zilhao2013}%
  \BibitemOpen
  \bibfield  {author} {\bibinfo {author} {\bibfnamefont {M.}~\bibnamefont
  {{Zilh{\~a}o}}}, \bibinfo {author} {\bibfnamefont {V.}~\bibnamefont
  {{Cardoso}}}, \bibinfo {author} {\bibfnamefont {C.}~\bibnamefont
  {{Herdeiro}}}, \bibinfo {author} {\bibfnamefont {L.}~\bibnamefont
  {{Lehner}}},\ and\ \bibinfo {author} {\bibfnamefont {U.}~\bibnamefont
  {{Sperhake}}},\ }\bibfield  {title} {\bibinfo {title} {{Collisions of
  oppositely charged black holes}},\ }\href
  {https://doi.org/10.1103/PhysRevD.89.044008} {\bibfield  {journal} {\bibinfo
  {journal} {\prd}\ }\textbf {\bibinfo {volume} {89}},\ \bibinfo {eid} {044008}
  (\bibinfo {year} {2014})},\ \Eprint {https://arxiv.org/abs/1311.6483}
  {arXiv:1311.6483 [gr-qc]} \BibitemShut {NoStop}%
\bibitem [{\citenamefont {{Bozzola}}\ and\ \citenamefont
  {{Paschalidis}}(2019)}]{Bozzola2019}%
  \BibitemOpen
  \bibfield  {author} {\bibinfo {author} {\bibfnamefont {G.}~\bibnamefont
  {{Bozzola}}}\ and\ \bibinfo {author} {\bibfnamefont {V.}~\bibnamefont
  {{Paschalidis}}},\ }\bibfield  {title} {\bibinfo {title} {{Initial data for
  general relativistic simulations of multiple electrically charged black holes
  with linear and angular momenta}},\ }\href
  {https://doi.org/10.1103/PhysRevD.99.104044} {\bibfield  {journal} {\bibinfo
  {journal} {\prd}\ }\textbf {\bibinfo {volume} {99}},\ \bibinfo {eid} {104044}
  (\bibinfo {year} {2019})},\ \Eprint {https://arxiv.org/abs/1903.01036}
  {arXiv:1903.01036 [gr-qc]} \BibitemShut {NoStop}%
\bibitem [{\citenamefont {Bowen}\ and\ \citenamefont
  {York}(1980)}]{Bowen:1980yu}%
  \BibitemOpen
  \bibfield  {author} {\bibinfo {author} {\bibfnamefont {J.~M.}\ \bibnamefont
  {Bowen}}\ and\ \bibinfo {author} {\bibfnamefont {J.}~\bibnamefont {York},
  \bibfnamefont {James~W.}},\ }\bibfield  {title} {\bibinfo {title} {{Time
  asymmetric initial data for black holes and black hole collisions}},\ }\href
  {https://doi.org/10.1103/PhysRevD.21.2047} {\bibfield  {journal} {\bibinfo
  {journal} {Phys. Rev. D}\ }\textbf {\bibinfo {volume} {21}},\ \bibinfo
  {pages} {2047} (\bibinfo {year} {1980})}\BibitemShut {NoStop}%
\bibitem [{\citenamefont {{Bowen}}(1985)}]{Bowen1985}%
  \BibitemOpen
  \bibfield  {author} {\bibinfo {author} {\bibfnamefont {J.~M.}\ \bibnamefont
  {{Bowen}}},\ }\bibfield  {title} {\bibinfo {title} {{Inversion symmetric
  initial data for N charged black holes.}},\ }\href
  {https://doi.org/10.1016/S0003-4916(85)80003-8} {\bibfield  {journal}
  {\bibinfo  {journal} {Annals of Physics}\ }\textbf {\bibinfo {volume}
  {165}},\ \bibinfo {pages} {17} (\bibinfo {year} {1985})}\BibitemShut
  {NoStop}%
\bibitem [{\citenamefont {{Alcubierre}}\ \emph {et~al.}(2009)\citenamefont
  {{Alcubierre}}, \citenamefont {{Degollado}},\ and\ \citenamefont
  {{Salgado}}}]{Alcubierre2009}%
  \BibitemOpen
  \bibfield  {author} {\bibinfo {author} {\bibfnamefont {M.}~\bibnamefont
  {{Alcubierre}}}, \bibinfo {author} {\bibfnamefont {J.~C.}\ \bibnamefont
  {{Degollado}}},\ and\ \bibinfo {author} {\bibfnamefont {M.}~\bibnamefont
  {{Salgado}}},\ }\bibfield  {title} {\bibinfo {title} {{Einstein-Maxwell
  system in 3+1 form and initial data for multiple charged black holes}},\
  }\href {https://doi.org/10.1103/PhysRevD.80.104022} {\bibfield  {journal}
  {\bibinfo  {journal} {\prd}\ }\textbf {\bibinfo {volume} {80}},\ \bibinfo
  {eid} {104022} (\bibinfo {year} {2009})},\ \Eprint
  {https://arxiv.org/abs/0907.1151} {arXiv:0907.1151 [gr-qc]} \BibitemShut
  {NoStop}%
\bibitem [{\citenamefont {{Witek}}\ and\ \citenamefont
  {{Zilh{\~a}o}}(2015)}]{canudacode}%
  \BibitemOpen
  \bibfield  {author} {\bibinfo {author} {\bibfnamefont {H.}~\bibnamefont
  {{Witek}}}\ and\ \bibinfo {author} {\bibfnamefont {M.}~\bibnamefont
  {{Zilh{\~a}o}}},\ }\href {https://bitbucket.org/canuda/} {\bibinfo {title}
  {Canuda code}} (\bibinfo {year} {2015}),\ \bibinfo {note}
  {\url{https://bitbucket.org/canuda/}}\BibitemShut {NoStop}%
\bibitem [{\citenamefont {Witek}\ \emph {et~al.}(2020)\citenamefont {Witek},
  \citenamefont {Zilhao}, \citenamefont {Ficarra},\ and\ \citenamefont
  {Elley}}]{canuda}%
  \BibitemOpen
  \bibfield  {author} {\bibinfo {author} {\bibfnamefont {H.}~\bibnamefont
  {Witek}}, \bibinfo {author} {\bibfnamefont {M.}~\bibnamefont {Zilhao}},
  \bibinfo {author} {\bibfnamefont {G.}~\bibnamefont {Ficarra}},\ and\ \bibinfo
  {author} {\bibfnamefont {M.}~\bibnamefont {Elley}},\ }\href
  {https://doi.org/10.5281/zenodo.3565475} {\bibinfo {title} {{Canuda: a public
  numerical relativity library to probe fundamental physics}}} (\bibinfo {year}
  {2020})\BibitemShut {NoStop}%
\bibitem [{\citenamefont {{Sperhake}}(2007)}]{Sperhake2007}%
  \BibitemOpen
  \bibfield  {author} {\bibinfo {author} {\bibfnamefont {U.}~\bibnamefont
  {{Sperhake}}},\ }\bibfield  {title} {\bibinfo {title} {{Binary black-hole
  evolutions of excision and puncture data}},\ }\href
  {https://doi.org/10.1103/PhysRevD.76.104015} {\bibfield  {journal} {\bibinfo
  {journal} {\prd}\ }\textbf {\bibinfo {volume} {76}},\ \bibinfo {eid} {104015}
  (\bibinfo {year} {2007})},\ \Eprint {https://arxiv.org/abs/gr-qc/0606079}
  {gr-qc/0606079} \BibitemShut {NoStop}%
\bibitem [{\citenamefont {{Zilh{\~a}o}}\ \emph {et~al.}(2015)\citenamefont
  {{Zilh{\~a}o}}, \citenamefont {{Witek}},\ and\ \citenamefont
  {{Cardoso}}}]{Zilhao2015}%
  \BibitemOpen
  \bibfield  {author} {\bibinfo {author} {\bibfnamefont {M.}~\bibnamefont
  {{Zilh{\~a}o}}}, \bibinfo {author} {\bibfnamefont {H.}~\bibnamefont
  {{Witek}}},\ and\ \bibinfo {author} {\bibfnamefont {V.}~\bibnamefont
  {{Cardoso}}},\ }\bibfield  {title} {\bibinfo {title} {{Nonlinear interactions
  between black holes and Proca fields}},\ }\href
  {https://doi.org/10.1088/0264-9381/32/23/234003} {\bibfield  {journal}
  {\bibinfo  {journal} {\cqg}\ }\textbf {\bibinfo {volume} {32}},\ \bibinfo
  {eid} {234003} (\bibinfo {year} {2015})},\ \Eprint
  {https://arxiv.org/abs/1505.00797} {arXiv:1505.00797 [gr-qc]} \BibitemShut
  {NoStop}%
\bibitem [{\citenamefont {Shibata}\ and\ \citenamefont
  {Nakamura}(1995)}]{Shibata1995}%
  \BibitemOpen
  \bibfield  {author} {\bibinfo {author} {\bibfnamefont {M.}~\bibnamefont
  {Shibata}}\ and\ \bibinfo {author} {\bibfnamefont {T.}~\bibnamefont
  {Nakamura}},\ }\bibfield  {title} {\bibinfo {title} {Evolution of
  three-dimensional gravitational waves: Harmonic slicing case},\ }\href
  {https://doi.org/10.1103/PhysRevD.52.5428} {\bibfield  {journal} {\bibinfo
  {journal} {Phys. Rev. D}\ }\textbf {\bibinfo {volume} {52}},\ \bibinfo
  {pages} {5428} (\bibinfo {year} {1995})}\BibitemShut {NoStop}%
\bibitem [{\citenamefont {Baumgarte}\ and\ \citenamefont
  {Shapiro}(1998)}]{Baumgarte1998}%
  \BibitemOpen
  \bibfield  {author} {\bibinfo {author} {\bibfnamefont {T.~W.}\ \bibnamefont
  {Baumgarte}}\ and\ \bibinfo {author} {\bibfnamefont {S.~L.}\ \bibnamefont
  {Shapiro}},\ }\bibfield  {title} {\bibinfo {title} {Numerical integration of
  einstein's field equations},\ }\href
  {https://doi.org/10.1103/PhysRevD.59.024007} {\bibfield  {journal} {\bibinfo
  {journal} {Phys. Rev. D}\ }\textbf {\bibinfo {volume} {59}},\ \bibinfo
  {pages} {024007} (\bibinfo {year} {1998})}\BibitemShut {NoStop}%
\bibitem [{\citenamefont {Schnetter}\ \emph {et~al.}(2004)\citenamefont
  {Schnetter}, \citenamefont {Hawley},\ and\ \citenamefont
  {Hawke}}]{Schnetter:2003rb}%
  \BibitemOpen
  \bibfield  {author} {\bibinfo {author} {\bibfnamefont {E.}~\bibnamefont
  {Schnetter}}, \bibinfo {author} {\bibfnamefont {S.~H.}\ \bibnamefont
  {Hawley}},\ and\ \bibinfo {author} {\bibfnamefont {I.}~\bibnamefont
  {Hawke}},\ }\bibfield  {title} {\bibinfo {title} {{Evolutions in 3-D
  numerical relativity using fixed mesh refinement}},\ }\href
  {https://doi.org/10.1088/0264-9381/21/6/014} {\bibfield  {journal} {\bibinfo
  {journal} {Class. Quantum Grav.}\ }\textbf {\bibinfo {volume} {21}},\
  \bibinfo {pages} {1465} (\bibinfo {year} {2004})},\ \Eprint
  {https://arxiv.org/abs/arXiv:gr-qc/0310042} {arXiv:gr-qc/0310042}
  \BibitemShut {NoStop}%
\bibitem [{\citenamefont {{Newman}}\ and\ \citenamefont
  {{Penrose}}(1962)}]{Newman1962}%
  \BibitemOpen
  \bibfield  {author} {\bibinfo {author} {\bibfnamefont {E.}~\bibnamefont
  {{Newman}}}\ and\ \bibinfo {author} {\bibfnamefont {R.}~\bibnamefont
  {{Penrose}}},\ }\bibfield  {title} {\bibinfo {title} {{An Approach to
  Gravitational Radiation by a Method of Spin Coefficients}},\ }\href
  {https://doi.org/10.1063/1.1724257} {\bibfield  {journal} {\bibinfo
  {journal} {Journal of Mathematical Physics}\ }\textbf {\bibinfo {volume}
  {3}},\ \bibinfo {pages} {566} (\bibinfo {year} {1962})}\BibitemShut {NoStop}%
\bibitem [{\citenamefont {{Reisswig}}\ and\ \citenamefont
  {{Pollney}}(2011)}]{Reisswig2011}%
  \BibitemOpen
  \bibfield  {author} {\bibinfo {author} {\bibfnamefont {C.}~\bibnamefont
  {{Reisswig}}}\ and\ \bibinfo {author} {\bibfnamefont {D.}~\bibnamefont
  {{Pollney}}},\ }\bibfield  {title} {\bibinfo {title} {{Notes on the
  integration of numerical relativity waveforms}},\ }\href
  {https://doi.org/10.1088/0264-9381/28/19/195015} {\bibfield  {journal}
  {\bibinfo  {journal} {\cqg}\ }\textbf {\bibinfo {volume} {28}},\ \bibinfo
  {eid} {195015} (\bibinfo {year} {2011})},\ \Eprint
  {https://arxiv.org/abs/1006.1632} {arXiv:1006.1632 [gr-qc]} \BibitemShut
  {NoStop}%
\bibitem [{\citenamefont {{Abbott}}\ \emph
  {et~al.}(2016{\natexlab{b}})\citenamefont {{Abbott}}, \citenamefont
  {{Abbott}}, \citenamefont {{Abbott}}, \citenamefont {{Abernathy}},
  \citenamefont {{Acernese}}, \citenamefont {{Ackley}}, \citenamefont
  {{Adams}}, \citenamefont {{Adams}}, \citenamefont {{Addesso}}, \citenamefont
  {{Adhikari}},\ and\ \citenamefont {et~al.}}]{Abbott2016d}%
  \BibitemOpen
  \bibfield  {author} {\bibinfo {author} {\bibfnamefont {B.~P.}\ \bibnamefont
  {{Abbott}}}, \bibinfo {author} {\bibfnamefont {R.}~\bibnamefont {{Abbott}}},
  \bibinfo {author} {\bibfnamefont {T.~D.}\ \bibnamefont {{Abbott}}}, \bibinfo
  {author} {\bibfnamefont {M.~R.}\ \bibnamefont {{Abernathy}}}, \bibinfo
  {author} {\bibfnamefont {F.}~\bibnamefont {{Acernese}}}, \bibinfo {author}
  {\bibfnamefont {K.}~\bibnamefont {{Ackley}}}, \bibinfo {author}
  {\bibfnamefont {C.}~\bibnamefont {{Adams}}}, \bibinfo {author} {\bibfnamefont
  {T.}~\bibnamefont {{Adams}}}, \bibinfo {author} {\bibfnamefont
  {P.}~\bibnamefont {{Addesso}}}, \bibinfo {author} {\bibfnamefont {R.~X.}\
  \bibnamefont {{Adhikari}}},\ and\ \bibinfo {author} {\bibnamefont {et~al.}},\
  }\bibfield  {title} {\bibinfo {title} {{Properties of the Binary Black Hole
  Merger GW150914}},\ }\href {https://doi.org/10.1103/PhysRevLett.116.241102}
  {\bibfield  {journal} {\bibinfo  {journal} {Physical Review Letters}\
  }\textbf {\bibinfo {volume} {116}},\ \bibinfo {eid} {241102} (\bibinfo {year}
  {2016}{\natexlab{b}})},\ \Eprint {https://arxiv.org/abs/1602.03840}
  {arXiv:1602.03840 [gr-qc]} \BibitemShut {NoStop}%
\bibitem [{\citenamefont {{Damour}}\ \emph {et~al.}(1998)\citenamefont
  {{Damour}}, \citenamefont {{Iyer}},\ and\ \citenamefont
  {{Sathyaprakash}}}]{Damour1998}%
  \BibitemOpen
  \bibfield  {author} {\bibinfo {author} {\bibfnamefont {T.}~\bibnamefont
  {{Damour}}}, \bibinfo {author} {\bibfnamefont {B.~R.}\ \bibnamefont
  {{Iyer}}},\ and\ \bibinfo {author} {\bibfnamefont {B.~S.}\ \bibnamefont
  {{Sathyaprakash}}},\ }\bibfield  {title} {\bibinfo {title} {{Improved filters
  for gravitational waves from inspiraling compact binaries}},\ }\href
  {https://doi.org/10.1103/PhysRevD.57.885} {\bibfield  {journal} {\bibinfo
  {journal} {\prd}\ }\textbf {\bibinfo {volume} {57}},\ \bibinfo {pages} {885}
  (\bibinfo {year} {1998})},\ \Eprint {https://arxiv.org/abs/gr-qc/9708034}
  {gr-qc/9708034} \BibitemShut {NoStop}%
\bibitem [{\citenamefont {{Planck Collaboration}}\ \emph
  {et~al.}(2018)\citenamefont {{Planck Collaboration}}, \citenamefont
  {{Aghanim}}, \citenamefont {{Akrami}}, \citenamefont {{Ashdown}},
  \citenamefont {{Aumont}}, \citenamefont {{Baccigalupi}}, \citenamefont
  {{Ballardini}}, \citenamefont {{Banday}}, \citenamefont {{Barreiro}},
  \citenamefont {{Bartolo}}, \citenamefont {{Basak}}, \citenamefont {{Battye}},
  \citenamefont {{Benabed}}, \citenamefont {{Bernard}}, \citenamefont
  {{Bersanelli}}, \citenamefont {{Bielewicz}}, \citenamefont {{Bock}},
  \citenamefont {{Bond}}, \citenamefont {{Borrill}}, \citenamefont {{Bouchet}},
  \citenamefont {{Boulanger}}, \citenamefont {{Bucher}}, \citenamefont
  {{Burigana}}, \citenamefont {{Butler}}, \citenamefont {{Calabrese}},
  \citenamefont {{Cardoso}}, \citenamefont {{Carron}}, \citenamefont
  {{Challinor}}, \citenamefont {{Chiang}}, \citenamefont {{Chluba}},
  \citenamefont {{Colombo}}, \citenamefont {{Combet}}, \citenamefont
  {{Contreras}}, \citenamefont {{Crill}}, \citenamefont {{Cuttaia}},
  \citenamefont {{de Bernardis}}, \citenamefont {{de Zotti}}, \citenamefont
  {{Delabrouille}}, \citenamefont {{Delouis}}, \citenamefont {{Di Valentino}},
  \citenamefont {{Diego}}, \citenamefont {{Dor{\'e}}}, \citenamefont
  {{Douspis}}, \citenamefont {{Ducout}}, \citenamefont {{Dupac}}, \citenamefont
  {{Dusini}}, \citenamefont {{Efstathiou}}, \citenamefont {{Elsner}},
  \citenamefont {{En{\ss}lin}}, \citenamefont {{Eriksen}}, \citenamefont
  {{Fantaye}}, \citenamefont {{Farhang}}, \citenamefont {{Fergusson}},
  \citenamefont {{Fernandez-Cobos}}, \citenamefont {{Finelli}}, \citenamefont
  {{Forastieri}}, \citenamefont {{Frailis}}, \citenamefont {{Fraisse}},
  \citenamefont {{Franceschi}}, \citenamefont {{Frolov}}, \citenamefont
  {{Galeotta}}, \citenamefont {{Galli}}, \citenamefont {{Ganga}}, \citenamefont
  {{G{\'e}nova-Santos}}, \citenamefont {{Gerbino}}, \citenamefont {{Ghosh}},
  \citenamefont {{Gonz{\'a}lez-Nuevo}}, \citenamefont {{G{\'o}rski}},
  \citenamefont {{Gratton}}, \citenamefont {{Gruppuso}}, \citenamefont
  {{Gudmundsson}}, \citenamefont {{Hamann}}, \citenamefont {{Handley}},
  \citenamefont {{Hansen}}, \citenamefont {{Herranz}}, \citenamefont
  {{Hildebrandt}}, \citenamefont {{Hivon}}, \citenamefont {{Huang}},
  \citenamefont {{Jaffe}}, \citenamefont {{Jones}}, \citenamefont {{Karakci}},
  \citenamefont {{Keih{\"a}nen}}, \citenamefont {{Keskitalo}}, \citenamefont
  {{Kiiveri}}, \citenamefont {{Kim}}, \citenamefont {{Kisner}}, \citenamefont
  {{Knox}}, \citenamefont {{Krachmalnicoff}}, \citenamefont {{Kunz}},
  \citenamefont {{Kurki-Suonio}}, \citenamefont {{Lagache}}, \citenamefont
  {{Lamarre}}, \citenamefont {{Lasenby}}, \citenamefont {{Lattanzi}},
  \citenamefont {{Lawrence}}, \citenamefont {{Le Jeune}}, \citenamefont
  {{Lemos}}, \citenamefont {{Lesgourgues}}, \citenamefont {{Levrier}},
  \citenamefont {{Lewis}}, \citenamefont {{Liguori}}, \citenamefont {{Lilje}},
  \citenamefont {{Lilley}}, \citenamefont {{Lindholm}}, \citenamefont
  {{L{\'o}pez-Caniego}}, \citenamefont {{Lubin}}, \citenamefont {{Ma}},
  \citenamefont {{Mac{\'\i}as-P{\'e}rez}}, \citenamefont {{Maggio}},
  \citenamefont {{Maino}}, \citenamefont {{Mandolesi}}, \citenamefont
  {{Mangilli}}, \citenamefont {{Marcos-Caballero}}, \citenamefont {{Maris}},
  \citenamefont {{Martin}}, \citenamefont {{Martinelli}}, \citenamefont
  {{Mart{\'\i}nez-Gonz{\'a}lez}}, \citenamefont {{Matarrese}}, \citenamefont
  {{Mauri}}, \citenamefont {{McEwen}}, \citenamefont {{Meinhold}},
  \citenamefont {{Melchiorri}}, \citenamefont {{Mennella}}, \citenamefont
  {{Migliaccio}}, \citenamefont {{Millea}}, \citenamefont {{Mitra}},
  \citenamefont {{Miville-Desch{\^e}nes}}, \citenamefont {{Molinari}},
  \citenamefont {{Montier}}, \citenamefont {{Morgante}}, \citenamefont
  {{Moss}}, \citenamefont {{Natoli}}, \citenamefont {{N{\o}rgaard-Nielsen}},
  \citenamefont {{Pagano}}, \citenamefont {{Paoletti}}, \citenamefont
  {{Partridge}}, \citenamefont {{Patanchon}}, \citenamefont {{Peiris}},
  \citenamefont {{Perrotta}}, \citenamefont {{Pettorino}}, \citenamefont
  {{Piacentini}}, \citenamefont {{Polastri}}, \citenamefont {{Polenta}},
  \citenamefont {{Puget}}, \citenamefont {{Rachen}}, \citenamefont
  {{Reinecke}}, \citenamefont {{Remazeilles}}, \citenamefont {{Renzi}},
  \citenamefont {{Rocha}}, \citenamefont {{Rosset}}, \citenamefont {{Roudier}},
  \citenamefont {{Rubi{\~n}o-Mart{\'\i}n}}, \citenamefont {{Ruiz-Granados}},
  \citenamefont {{Salvati}}, \citenamefont {{Sandri}}, \citenamefont
  {{Savelainen}}, \citenamefont {{Scott}}, \citenamefont {{Shellard}},
  \citenamefont {{Sirignano}}, \citenamefont {{Sirri}}, \citenamefont
  {{Spencer}}, \citenamefont {{Sunyaev}}, \citenamefont {{Suur-Uski}},
  \citenamefont {{Tauber}}, \citenamefont {{Tavagnacco}}, \citenamefont
  {{Tenti}}, \citenamefont {{Toffolatti}}, \citenamefont {{Tomasi}},
  \citenamefont {{Trombetti}}, \citenamefont {{Valenziano}}, \citenamefont
  {{Valiviita}}, \citenamefont {{Van Tent}}, \citenamefont {{Vibert}},
  \citenamefont {{Vielva}}, \citenamefont {{Villa}}, \citenamefont
  {{Vittorio}}, \citenamefont {{Wand elt}}, \citenamefont {{Wehus}},
  \citenamefont {{White}}, \citenamefont {{White}}, \citenamefont {{Zacchei}},\
  and\ \citenamefont {{Zonca}}}]{Planck2018}%
  \BibitemOpen
  \bibfield  {author} {\bibinfo {author} {\bibnamefont {{Planck
  Collaboration}}}, \bibinfo {author} {\bibfnamefont {N.}~\bibnamefont
  {{Aghanim}}}, \bibinfo {author} {\bibfnamefont {Y.}~\bibnamefont {{Akrami}}},
  \bibinfo {author} {\bibfnamefont {M.}~\bibnamefont {{Ashdown}}}, \bibinfo
  {author} {\bibfnamefont {J.}~\bibnamefont {{Aumont}}}, \bibinfo {author}
  {\bibfnamefont {C.}~\bibnamefont {{Baccigalupi}}}, \bibinfo {author}
  {\bibfnamefont {M.}~\bibnamefont {{Ballardini}}}, \bibinfo {author}
  {\bibfnamefont {A.~J.}\ \bibnamefont {{Banday}}}, \bibinfo {author}
  {\bibfnamefont {R.~B.}\ \bibnamefont {{Barreiro}}}, \bibinfo {author}
  {\bibfnamefont {N.}~\bibnamefont {{Bartolo}}}, \bibinfo {author}
  {\bibfnamefont {S.}~\bibnamefont {{Basak}}}, \bibinfo {author} {\bibfnamefont
  {R.}~\bibnamefont {{Battye}}}, \bibinfo {author} {\bibfnamefont
  {K.}~\bibnamefont {{Benabed}}}, \bibinfo {author} {\bibfnamefont {J.~P.}\
  \bibnamefont {{Bernard}}}, \bibinfo {author} {\bibfnamefont {M.}~\bibnamefont
  {{Bersanelli}}}, \bibinfo {author} {\bibfnamefont {P.}~\bibnamefont
  {{Bielewicz}}}, \bibinfo {author} {\bibfnamefont {J.~J.}\ \bibnamefont
  {{Bock}}}, \bibinfo {author} {\bibfnamefont {J.~R.}\ \bibnamefont {{Bond}}},
  \bibinfo {author} {\bibfnamefont {J.}~\bibnamefont {{Borrill}}}, \bibinfo
  {author} {\bibfnamefont {F.~R.}\ \bibnamefont {{Bouchet}}}, \bibinfo {author}
  {\bibfnamefont {F.}~\bibnamefont {{Boulanger}}}, \bibinfo {author}
  {\bibfnamefont {M.}~\bibnamefont {{Bucher}}}, \bibinfo {author}
  {\bibfnamefont {C.}~\bibnamefont {{Burigana}}}, \bibinfo {author}
  {\bibfnamefont {R.~C.}\ \bibnamefont {{Butler}}}, \bibinfo {author}
  {\bibfnamefont {E.}~\bibnamefont {{Calabrese}}}, \bibinfo {author}
  {\bibfnamefont {J.~F.}\ \bibnamefont {{Cardoso}}}, \bibinfo {author}
  {\bibfnamefont {J.}~\bibnamefont {{Carron}}}, \bibinfo {author}
  {\bibfnamefont {A.}~\bibnamefont {{Challinor}}}, \bibinfo {author}
  {\bibfnamefont {H.~C.}\ \bibnamefont {{Chiang}}}, \bibinfo {author}
  {\bibfnamefont {J.}~\bibnamefont {{Chluba}}}, \bibinfo {author}
  {\bibfnamefont {L.~P.~L.}\ \bibnamefont {{Colombo}}}, \bibinfo {author}
  {\bibfnamefont {C.}~\bibnamefont {{Combet}}}, \bibinfo {author}
  {\bibfnamefont {D.}~\bibnamefont {{Contreras}}}, \bibinfo {author}
  {\bibfnamefont {B.~P.}\ \bibnamefont {{Crill}}}, \bibinfo {author}
  {\bibfnamefont {F.}~\bibnamefont {{Cuttaia}}}, \bibinfo {author}
  {\bibfnamefont {P.}~\bibnamefont {{de Bernardis}}}, \bibinfo {author}
  {\bibfnamefont {G.}~\bibnamefont {{de Zotti}}}, \bibinfo {author}
  {\bibfnamefont {J.}~\bibnamefont {{Delabrouille}}}, \bibinfo {author}
  {\bibfnamefont {J.~M.}\ \bibnamefont {{Delouis}}}, \bibinfo {author}
  {\bibfnamefont {E.}~\bibnamefont {{Di Valentino}}}, \bibinfo {author}
  {\bibfnamefont {J.~M.}\ \bibnamefont {{Diego}}}, \bibinfo {author}
  {\bibfnamefont {O.}~\bibnamefont {{Dor{\'e}}}}, \bibinfo {author}
  {\bibfnamefont {M.}~\bibnamefont {{Douspis}}}, \bibinfo {author}
  {\bibfnamefont {A.}~\bibnamefont {{Ducout}}}, \bibinfo {author}
  {\bibfnamefont {X.}~\bibnamefont {{Dupac}}}, \bibinfo {author} {\bibfnamefont
  {S.}~\bibnamefont {{Dusini}}}, \bibinfo {author} {\bibfnamefont
  {G.}~\bibnamefont {{Efstathiou}}}, \bibinfo {author} {\bibfnamefont
  {F.}~\bibnamefont {{Elsner}}}, \bibinfo {author} {\bibfnamefont {T.~A.}\
  \bibnamefont {{En{\ss}lin}}}, \bibinfo {author} {\bibfnamefont {H.~K.}\
  \bibnamefont {{Eriksen}}}, \bibinfo {author} {\bibfnamefont {Y.}~\bibnamefont
  {{Fantaye}}}, \bibinfo {author} {\bibfnamefont {M.}~\bibnamefont
  {{Farhang}}}, \bibinfo {author} {\bibfnamefont {J.}~\bibnamefont
  {{Fergusson}}}, \bibinfo {author} {\bibfnamefont {R.}~\bibnamefont
  {{Fernandez-Cobos}}}, \bibinfo {author} {\bibfnamefont {F.}~\bibnamefont
  {{Finelli}}}, \bibinfo {author} {\bibfnamefont {F.}~\bibnamefont
  {{Forastieri}}}, \bibinfo {author} {\bibfnamefont {M.}~\bibnamefont
  {{Frailis}}}, \bibinfo {author} {\bibfnamefont {A.~A.}\ \bibnamefont
  {{Fraisse}}}, \bibinfo {author} {\bibfnamefont {E.}~\bibnamefont
  {{Franceschi}}}, \bibinfo {author} {\bibfnamefont {A.}~\bibnamefont
  {{Frolov}}}, \bibinfo {author} {\bibfnamefont {S.}~\bibnamefont
  {{Galeotta}}}, \bibinfo {author} {\bibfnamefont {S.}~\bibnamefont {{Galli}}},
  \bibinfo {author} {\bibfnamefont {K.}~\bibnamefont {{Ganga}}}, \bibinfo
  {author} {\bibfnamefont {R.~T.}\ \bibnamefont {{G{\'e}nova-Santos}}},
  \bibinfo {author} {\bibfnamefont {M.}~\bibnamefont {{Gerbino}}}, \bibinfo
  {author} {\bibfnamefont {T.}~\bibnamefont {{Ghosh}}}, \bibinfo {author}
  {\bibfnamefont {J.}~\bibnamefont {{Gonz{\'a}lez-Nuevo}}}, \bibinfo {author}
  {\bibfnamefont {K.~M.}\ \bibnamefont {{G{\'o}rski}}}, \bibinfo {author}
  {\bibfnamefont {S.}~\bibnamefont {{Gratton}}}, \bibinfo {author}
  {\bibfnamefont {A.}~\bibnamefont {{Gruppuso}}}, \bibinfo {author}
  {\bibfnamefont {J.~E.}\ \bibnamefont {{Gudmundsson}}}, \bibinfo {author}
  {\bibfnamefont {J.}~\bibnamefont {{Hamann}}}, \bibinfo {author}
  {\bibfnamefont {W.}~\bibnamefont {{Handley}}}, \bibinfo {author}
  {\bibfnamefont {F.~K.}\ \bibnamefont {{Hansen}}}, \bibinfo {author}
  {\bibfnamefont {D.}~\bibnamefont {{Herranz}}}, \bibinfo {author}
  {\bibfnamefont {S.~R.}\ \bibnamefont {{Hildebrandt}}}, \bibinfo {author}
  {\bibfnamefont {E.}~\bibnamefont {{Hivon}}}, \bibinfo {author} {\bibfnamefont
  {Z.}~\bibnamefont {{Huang}}}, \bibinfo {author} {\bibfnamefont {A.~H.}\
  \bibnamefont {{Jaffe}}}, \bibinfo {author} {\bibfnamefont {W.~C.}\
  \bibnamefont {{Jones}}}, \bibinfo {author} {\bibfnamefont {A.}~\bibnamefont
  {{Karakci}}}, \bibinfo {author} {\bibfnamefont {E.}~\bibnamefont
  {{Keih{\"a}nen}}}, \bibinfo {author} {\bibfnamefont {R.}~\bibnamefont
  {{Keskitalo}}}, \bibinfo {author} {\bibfnamefont {K.}~\bibnamefont
  {{Kiiveri}}}, \bibinfo {author} {\bibfnamefont {J.}~\bibnamefont {{Kim}}},
  \bibinfo {author} {\bibfnamefont {T.~S.}\ \bibnamefont {{Kisner}}}, \bibinfo
  {author} {\bibfnamefont {L.}~\bibnamefont {{Knox}}}, \bibinfo {author}
  {\bibfnamefont {N.}~\bibnamefont {{Krachmalnicoff}}}, \bibinfo {author}
  {\bibfnamefont {M.}~\bibnamefont {{Kunz}}}, \bibinfo {author} {\bibfnamefont
  {H.}~\bibnamefont {{Kurki-Suonio}}}, \bibinfo {author} {\bibfnamefont
  {G.}~\bibnamefont {{Lagache}}}, \bibinfo {author} {\bibfnamefont {J.~M.}\
  \bibnamefont {{Lamarre}}}, \bibinfo {author} {\bibfnamefont {A.}~\bibnamefont
  {{Lasenby}}}, \bibinfo {author} {\bibfnamefont {M.}~\bibnamefont
  {{Lattanzi}}}, \bibinfo {author} {\bibfnamefont {C.~R.}\ \bibnamefont
  {{Lawrence}}}, \bibinfo {author} {\bibfnamefont {M.}~\bibnamefont {{Le
  Jeune}}}, \bibinfo {author} {\bibfnamefont {P.}~\bibnamefont {{Lemos}}},
  \bibinfo {author} {\bibfnamefont {J.}~\bibnamefont {{Lesgourgues}}}, \bibinfo
  {author} {\bibfnamefont {F.}~\bibnamefont {{Levrier}}}, \bibinfo {author}
  {\bibfnamefont {A.}~\bibnamefont {{Lewis}}}, \bibinfo {author} {\bibfnamefont
  {M.}~\bibnamefont {{Liguori}}}, \bibinfo {author} {\bibfnamefont {P.~B.}\
  \bibnamefont {{Lilje}}}, \bibinfo {author} {\bibfnamefont {M.}~\bibnamefont
  {{Lilley}}}, \bibinfo {author} {\bibfnamefont {V.}~\bibnamefont
  {{Lindholm}}}, \bibinfo {author} {\bibfnamefont {M.}~\bibnamefont
  {{L{\'o}pez-Caniego}}}, \bibinfo {author} {\bibfnamefont {P.~M.}\
  \bibnamefont {{Lubin}}}, \bibinfo {author} {\bibfnamefont {Y.~Z.}\
  \bibnamefont {{Ma}}}, \bibinfo {author} {\bibfnamefont {J.~F.}\ \bibnamefont
  {{Mac{\'\i}as-P{\'e}rez}}}, \bibinfo {author} {\bibfnamefont
  {G.}~\bibnamefont {{Maggio}}}, \bibinfo {author} {\bibfnamefont
  {D.}~\bibnamefont {{Maino}}}, \bibinfo {author} {\bibfnamefont
  {N.}~\bibnamefont {{Mandolesi}}}, \bibinfo {author} {\bibfnamefont
  {A.}~\bibnamefont {{Mangilli}}}, \bibinfo {author} {\bibfnamefont
  {A.}~\bibnamefont {{Marcos-Caballero}}}, \bibinfo {author} {\bibfnamefont
  {M.}~\bibnamefont {{Maris}}}, \bibinfo {author} {\bibfnamefont {P.~G.}\
  \bibnamefont {{Martin}}}, \bibinfo {author} {\bibfnamefont {M.}~\bibnamefont
  {{Martinelli}}}, \bibinfo {author} {\bibfnamefont {E.}~\bibnamefont
  {{Mart{\'\i}nez-Gonz{\'a}lez}}}, \bibinfo {author} {\bibfnamefont
  {S.}~\bibnamefont {{Matarrese}}}, \bibinfo {author} {\bibfnamefont
  {N.}~\bibnamefont {{Mauri}}}, \bibinfo {author} {\bibfnamefont {J.~D.}\
  \bibnamefont {{McEwen}}}, \bibinfo {author} {\bibfnamefont {P.~R.}\
  \bibnamefont {{Meinhold}}}, \bibinfo {author} {\bibfnamefont
  {A.}~\bibnamefont {{Melchiorri}}}, \bibinfo {author} {\bibfnamefont
  {A.}~\bibnamefont {{Mennella}}}, \bibinfo {author} {\bibfnamefont
  {M.}~\bibnamefont {{Migliaccio}}}, \bibinfo {author} {\bibfnamefont
  {M.}~\bibnamefont {{Millea}}}, \bibinfo {author} {\bibfnamefont
  {S.}~\bibnamefont {{Mitra}}}, \bibinfo {author} {\bibfnamefont {M.~A.}\
  \bibnamefont {{Miville-Desch{\^e}nes}}}, \bibinfo {author} {\bibfnamefont
  {D.}~\bibnamefont {{Molinari}}}, \bibinfo {author} {\bibfnamefont
  {L.}~\bibnamefont {{Montier}}}, \bibinfo {author} {\bibfnamefont
  {G.}~\bibnamefont {{Morgante}}}, \bibinfo {author} {\bibfnamefont
  {A.}~\bibnamefont {{Moss}}}, \bibinfo {author} {\bibfnamefont
  {P.}~\bibnamefont {{Natoli}}}, \bibinfo {author} {\bibfnamefont {H.~U.}\
  \bibnamefont {{N{\o}rgaard-Nielsen}}}, \bibinfo {author} {\bibfnamefont
  {L.}~\bibnamefont {{Pagano}}}, \bibinfo {author} {\bibfnamefont
  {D.}~\bibnamefont {{Paoletti}}}, \bibinfo {author} {\bibfnamefont
  {B.}~\bibnamefont {{Partridge}}}, \bibinfo {author} {\bibfnamefont
  {G.}~\bibnamefont {{Patanchon}}}, \bibinfo {author} {\bibfnamefont {H.~V.}\
  \bibnamefont {{Peiris}}}, \bibinfo {author} {\bibfnamefont {F.}~\bibnamefont
  {{Perrotta}}}, \bibinfo {author} {\bibfnamefont {V.}~\bibnamefont
  {{Pettorino}}}, \bibinfo {author} {\bibfnamefont {F.}~\bibnamefont
  {{Piacentini}}}, \bibinfo {author} {\bibfnamefont {L.}~\bibnamefont
  {{Polastri}}}, \bibinfo {author} {\bibfnamefont {G.}~\bibnamefont
  {{Polenta}}}, \bibinfo {author} {\bibfnamefont {J.~L.}\ \bibnamefont
  {{Puget}}}, \bibinfo {author} {\bibfnamefont {J.~P.}\ \bibnamefont
  {{Rachen}}}, \bibinfo {author} {\bibfnamefont {M.}~\bibnamefont
  {{Reinecke}}}, \bibinfo {author} {\bibfnamefont {M.}~\bibnamefont
  {{Remazeilles}}}, \bibinfo {author} {\bibfnamefont {A.}~\bibnamefont
  {{Renzi}}}, \bibinfo {author} {\bibfnamefont {G.}~\bibnamefont {{Rocha}}},
  \bibinfo {author} {\bibfnamefont {C.}~\bibnamefont {{Rosset}}}, \bibinfo
  {author} {\bibfnamefont {G.}~\bibnamefont {{Roudier}}}, \bibinfo {author}
  {\bibfnamefont {J.~A.}\ \bibnamefont {{Rubi{\~n}o-Mart{\'\i}n}}}, \bibinfo
  {author} {\bibfnamefont {B.}~\bibnamefont {{Ruiz-Granados}}}, \bibinfo
  {author} {\bibfnamefont {L.}~\bibnamefont {{Salvati}}}, \bibinfo {author}
  {\bibfnamefont {M.}~\bibnamefont {{Sandri}}}, \bibinfo {author}
  {\bibfnamefont {M.}~\bibnamefont {{Savelainen}}}, \bibinfo {author}
  {\bibfnamefont {D.}~\bibnamefont {{Scott}}}, \bibinfo {author} {\bibfnamefont
  {E.~P.~S.}\ \bibnamefont {{Shellard}}}, \bibinfo {author} {\bibfnamefont
  {C.}~\bibnamefont {{Sirignano}}}, \bibinfo {author} {\bibfnamefont
  {G.}~\bibnamefont {{Sirri}}}, \bibinfo {author} {\bibfnamefont {L.~D.}\
  \bibnamefont {{Spencer}}}, \bibinfo {author} {\bibfnamefont {R.}~\bibnamefont
  {{Sunyaev}}}, \bibinfo {author} {\bibfnamefont {A.~S.}\ \bibnamefont
  {{Suur-Uski}}}, \bibinfo {author} {\bibfnamefont {J.~A.}\ \bibnamefont
  {{Tauber}}}, \bibinfo {author} {\bibfnamefont {D.}~\bibnamefont
  {{Tavagnacco}}}, \bibinfo {author} {\bibfnamefont {M.}~\bibnamefont
  {{Tenti}}}, \bibinfo {author} {\bibfnamefont {L.}~\bibnamefont
  {{Toffolatti}}}, \bibinfo {author} {\bibfnamefont {M.}~\bibnamefont
  {{Tomasi}}}, \bibinfo {author} {\bibfnamefont {T.}~\bibnamefont
  {{Trombetti}}}, \bibinfo {author} {\bibfnamefont {L.}~\bibnamefont
  {{Valenziano}}}, \bibinfo {author} {\bibfnamefont {J.}~\bibnamefont
  {{Valiviita}}}, \bibinfo {author} {\bibfnamefont {B.}~\bibnamefont {{Van
  Tent}}}, \bibinfo {author} {\bibfnamefont {L.}~\bibnamefont {{Vibert}}},
  \bibinfo {author} {\bibfnamefont {P.}~\bibnamefont {{Vielva}}}, \bibinfo
  {author} {\bibfnamefont {F.}~\bibnamefont {{Villa}}}, \bibinfo {author}
  {\bibfnamefont {N.}~\bibnamefont {{Vittorio}}}, \bibinfo {author}
  {\bibfnamefont {B.~D.}\ \bibnamefont {{Wand elt}}}, \bibinfo {author}
  {\bibfnamefont {I.~K.}\ \bibnamefont {{Wehus}}}, \bibinfo {author}
  {\bibfnamefont {M.}~\bibnamefont {{White}}}, \bibinfo {author} {\bibfnamefont
  {S.~D.~M.}\ \bibnamefont {{White}}}, \bibinfo {author} {\bibfnamefont
  {A.}~\bibnamefont {{Zacchei}}},\ and\ \bibinfo {author} {\bibfnamefont
  {A.}~\bibnamefont {{Zonca}}},\ }\bibfield  {title} {\bibinfo {title} {{Planck
  2018 results. VI. Cosmological parameters}},\ }\href@noop {} {\bibfield
  {journal} {\bibinfo  {journal} {arXiv e-prints}\ ,\ \bibinfo {eid}
  {arXiv:1807.06209}} (\bibinfo {year} {2018})},\ \Eprint
  {https://arxiv.org/abs/1807.06209} {arXiv:1807.06209 [astro-ph.CO]}
  \BibitemShut {NoStop}%
\bibitem [{\citenamefont {{LIGO Scientific Collaboration}}(2009)}]{ligozdhp}%
  \BibitemOpen
  \bibfield  {author} {\bibinfo {author} {\bibnamefont {{LIGO Scientific
  Collaboration}}},\ }\href {https://dcc.ligo.org/LIGO-T0900288/public}
  {\bibinfo {title} {Advanced ligo anticipated sensitivity curves}} (\bibinfo
  {year} {2009})\BibitemShut {NoStop}%
\bibitem [{\citenamefont {{Abbott}}\ \emph
  {et~al.}(2016{\natexlab{c}})\citenamefont {{Abbott}}, \citenamefont
  {{Abbott}}, \citenamefont {{Abbott}}, \citenamefont {{Abernathy}},
  \citenamefont {{Acernese}}, \citenamefont {{Ackley}}, \citenamefont
  {{Adams}}, \citenamefont {{Adams}}, \citenamefont {{Addesso}}, \citenamefont
  {{Adhikari}},\ and\ \citenamefont {et~al.}}]{Abbott2016o}%
  \BibitemOpen
  \bibfield  {author} {\bibinfo {author} {\bibfnamefont {B.~P.}\ \bibnamefont
  {{Abbott}}}, \bibinfo {author} {\bibfnamefont {R.}~\bibnamefont {{Abbott}}},
  \bibinfo {author} {\bibfnamefont {T.~D.}\ \bibnamefont {{Abbott}}}, \bibinfo
  {author} {\bibfnamefont {M.~R.}\ \bibnamefont {{Abernathy}}}, \bibinfo
  {author} {\bibfnamefont {F.}~\bibnamefont {{Acernese}}}, \bibinfo {author}
  {\bibfnamefont {K.}~\bibnamefont {{Ackley}}}, \bibinfo {author}
  {\bibfnamefont {C.}~\bibnamefont {{Adams}}}, \bibinfo {author} {\bibfnamefont
  {T.}~\bibnamefont {{Adams}}}, \bibinfo {author} {\bibfnamefont
  {P.}~\bibnamefont {{Addesso}}}, \bibinfo {author} {\bibfnamefont {R.~X.}\
  \bibnamefont {{Adhikari}}},\ and\ \bibinfo {author} {\bibnamefont {et~al.}},\
  }\bibfield  {title} {\bibinfo {title} {{Tests of General Relativity with
  GW150914}},\ }\href {https://doi.org/10.1103/PhysRevLett.116.221101}
  {\bibfield  {journal} {\bibinfo  {journal} {Physical Review Letters}\
  }\textbf {\bibinfo {volume} {116}},\ \bibinfo {eid} {221101} (\bibinfo {year}
  {2016}{\natexlab{c}})},\ \Eprint {https://arxiv.org/abs/1602.03841}
  {arXiv:1602.03841 [gr-qc]} \BibitemShut {NoStop}%
\bibitem [{\citenamefont {{Jaiakson}}\ \emph {et~al.}(2017)\citenamefont
  {{Jaiakson}}, \citenamefont {{Chatrabhuti}}, \citenamefont {{Evnin}},\ and\
  \citenamefont {{Lehner}}}]{Jaiakson2017}%
  \BibitemOpen
  \bibfield  {author} {\bibinfo {author} {\bibfnamefont {P.}~\bibnamefont
  {{Jaiakson}}}, \bibinfo {author} {\bibfnamefont {A.}~\bibnamefont
  {{Chatrabhuti}}}, \bibinfo {author} {\bibfnamefont {O.}~\bibnamefont
  {{Evnin}}},\ and\ \bibinfo {author} {\bibfnamefont {L.}~\bibnamefont
  {{Lehner}}},\ }\bibfield  {title} {\bibinfo {title} {{Black hole merger
  estimates in Einstein-Maxwell and Einstein-Maxwell-dilaton gravity}},\ }\href
  {https://doi.org/10.1103/PhysRevD.96.044031} {\bibfield  {journal} {\bibinfo
  {journal} {\prd}\ }\textbf {\bibinfo {volume} {96}},\ \bibinfo {eid} {044031}
  (\bibinfo {year} {2017})},\ \Eprint {https://arxiv.org/abs/1706.06519}
  {arXiv:1706.06519 [gr-qc]} \BibitemShut {NoStop}%
\bibitem [{\citenamefont {{Moffat}}(2015)}]{Moffat2015}%
  \BibitemOpen
  \bibfield  {author} {\bibinfo {author} {\bibfnamefont {J.~W.}\ \bibnamefont
  {{Moffat}}},\ }\bibfield  {title} {\bibinfo {title} {{Black holes in modified
  gravity (MOG)}},\ }\href {https://doi.org/10.1140/epjc/s10052-015-3405-x}
  {\bibfield  {journal} {\bibinfo  {journal} {European Physical Journal C}\
  }\textbf {\bibinfo {volume} {75}},\ \bibinfo {eid} {175} (\bibinfo {year}
  {2015})},\ \Eprint {https://arxiv.org/abs/1412.5424} {arXiv:1412.5424
  [gr-qc]} \BibitemShut {NoStop}%
\bibitem [{\citenamefont {Varma}\ \emph {et~al.}(2019)\citenamefont {Varma},
  \citenamefont {Field}, \citenamefont {Scheel}, \citenamefont {Blackman},
  \citenamefont {Kidder},\ and\ \citenamefont {Pfeiffer}}]{Varma:2018mmi}%
  \BibitemOpen
  \bibfield  {author} {\bibinfo {author} {\bibfnamefont {V.}~\bibnamefont
  {Varma}}, \bibinfo {author} {\bibfnamefont {S.~E.}\ \bibnamefont {Field}},
  \bibinfo {author} {\bibfnamefont {M.~A.}\ \bibnamefont {Scheel}}, \bibinfo
  {author} {\bibfnamefont {J.}~\bibnamefont {Blackman}}, \bibinfo {author}
  {\bibfnamefont {L.~E.}\ \bibnamefont {Kidder}},\ and\ \bibinfo {author}
  {\bibfnamefont {H.~P.}\ \bibnamefont {Pfeiffer}},\ }\bibfield  {title}
  {\bibinfo {title} {{Surrogate model of hybridized numerical relativity binary
  black hole waveforms}},\ }\href {https://doi.org/10.1103/PhysRevD.99.064045}
  {\bibfield  {journal} {\bibinfo  {journal} {Phys. Rev. D}\ }\textbf {\bibinfo
  {volume} {99}},\ \bibinfo {pages} {064045} (\bibinfo {year} {2019})},\
  \Eprint {https://arxiv.org/abs/1812.07865} {arXiv:1812.07865 [gr-qc]}
  \BibitemShut {NoStop}%
\bibitem [{\citenamefont {{Pfeiffer}}\ \emph {et~al.}(2007)\citenamefont
  {{Pfeiffer}}, \citenamefont {{Brown}}, \citenamefont {{Kidder}},
  \citenamefont {{Lindblom}}, \citenamefont {{Lovelace}},\ and\ \citenamefont
  {{Scheel}}}]{Pfeiffer2007}%
  \BibitemOpen
  \bibfield  {author} {\bibinfo {author} {\bibfnamefont {H.~P.}\ \bibnamefont
  {{Pfeiffer}}}, \bibinfo {author} {\bibfnamefont {D.~A.}\ \bibnamefont
  {{Brown}}}, \bibinfo {author} {\bibfnamefont {L.~E.}\ \bibnamefont
  {{Kidder}}}, \bibinfo {author} {\bibfnamefont {L.}~\bibnamefont
  {{Lindblom}}}, \bibinfo {author} {\bibfnamefont {G.}~\bibnamefont
  {{Lovelace}}},\ and\ \bibinfo {author} {\bibfnamefont {M.~A.}\ \bibnamefont
  {{Scheel}}},\ }\bibfield  {title} {\bibinfo {title} {{Reducing orbital
  eccentricity in binary black hole simulations}},\ }\href
  {https://doi.org/10.1088/0264-9381/24/12/S06} {\bibfield  {journal} {\bibinfo
   {journal} {Classical and Quantum Gravity}\ }\textbf {\bibinfo {volume}
  {24}},\ \bibinfo {pages} {S59} (\bibinfo {year} {2007})},\ \Eprint
  {https://arxiv.org/abs/gr-qc/0702106} {arXiv:gr-qc/0702106 [gr-qc]}
  \BibitemShut {NoStop}%
\bibitem [{\citenamefont {{Tsokaros}}\ \emph {et~al.}(2019)\citenamefont
  {{Tsokaros}}, \citenamefont {{Ruiz}}, \citenamefont {{Paschalidis}},
  \citenamefont {{Shapiro}},\ and\ \citenamefont {{Ury{\= u}}}}]{Tsokaros2019}%
  \BibitemOpen
  \bibfield  {author} {\bibinfo {author} {\bibfnamefont {A.}~\bibnamefont
  {{Tsokaros}}}, \bibinfo {author} {\bibfnamefont {M.}~\bibnamefont {{Ruiz}}},
  \bibinfo {author} {\bibfnamefont {V.}~\bibnamefont {{Paschalidis}}}, \bibinfo
  {author} {\bibfnamefont {S.~L.}\ \bibnamefont {{Shapiro}}},\ and\ \bibinfo
  {author} {\bibfnamefont {K.}~\bibnamefont {{Ury{\= u}}}},\ }\bibfield
  {title} {\bibinfo {title} {{Effect of spin on the inspiral of binary neutron
  stars}},\ }\href {https://doi.org/10.1103/PhysRevD.100.024061} {\bibfield
  {journal} {\bibinfo  {journal} {\prd}\ }\textbf {\bibinfo {volume} {100}},\
  \bibinfo {eid} {024061} (\bibinfo {year} {2019})},\ \Eprint
  {https://arxiv.org/abs/1906.00011} {arXiv:1906.00011 [gr-qc]} \BibitemShut
  {NoStop}%
\bibitem [{\citenamefont {Thornburg}(1996)}]{Thornburg:1995cp}%
  \BibitemOpen
  \bibfield  {author} {\bibinfo {author} {\bibfnamefont {J.}~\bibnamefont
  {Thornburg}},\ }\bibfield  {title} {\bibinfo {title} {{Finding apparent
  horizons in numerical relativity}},\ }\href
  {https://doi.org/10.1103/PhysRevD.54.4899} {\bibfield  {journal} {\bibinfo
  {journal} {Phys. Rev. D}\ }\textbf {\bibinfo {volume} {54}},\ \bibinfo
  {pages} {4899} (\bibinfo {year} {1996})},\ \Eprint
  {https://arxiv.org/abs/arXiv:gr-qc/9508014} {arXiv:gr-qc/9508014}
  \BibitemShut {NoStop}%
%%CITATION = GR-QC/9508014;%%
\bibitem [{\citenamefont {Thornburg}(2004)}]{Thornburg:2003sf}%
  \BibitemOpen
  \bibfield  {author} {\bibinfo {author} {\bibfnamefont {J.}~\bibnamefont
  {Thornburg}},\ }\bibfield  {title} {\bibinfo {title} {{A Fast
  Apparent-Horizon Finder for 3-Dimensional Cartesian Grids in Numerical
  Relativity}},\ }\href {https://doi.org/10.1088/0264-9381/21/2/026} {\bibfield
   {journal} {\bibinfo  {journal} {Class. Quantum Grav.}\ }\textbf {\bibinfo
  {volume} {21}},\ \bibinfo {pages} {743} (\bibinfo {year} {2004})},\ \Eprint
  {https://arxiv.org/abs/arXiv:gr-qc/0306056} {arXiv:gr-qc/0306056}
  \BibitemShut {NoStop}%
%%CITATION = GR-QC/0306056;%%
\bibitem [{\citenamefont {{Ashtekar}}\ \emph {et~al.}(2000)\citenamefont
  {{Ashtekar}}, \citenamefont {{Beetle}}, \citenamefont {{Dreyer}},
  \citenamefont {{Fairhurst}}, \citenamefont {{Krishnan}}, \citenamefont
  {{Lewandowski}},\ and\ \citenamefont {{Wi{\'s}niewski}}}]{Ashtekar2000}%
  \BibitemOpen
  \bibfield  {author} {\bibinfo {author} {\bibfnamefont {A.}~\bibnamefont
  {{Ashtekar}}}, \bibinfo {author} {\bibfnamefont {C.}~\bibnamefont
  {{Beetle}}}, \bibinfo {author} {\bibfnamefont {O.}~\bibnamefont {{Dreyer}}},
  \bibinfo {author} {\bibfnamefont {S.}~\bibnamefont {{Fairhurst}}}, \bibinfo
  {author} {\bibfnamefont {B.}~\bibnamefont {{Krishnan}}}, \bibinfo {author}
  {\bibfnamefont {J.}~\bibnamefont {{Lewandowski}}},\ and\ \bibinfo {author}
  {\bibfnamefont {J.}~\bibnamefont {{Wi{\'s}niewski}}},\ }\bibfield  {title}
  {\bibinfo {title} {{Generic Isolated Horizons and Their Applications}},\
  }\href {https://doi.org/10.1103/PhysRevLett.85.3564} {\bibfield  {journal}
  {\bibinfo  {journal} {\prl}\ }\textbf {\bibinfo {volume} {85}},\ \bibinfo
  {pages} {3564} (\bibinfo {year} {2000})},\ \Eprint
  {https://arxiv.org/abs/gr-qc/0006006} {gr-qc/0006006} \BibitemShut {NoStop}%
\bibitem [{\citenamefont {{Ashtekar}}\ and\ \citenamefont
  {{Krishnan}}(2004)}]{Ashtekar2004}%
  \BibitemOpen
  \bibfield  {author} {\bibinfo {author} {\bibfnamefont {A.}~\bibnamefont
  {{Ashtekar}}}\ and\ \bibinfo {author} {\bibfnamefont {B.}~\bibnamefont
  {{Krishnan}}},\ }\bibfield  {title} {\bibinfo {title} {{Isolated and
  Dynamical Horizons and Their Applications}},\ }\href
  {https://doi.org/10.12942/lrr-2004-10} {\bibfield  {journal} {\bibinfo
  {journal} {Living Reviews in Relativity}\ }\textbf {\bibinfo {volume} {7}},\
  \bibinfo {eid} {10} (\bibinfo {year} {2004})},\ \Eprint
  {https://arxiv.org/abs/gr-qc/0407042} {gr-qc/0407042} \BibitemShut {NoStop}%
\bibitem [{\citenamefont {Dreyer}\ \emph {et~al.}(2003)\citenamefont {Dreyer},
  \citenamefont {Krishnan}, \citenamefont {Shoemaker},\ and\ \citenamefont
  {Schnetter}}]{Dreyer:2002mx}%
  \BibitemOpen
  \bibfield  {author} {\bibinfo {author} {\bibfnamefont {O.}~\bibnamefont
  {Dreyer}}, \bibinfo {author} {\bibfnamefont {B.}~\bibnamefont {Krishnan}},
  \bibinfo {author} {\bibfnamefont {D.}~\bibnamefont {Shoemaker}},\ and\
  \bibinfo {author} {\bibfnamefont {E.}~\bibnamefont {Schnetter}},\ }\bibfield
  {title} {\bibinfo {title} {{Introduction to isolated horizons in numerical
  relativity}},\ }\href {https://doi.org/10.1103/PhysRevD.67.024018} {\bibfield
   {journal} {\bibinfo  {journal} {Phys. Rev. D}\ }\textbf {\bibinfo {volume}
  {67}},\ \bibinfo {pages} {024018} (\bibinfo {year} {2003})},\ \Eprint
  {https://arxiv.org/abs/arXiv:gr-qc/0206008} {arXiv:gr-qc/0206008}
  \BibitemShut {NoStop}%
\bibitem [{\citenamefont {Alcubierre}\ \emph {et~al.}(2003)\citenamefont
  {Alcubierre}, \citenamefont {Br{\"u}gmann}, \citenamefont {Diener},
  \citenamefont {Koppitz}, \citenamefont {Pollney}, \citenamefont {Seidel},\
  and\ \citenamefont {Takahashi}}]{Alcubierre:2002kk}%
  \BibitemOpen
  \bibfield  {author} {\bibinfo {author} {\bibfnamefont {M.}~\bibnamefont
  {Alcubierre}}, \bibinfo {author} {\bibfnamefont {B.}~\bibnamefont
  {Br{\"u}gmann}}, \bibinfo {author} {\bibfnamefont {P.}~\bibnamefont
  {Diener}}, \bibinfo {author} {\bibfnamefont {M.}~\bibnamefont {Koppitz}},
  \bibinfo {author} {\bibfnamefont {D.}~\bibnamefont {Pollney}}, \bibinfo
  {author} {\bibfnamefont {E.}~\bibnamefont {Seidel}},\ and\ \bibinfo {author}
  {\bibfnamefont {R.}~\bibnamefont {Takahashi}},\ }\bibfield  {title} {\bibinfo
  {title} {{Gauge conditions for long term numerical black hole evolutions
  without excision}},\ }\href {https://doi.org/10.1103/PhysRevD.67.084023}
  {\bibfield  {journal} {\bibinfo  {journal} {Phys. Rev. D}\ }\textbf {\bibinfo
  {volume} {67}},\ \bibinfo {pages} {084023} (\bibinfo {year} {2003})},\
  \Eprint {https://arxiv.org/abs/arXiv:gr-qc/0206072} {arXiv:gr-qc/0206072}
  \BibitemShut {NoStop}%
\bibitem [{\citenamefont {{van Meter}}\ \emph {et~al.}(2006)\citenamefont {{van
  Meter}}, \citenamefont {{Baker}}, \citenamefont {{Koppitz}},\ and\
  \citenamefont {{Choi}}}]{van-Meter2006}%
  \BibitemOpen
  \bibfield  {author} {\bibinfo {author} {\bibfnamefont {J.~R.}\ \bibnamefont
  {{van Meter}}}, \bibinfo {author} {\bibfnamefont {J.~G.}\ \bibnamefont
  {{Baker}}}, \bibinfo {author} {\bibfnamefont {M.}~\bibnamefont {{Koppitz}}},\
  and\ \bibinfo {author} {\bibfnamefont {D.-I.}\ \bibnamefont {{Choi}}},\
  }\bibfield  {title} {\bibinfo {title} {{How to move a black hole without
  excision: Gauge conditions for the numerical evolution of a moving
  puncture}},\ }\href {https://doi.org/10.1103/PhysRevD.73.124011} {\bibfield
  {journal} {\bibinfo  {journal} {\prd}\ }\textbf {\bibinfo {volume} {73}},\
  \bibinfo {eid} {124011} (\bibinfo {year} {2006})},\ \Eprint
  {https://arxiv.org/abs/gr-qc/0605030} {gr-qc/0605030} \BibitemShut {NoStop}%
\bibitem [{\citenamefont {Hinder}\ \emph {et~al.}(2014)\citenamefont {Hinder}
  \emph {et~al.}}]{Hinder:2013oqa}%
  \BibitemOpen
  \bibfield  {author} {\bibinfo {author} {\bibfnamefont {I.}~\bibnamefont
  {Hinder}} \emph {et~al.},\ }\bibfield  {title} {\bibinfo {title}
  {{Error-analysis and comparison to analytical models of numerical waveforms
  produced by the NRAR Collaboration}},\ }\href
  {https://doi.org/10.1088/0264-9381/31/2/025012} {\bibfield  {journal}
  {\bibinfo  {journal} {Class. Quant. Grav.}\ }\textbf {\bibinfo {volume}
  {31}},\ \bibinfo {pages} {025012} (\bibinfo {year} {2014})},\ \Eprint
  {https://arxiv.org/abs/1307.5307} {arXiv:1307.5307 [gr-qc]} \BibitemShut
  {NoStop}%
%%CITATION = ARXIV:1307.5307;%%
\bibitem [{\citenamefont {Kreiss}\ and\ \citenamefont
  {Oliger}(1973)}]{Kreiss:1973aa}%
  \BibitemOpen
  \bibfield  {author} {\bibinfo {author} {\bibfnamefont {H.}~\bibnamefont
  {Kreiss}}\ and\ \bibinfo {author} {\bibfnamefont {J.}~\bibnamefont
  {Oliger}},\ }\href@noop {} {\emph {\bibinfo {title} {{Methods for the
  Approximate Solution of Time Dependent Problems}}}},\ \bibinfo {series}
  {Global Atmospheric Research Programme (GARP): GARP Publication Series},
  Vol.~\bibinfo {volume} {10}\ (\bibinfo  {publisher} {GARP Publication},\
  \bibinfo {year} {1973})\BibitemShut {NoStop}%
\bibitem [{\citenamefont {{Brandt}}\ and\ \citenamefont
  {{Seidel}}(1995)}]{Brandt1995}%
  \BibitemOpen
  \bibfield  {author} {\bibinfo {author} {\bibfnamefont {S.~R.}\ \bibnamefont
  {{Brandt}}}\ and\ \bibinfo {author} {\bibfnamefont {E.}~\bibnamefont
  {{Seidel}}},\ }\bibfield  {title} {\bibinfo {title} {{Evolution of distorted
  rotating black holes. II. Dynamics and analysis}},\ }\href
  {https://doi.org/10.1103/PhysRevD.52.870} {\bibfield  {journal} {\bibinfo
  {journal} {\prd}\ }\textbf {\bibinfo {volume} {52}},\ \bibinfo {pages} {870}
  (\bibinfo {year} {1995})},\ \Eprint {https://arxiv.org/abs/gr-qc/9412073}
  {gr-qc/9412073} \BibitemShut {NoStop}%
\bibitem [{\citenamefont {{Gleiser}}\ \emph {et~al.}(1998)\citenamefont
  {{Gleiser}}, \citenamefont {{Nicasio}}, \citenamefont {{Price}},\ and\
  \citenamefont {{Pullin}}}]{Gleiser1998}%
  \BibitemOpen
  \bibfield  {author} {\bibinfo {author} {\bibfnamefont {R.~J.}\ \bibnamefont
  {{Gleiser}}}, \bibinfo {author} {\bibfnamefont {C.~O.}\ \bibnamefont
  {{Nicasio}}}, \bibinfo {author} {\bibfnamefont {R.~H.}\ \bibnamefont
  {{Price}}},\ and\ \bibinfo {author} {\bibfnamefont {J.}~\bibnamefont
  {{Pullin}}},\ }\bibfield  {title} {\bibinfo {title} {{Evolving the Bowen-York
  initial data for spinning black holes}},\ }\href
  {https://doi.org/10.1103/PhysRevD.57.3401} {\bibfield  {journal} {\bibinfo
  {journal} {\prd}\ }\textbf {\bibinfo {volume} {57}},\ \bibinfo {pages} {3401}
  (\bibinfo {year} {1998})},\ \Eprint {https://arxiv.org/abs/gr-qc/9710096}
  {gr-qc/9710096} \BibitemShut {NoStop}%
\bibitem [{\citenamefont {{Harry}}\ and\ \citenamefont
  {{Fairhurst}}(2011)}]{Harry2011}%
  \BibitemOpen
  \bibfield  {author} {\bibinfo {author} {\bibfnamefont {I.~W.}\ \bibnamefont
  {{Harry}}}\ and\ \bibinfo {author} {\bibfnamefont {S.}~\bibnamefont
  {{Fairhurst}}},\ }\bibfield  {title} {\bibinfo {title} {{Targeted coherent
  search for gravitational waves from compact binary coalescences}},\ }\href
  {https://doi.org/10.1103/PhysRevD.83.084002} {\bibfield  {journal} {\bibinfo
  {journal} {\prd}\ }\textbf {\bibinfo {volume} {83}},\ \bibinfo {eid} {084002}
  (\bibinfo {year} {2011})},\ \Eprint {https://arxiv.org/abs/1012.4939}
  {arXiv:1012.4939 [gr-qc]} \BibitemShut {NoStop}%
\bibitem [{\citenamefont {Collaboration}(2016)}]{GW150914}%
  \BibitemOpen
  \bibfield  {author} {\bibinfo {author} {\bibfnamefont {L.-V.}\ \bibnamefont
  {Collaboration}},\ }\href {https://www.gw-openscience.org/events/GW150914/}
  {\bibinfo {title} {Data release for event gw150914}} (\bibinfo {year}
  {2016})\BibitemShut {NoStop}%
\bibitem [{\citenamefont {{Ruiz}}\ \emph {et~al.}(2008)\citenamefont {{Ruiz}},
  \citenamefont {{Alcubierre}}, \citenamefont {{N{\'u}{\~n}ez}},\ and\
  \citenamefont {{Takahashi}}}]{Ruiz2008}%
  \BibitemOpen
  \bibfield  {author} {\bibinfo {author} {\bibfnamefont {M.}~\bibnamefont
  {{Ruiz}}}, \bibinfo {author} {\bibfnamefont {M.}~\bibnamefont
  {{Alcubierre}}}, \bibinfo {author} {\bibfnamefont {D.}~\bibnamefont
  {{N{\'u}{\~n}ez}}},\ and\ \bibinfo {author} {\bibfnamefont {R.}~\bibnamefont
  {{Takahashi}}},\ }\bibfield  {title} {\bibinfo {title} {{Multiple expansions
  for energy and momenta carried by gravitational waves}},\ }\href
  {https://doi.org/10.1007/s10714-007-0570-8} {\bibfield  {journal} {\bibinfo
  {journal} {General Relativity and Gravitation}\ }\textbf {\bibinfo {volume}
  {40}},\ \bibinfo {pages} {1705} (\bibinfo {year} {2008})},\ \Eprint
  {https://arxiv.org/abs/0707.4654} {arXiv:0707.4654 [gr-qc]} \BibitemShut
  {NoStop}%
\bibitem [{\citenamefont {{Ashtekar}}\ and\ \citenamefont
  {{Bonga}}(2017)}]{Ashtekar2017}%
  \BibitemOpen
  \bibfield  {author} {\bibinfo {author} {\bibfnamefont {A.}~\bibnamefont
  {{Ashtekar}}}\ and\ \bibinfo {author} {\bibfnamefont {B.}~\bibnamefont
  {{Bonga}}},\ }\bibfield  {title} {\bibinfo {title} {{On the ambiguity in the
  notion of transverse traceless modes of gravitational waves}},\ }\href
  {https://doi.org/10.1007/s10714-017-2290-z} {\bibfield  {journal} {\bibinfo
  {journal} {General Relativity and Gravitation}\ }\textbf {\bibinfo {volume}
  {49}},\ \bibinfo {eid} {122} (\bibinfo {year} {2017})},\ \Eprint
  {https://arxiv.org/abs/1707.09914} {arXiv:1707.09914 [gr-qc]} \BibitemShut
  {NoStop}%
\bibitem [{\citenamefont {{Zlochower}}\ \emph {et~al.}(2012)\citenamefont
  {{Zlochower}}, \citenamefont {{Ponce}},\ and\ \citenamefont
  {{Lousto}}}]{Zlochower2012}%
  \BibitemOpen
  \bibfield  {author} {\bibinfo {author} {\bibfnamefont {Y.}~\bibnamefont
  {{Zlochower}}}, \bibinfo {author} {\bibfnamefont {M.}~\bibnamefont
  {{Ponce}}},\ and\ \bibinfo {author} {\bibfnamefont {C.~O.}\ \bibnamefont
  {{Lousto}}},\ }\bibfield  {title} {\bibinfo {title} {{Accuracy issues for
  numerical waveforms}},\ }\href {https://doi.org/10.1103/PhysRevD.86.104056}
  {\bibfield  {journal} {\bibinfo  {journal} {\prd}\ }\textbf {\bibinfo
  {volume} {86}},\ \bibinfo {eid} {104056} (\bibinfo {year} {2012})},\ \Eprint
  {https://arxiv.org/abs/1208.5494} {arXiv:1208.5494 [gr-qc]} \BibitemShut
  {NoStop}%
\bibitem [{\citenamefont {Etienne}\ \emph {et~al.}(2014)\citenamefont
  {Etienne}, \citenamefont {Baker}, \citenamefont {Paschalidis}, \citenamefont
  {Kelly},\ and\ \citenamefont {Shapiro}}]{Etienne:2014tia}%
  \BibitemOpen
  \bibfield  {author} {\bibinfo {author} {\bibfnamefont {Z.~B.}\ \bibnamefont
  {Etienne}}, \bibinfo {author} {\bibfnamefont {J.~G.}\ \bibnamefont {Baker}},
  \bibinfo {author} {\bibfnamefont {V.}~\bibnamefont {Paschalidis}}, \bibinfo
  {author} {\bibfnamefont {B.~J.}\ \bibnamefont {Kelly}},\ and\ \bibinfo
  {author} {\bibfnamefont {S.~L.}\ \bibnamefont {Shapiro}},\ }\bibfield
  {title} {\bibinfo {title} {{Improved Moving Puncture Gauge Conditions for
  Compact Binary Evolutions}},\ }\href
  {https://doi.org/10.1103/PhysRevD.90.064032} {\bibfield  {journal} {\bibinfo
  {journal} {Phys. Rev. D}\ }\textbf {\bibinfo {volume} {90}},\ \bibinfo
  {pages} {064032} (\bibinfo {year} {2014})},\ \Eprint
  {https://arxiv.org/abs/1404.6523} {arXiv:1404.6523 [astro-ph.HE]}
  \BibitemShut {NoStop}%
\bibitem [{\citenamefont {{Huerta}}\ \emph {et~al.}(2014)\citenamefont
  {{Huerta}}, \citenamefont {{Kumar}}, \citenamefont {{McWilliams}},
  \citenamefont {{O'Shaughnessy}},\ and\ \citenamefont {{Yunes}}}]{Huerta2014}%
  \BibitemOpen
  \bibfield  {author} {\bibinfo {author} {\bibfnamefont {E.~A.}\ \bibnamefont
  {{Huerta}}}, \bibinfo {author} {\bibfnamefont {P.}~\bibnamefont {{Kumar}}},
  \bibinfo {author} {\bibfnamefont {S.~T.}\ \bibnamefont {{McWilliams}}},
  \bibinfo {author} {\bibfnamefont {R.}~\bibnamefont {{O'Shaughnessy}}},\ and\
  \bibinfo {author} {\bibfnamefont {N.}~\bibnamefont {{Yunes}}},\ }\bibfield
  {title} {\bibinfo {title} {{Accurate and efficient waveforms for compact
  binaries on eccentric orbits}},\ }\href
  {https://doi.org/10.1103/PhysRevD.90.084016} {\bibfield  {journal} {\bibinfo
  {journal} {\prd}\ }\textbf {\bibinfo {volume} {90}},\ \bibinfo {eid} {084016}
  (\bibinfo {year} {2014})},\ \Eprint {https://arxiv.org/abs/1408.3406}
  {arXiv:1408.3406 [gr-qc]} \BibitemShut {NoStop}%
\bibitem [{\citenamefont {Nitz}\ \emph {et~al.}(2019)\citenamefont {Nitz},
  \citenamefont {Harry}, \citenamefont {Brown}, \citenamefont {Biwer},
  \citenamefont {Willis}, \citenamefont {Canton}, \citenamefont {Capano},
  \citenamefont {Pekowsky}, \citenamefont {Dent}, \citenamefont {Williamson},
  \citenamefont {Cabero}, \citenamefont {De}, \citenamefont {Davies},
  \citenamefont {Macleod}, \citenamefont {Machenschalk}, \citenamefont {Kumar},
  \citenamefont {Reyes}, \citenamefont {Massinger}, \citenamefont {Pannarale},
  \citenamefont {Tápai}, \citenamefont {dfinstad}, \citenamefont {Fairhurst},
  \citenamefont {Khan}, \citenamefont {Nielsen}, \citenamefont {shasvath},
  \citenamefont {Kumar}, \citenamefont {idorrington92}, \citenamefont {Singer},
  \citenamefont {Gabbard},\ and\ \citenamefont {Gadre}}]{PyCBC}%
  \BibitemOpen
  \bibfield  {author} {\bibinfo {author} {\bibfnamefont {A.}~\bibnamefont
  {Nitz}}, \bibinfo {author} {\bibfnamefont {I.}~\bibnamefont {Harry}},
  \bibinfo {author} {\bibfnamefont {D.}~\bibnamefont {Brown}}, \bibinfo
  {author} {\bibfnamefont {C.~M.}\ \bibnamefont {Biwer}}, \bibinfo {author}
  {\bibfnamefont {J.}~\bibnamefont {Willis}}, \bibinfo {author} {\bibfnamefont
  {T.~D.}\ \bibnamefont {Canton}}, \bibinfo {author} {\bibfnamefont
  {C.}~\bibnamefont {Capano}}, \bibinfo {author} {\bibfnamefont
  {L.}~\bibnamefont {Pekowsky}}, \bibinfo {author} {\bibfnamefont
  {T.}~\bibnamefont {Dent}}, \bibinfo {author} {\bibfnamefont {A.~R.}\
  \bibnamefont {Williamson}}, \bibinfo {author} {\bibfnamefont
  {M.}~\bibnamefont {Cabero}}, \bibinfo {author} {\bibfnamefont
  {S.}~\bibnamefont {De}}, \bibinfo {author} {\bibfnamefont {G.}~\bibnamefont
  {Davies}}, \bibinfo {author} {\bibfnamefont {D.}~\bibnamefont {Macleod}},
  \bibinfo {author} {\bibfnamefont {B.}~\bibnamefont {Machenschalk}}, \bibinfo
  {author} {\bibfnamefont {P.}~\bibnamefont {Kumar}}, \bibinfo {author}
  {\bibfnamefont {S.}~\bibnamefont {Reyes}}, \bibinfo {author} {\bibfnamefont
  {T.}~\bibnamefont {Massinger}}, \bibinfo {author} {\bibfnamefont
  {F.}~\bibnamefont {Pannarale}}, \bibinfo {author} {\bibfnamefont
  {M.}~\bibnamefont {Tápai}}, \bibinfo {author} {\bibnamefont {dfinstad}},
  \bibinfo {author} {\bibfnamefont {S.}~\bibnamefont {Fairhurst}}, \bibinfo
  {author} {\bibfnamefont {S.}~\bibnamefont {Khan}}, \bibinfo {author}
  {\bibfnamefont {A.}~\bibnamefont {Nielsen}}, \bibinfo {author} {\bibnamefont
  {shasvath}}, \bibinfo {author} {\bibfnamefont {S.}~\bibnamefont {Kumar}},
  \bibinfo {author} {\bibnamefont {idorrington92}}, \bibinfo {author}
  {\bibfnamefont {L.}~\bibnamefont {Singer}}, \bibinfo {author} {\bibfnamefont
  {H.}~\bibnamefont {Gabbard}},\ and\ \bibinfo {author} {\bibfnamefont
  {B.~U.~V.}\ \bibnamefont {Gadre}},\ }\href
  {https://doi.org/10.5281/zenodo.3483184} {\bibinfo {title} {gwastro/pycbc:
  Pycbc release v1.14.2}} (\bibinfo {year} {2019})\BibitemShut {NoStop}%
\bibitem [{\citenamefont {Biwer}\ \emph {et~al.}(2019)\citenamefont {Biwer},
  \citenamefont {Capano}, \citenamefont {De}, \citenamefont {Cabero},
  \citenamefont {Brown}, \citenamefont {Nitz},\ and\ \citenamefont
  {Raymond}}]{Biwer2019}%
  \BibitemOpen
  \bibfield  {author} {\bibinfo {author} {\bibfnamefont {C.~M.}\ \bibnamefont
  {Biwer}}, \bibinfo {author} {\bibfnamefont {C.~D.}\ \bibnamefont {Capano}},
  \bibinfo {author} {\bibfnamefont {S.}~\bibnamefont {De}}, \bibinfo {author}
  {\bibfnamefont {M.}~\bibnamefont {Cabero}}, \bibinfo {author} {\bibfnamefont
  {D.~A.}\ \bibnamefont {Brown}}, \bibinfo {author} {\bibfnamefont {A.~H.}\
  \bibnamefont {Nitz}},\ and\ \bibinfo {author} {\bibfnamefont
  {V.}~\bibnamefont {Raymond}},\ }\bibfield  {title} {\bibinfo {title} {{PyCBC}
  inference: A python-based parameter estimation toolkit for compact binary
  coalescence signals},\ }\href {https://doi.org/10.1088/1538-3873/aaef0b}
  {\bibfield  {journal} {\bibinfo  {journal} {Publications of the Astronomical
  Society of the Pacific}\ }\textbf {\bibinfo {volume} {131}},\ \bibinfo
  {pages} {024503} (\bibinfo {year} {2019})}\BibitemShut {NoStop}%
\bibitem [{\citenamefont {{Baird}}\ \emph {et~al.}(2013)\citenamefont
  {{Baird}}, \citenamefont {{Fairhurst}}, \citenamefont {{Hannam}},\ and\
  \citenamefont {{Murphy}}}]{Baird2013}%
  \BibitemOpen
  \bibfield  {author} {\bibinfo {author} {\bibfnamefont {E.}~\bibnamefont
  {{Baird}}}, \bibinfo {author} {\bibfnamefont {S.}~\bibnamefont
  {{Fairhurst}}}, \bibinfo {author} {\bibfnamefont {M.}~\bibnamefont
  {{Hannam}}},\ and\ \bibinfo {author} {\bibfnamefont {P.}~\bibnamefont
  {{Murphy}}},\ }\bibfield  {title} {\bibinfo {title} {{Degeneracy between mass
  and spin in black-hole-binary waveforms}},\ }\href
  {https://doi.org/10.1103/PhysRevD.87.024035} {\bibfield  {journal} {\bibinfo
  {journal} {\prd}\ }\textbf {\bibinfo {volume} {87}},\ \bibinfo {eid} {024035}
  (\bibinfo {year} {2013})},\ \Eprint {https://arxiv.org/abs/1211.0546}
  {arXiv:1211.0546 [gr-qc]} \BibitemShut {NoStop}%
\end{thebibliography}%

\end{document}